\documentclass[acmsmall,nonacm]{acmart}

\usepackage{booktabs}
\usepackage{lscape}
\usepackage{longtable}
\usepackage{subfigure}
\usepackage[flushleft]{threeparttable}

\AtBeginDocument{%
  \providecommand\BibTeX{{%
    \normalfont B\kern-0.5em{\scshape i\kern-0.25em b}\kern-0.8em\TeX}}}

\setcopyright{acmcopyright}
\copyrightyear{}
\acmYear{}
\acmDOI{}

\acmJournal{JACM}
\acmVolume{}
\acmNumber{}
\acmArticle{}
\acmMonth{}

\begin{document}
	
	\title{A Taxonomy of Live Migration Management in Cloud Computing}
	
	\author{TianZhang He}
	\authornote{The corresponding author.}
	\email{tianzhangh@student.unimelb.edu.au}
	\orcid{}
	\author{Rajkumar Buyya}
	\email{rbuyya@unimelb.edu.au}
	\affiliation{%
	\institution{Cloud Computing and Distributed Systems (CLOUDS) Laboratory,  School of
		Computing and Information Systems, The University of Melbourne}
	\city{Melbourne}
	\state{VIC}
	\country{Australia}
	\postcode{3000}}

	\renewcommand{\shortauthors}{TZ He and R. Buyya}
	
	\begin{abstract}
Cloud Data Centers have become the backbone infrastructure to provide services. With the emerging edge computing paradigm, computation and networking capabilities have been pushed from clouds to the edge to provide computation, intelligence, networking management with low end-to-end latency.
		Service migration across different computing nodes in edge and cloud computing becomes essential to guarantee the quality of service in the dynamic environment.
		Many studies have been conducted on the dynamic resource management involving migrating Virtual Machines to achieve various objectives, such as load balancing, consolidation, performance, energy-saving, and disaster recovery.
		Some have investigated to improve and predict the performance of single live migration of VM and container.
		Recently, several studies service migration in the edge-centric computing paradigms.
		However, there is a lack of surveys to focus on the live migration management in edge and cloud computing environments.
		We examine the characteristics of each field and conduct a migration management-centric taxonomy and survey. We also identify the gap and research opportunities to guarantee the performance of resource management with live migrations.
	\end{abstract}


	\maketitle
	
	\section{Introduction}
	The emergency of cloud computing has facilitated the dynamic provision in computing, networking, and storage resources to support the services on an on-demand basis~\cite{armbrust2010view,marston2011cloud}. 
	Traditionally, the process directly running on the operating systems is the foundational element to host the service by utilizing the resources. With the development of virtualization, Virtual Machines (VM), as one of the major virtualization technologies to host cloud services, can share computing, networking, and storage resources from the physical machines.
	On the other hand, the container is the emerging virtualization instance to support more elastic services framework due to its flexibility and small footprint~\cite{merkel2014docker,bernstein2014containers,joy2015performance}. Application providers can lease virtualized instances (VMs or containers) from cloud providers with various flavors under different Service Level Agreements (SLAs). Then, the VM or container managers initialize the instances and the cloud broker or orchestrator selects the feasible placement based on the available resources and the allocation policy.
	
	Under highly dynamic environments, cloud providers need to prevent the violation of the Service Level Agreement (SLA) and guarantee the Quality of Service (QoS), such as end-to-end delay, task processing time, etc. Therefore, there have been extensive works ~\cite{zhang2018survey,wang2018survey} focusing on dynamic resource management in performance, accessibility, energy, and economy in order to benefit both cloud computing subscribers and providers.
	Live migration of process, VM, container, or storage is the key feature to support the dynamic resource management in both edge and cloud computing environments. It can migrate and synchronize the running state of the instance, such as VM or container, from one host to another without disrupting the services~\cite{clark2005live}. Live migration provides a generic without any application-specific configuration and management.
	Many works have been focused on the different objectives of resource management through live migration in both cloud~\cite{manvi2014resource,jennings2015resource,zhang2018survey} and edge computing environments~\cite{machen2017live,wang2018survey,hong2019resource,rejiba2019survey}, such as load balancing, over-subscription, consolidation, networking, energy, disaster recovery, and maintenance for hardware and software updates. 
	
	Commercial cloud infrastructure and services providers, such as AWS, Azure, Google, IBM, RedHat, etc, have been integrating live VM and container migration~\cite{redhat-criu,google-e2,borg-criu,ruprecht2018vm}. 
	For example, to make the compute infrastructure cost-effective, reliable, and performant, Google Cloud Engine introduced dynamic resource management for E2 VMs through performance-aware live migration algorithms~\cite{google-e2}.
	Google has been adopted live VM and container migration into its cluster manager~\cite{verma2015large,borg-criu} for the purposes, such as higher priority task preemption, kernel, and firmware software updates, hardware updates, and reallocation for performance and availability. It manages all compute tasks and container clusters with up to tens of thousands of physical machines. A lower bound of 1,000,000 migrations monthly have been performed with 50 ms average downtime during the migration~\cite{ruprecht2018vm}.

	From cloud to edge computing, the processing resources and intelligence have been pushed to the edge of the network to facilitate time-critical services with higher bandwidth and lower latency~\cite{shi2016edge,hu2015mobile}. 
	With the combination of different paradigms, live migration can be performed between edge servers, physical hosts in the LAN network, or different data centers through the WAN~\cite{harney2007efficacy}.
	For example,
	consolidation through Live migrations between different physical hosts within a cloud data center can reduce the overall network communication latency and energy cost of both hosts and network devices~\cite{beloglazov2012managing}.
	Live migrations between data centers through WAN aims to optimize performance~\cite{nagin2011inter,lu2014clique,wood2014cloudnet,ge2014energy}, such as delay, jitter, packet loss rate, response time, as well as energy cost, running cost, regulation adoption, and evacuation before disasters~\cite{tsugawa2012use}.
	The mobility-induced migration~\cite{zhang2018survey, wang2018survey, rejiba2019survey} in edge computing is based on the user position and the coverage of each edge server and its base stations. When the position of the end-user change dramatically, the end-to-end latency will be suffered. As a result, the service may need to be migrated from the previous edge servers to the adjacent one.

	As the state transmission and synchronization is through the network, the performance of live migration heavily relies on the network resource, such as bandwidth and delay. The edge and cloud data center network has been established to provide data transmission for both live migration and service connectivity. However, with the expansion of edge and cloud computing, tens of thousands of nodes connect with each other, which makes it difficult to manage and configure the networking resource at scale. 
	To overcome the network topology complexity, Software-Defined Networking (SDN)~\cite{kirkpatrick2013software,kim2013improving,xia2014survey} is introduced to provide centralized networking management by separating the data and control plane in the traditional network devices. The SDN controllers can dynamically update the knowledge of the whole network topology through the southbound interfaces based on the OpenFlow protocol. It also provides northbound interfaces for high-level networking resource management such as flow routing, network separation, and bandwidth allocation. Therefore, SDN becomes the emerging solution in edge and cloud computing.
	For example, Google has presented its implementation of software-defined inter-data center WAN (B4)~\cite{jain2013b4} to showcase the Software Defined Networking at scale.

	\begin{figure}[t]
		\begin{minipage}{0.59\textwidth}
			\centering
			\includegraphics[width=\linewidth]{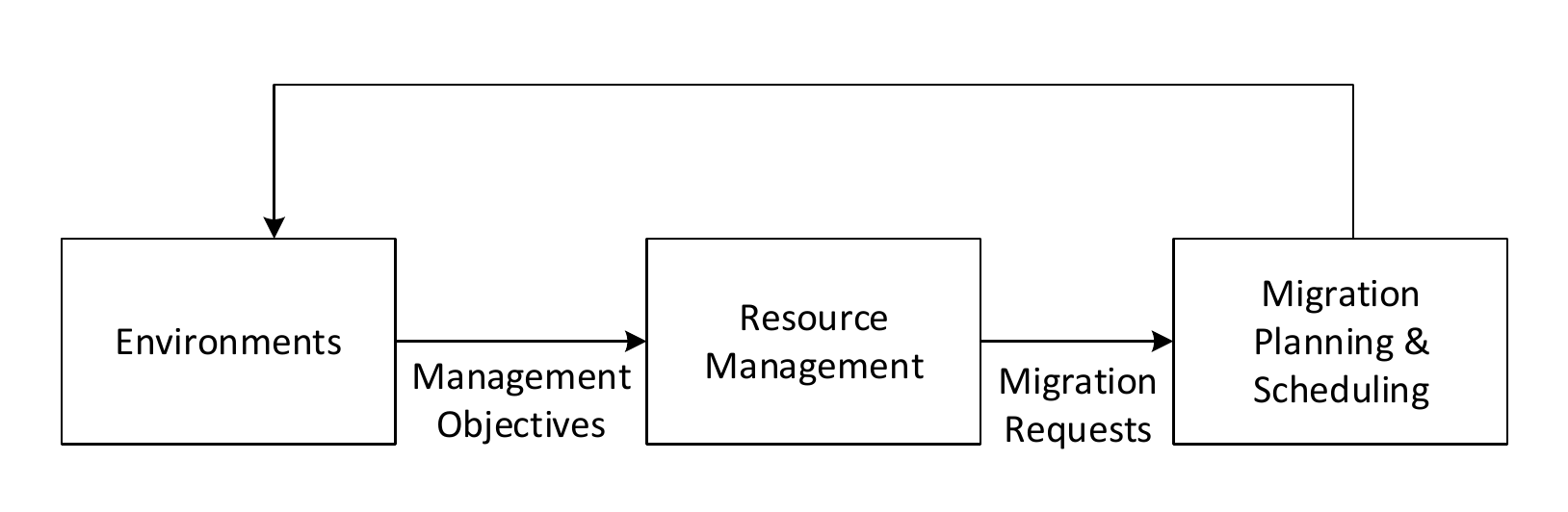}
			\caption{A general migration management framework}\label{fig: chapter1-management-framework}
		\end{minipage}\hfill
		\begin{minipage}{0.39\textwidth}
			\centering
			\includegraphics[width=\linewidth]{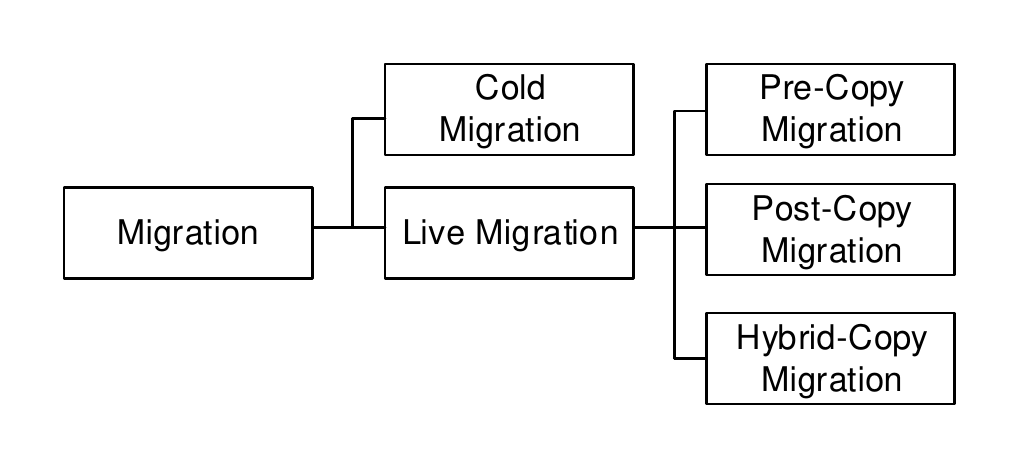}
			\caption{Categories of migration types}\label{fig: chapter1-migration-type}
		\end{minipage}
	\end{figure}
	
    This article focuses on the research of migration management during dynamic resource management in edge and cloud computing environments.
	Figure~\ref{fig: chapter1-management-framework} illustrates the general migration management workflow. Based on the various objectives, the resource management algorithms find the optimal placement by generating multiple live migrations. With the generated multiple migration requests, the migration planning and scheduling algorithms optimize the performance of multiple migrations, such as total and individual migration time and downtime, while minimizing the migration cost and overheads, such as migration influence on application QoS. On the other hand, the computing and networking resources are reallocated and affected by multiple migrations.
	
	For migration management, it is essential to minimize migration costs, maximize migration performance, while achieving the objectives of dynamic resource management.
	Since dynamic resource management requires multiple instance migrations to achieve the objectives, we emphasize migration management in the context of multiple migrations solutions and challenges.
	Based on the proposed taxonomy, we also review related state-of-art works in each category and identify the gaps and future research opportunities. 
	
	Although many surveys~\cite{strunk2012costs,shetty2012survey,xu2013managing,medina2014survey,li2015comparing,yamada2016survey,zhang2018survey,wang2018survey,noshy2018optimization,rejiba2019survey} of live migration have been presented in the contexts of performance, mechanism, optimization of single live migration, and general migration-based dynamic resource management, they only focus on the specific migration aspects and neglect the aspects of migration management including migration generation in resource policies and migration planning and scheduling algorithms. 
	Therefore, in this paper, we identify the following five aspects of migration management in both Edge and Cloud computing environments:
	\begin{itemize}
		\item migration performance and cost model
		\item resource management policy and migration generation
		\item migration planning and scheduling
		\item scheduling lifecycle management and orchestration
		\item evaluation method and platforms
	\end{itemize}
	
	The rest of this paper is organized as follows. 
	Section \ref{sect: background} presents the background of live migration and its running infrastructure. 
	In Section \ref{sect: management-tax}, we present the proposed taxonomy of migration management. Section~\ref{sect: tax-schedule} describes the details of essential aspects of migration planning and scheduling algorithms. We review the related works of migration management including migration generation and migration planning and scheduling in Section \ref{sect: review-management}. Evaluation methods and tools including simulation, emulation, and empirical platforms are introduced in Section~\ref{sect: evaluation-methods}. 
	Then, we analyze gaps within current studies in Section \ref{sect: gaps}. 
	Finally, we conclude the paper in Section \ref{sect: conclusion}.

	\section{Live Migration Background}\label{sect: background}
	
	Considering the scudding development of live migration virtualization and cloud computing paradigms, this section introduces the fundamentals of live migration of VM and container instances from three aspects: virtualization, migration span, and migration type. We use live instance migration to generalize both live VM migration and live container migration.

	\subsection{Virtualization}
	Virtual Machine and container are the two standard solutions for virtualized instances for live migrations. In this section, we introduce the runtime of VM and container and the memory tracing mechanism that supports the resource virtualization and isolation, which is the foundation for live migration.
	
	\paragraph{Hypervisor:}
	Hypervisor, known as Virtual Machine Monitor (VMM), is software that creates and runs VM by virtualizing host computer resources, such as memory and processors. 
	The hypervisor inserts shadow page tables underneath the running OS to log the dirty pages~\cite{clark2005live}. Then, the shadow tables are populated on demand by translating sections of guest page tables.
	By setting read-only mapping for all page-table entries (PTE) in the shadow tables, the hypervisor can trap the page fault when the guest OS tries to modify the memory page.
	In addition, libvirt, as an open-source toolkit for hypervisor management, is widely used in the development of cloud-based orchestration layer solutions. Integrating it with hypervisors, such as libvirt and KVM/QEMU, one can track the details of live migration through management user interface command \textit{virsh domjobinfo} including dirty page rate, expected downtime, iteration rounds, memory bandwidth, remaining memory, total memory size, etc.

	\paragraph{Container Runtime:}
	Container runtime is software that creates instances and manages instances on a compute node.
	Except for the most popular container runtime Docker, there are other container runtimes, such as containerd, CRI-O, runc, etc.
	CRIU~\cite{criu} is the de-facto software for live container migration. It relies on the \textit{ptrace} to seize the processes and injects the parasite code to dump the memory pages of the process into the image file from the address space of the process. Additional states, such as task states, register, opened files, credentials, are also gathered and stored in the dump files.
	CRIU creates checkpoint files for each connected child process during a process tree checkpointing. CRIU restores processes by using the information in the dump files during the checkpointing in the destination host.

	\subsection{Migration Types}
	Migration can be applied to different types of virtual resources, such as process, VM, and container is often referred to as the disk-less migration. Figure~\ref{fig: chapter1-migration-type} illustrates the categories of migration types. Generally, instance and storage migration can be categorized as cold or non-live migration and live migration. The live migration can be further categorized into pre-copy, post-copy, and hybrid migration. Based on granularity, the migration can be divided into single and multiple migrations.
	
	The design and continuous optimization and improvement of live migration mechanisms are striving to minimize downtime and live migration time.
	The \textit{downtime} is the time interval during the migration service is unavailable due to the need for synchronization.
	For a single migration, the \textit{migration time} refers to the time interval between the start of the pre-migration phase to the finish of post-migration phases that instance is running at the destination host. On the other hand, the \textit{total migration time} of multiple migrations is the time interval between the start of the first migration and the completion of the last migration.
	
	For the performance trade-off analysis, memory and storage transmission can be categorized into three phases:
	\begin{itemize}
		\item \textit{Push phase} where the instance is still running in the source host while memory pages and disk block or writing data are pushed through the network to the destination host.
		\item \textit{Stop-and-Copy phase} where the instance is stopped, and the memory pages or disk data is copied to the destination across the network. At the end of the phase, the instance will resume at the destination.
		\item \textit{Pull phase} where the new instance executes while pulling faulted memory pages when it is unavailable in the source from the source host across the network.
	\end{itemize}
	Based on the guideline of these three phases, single live migration can be categorized into three types: (1) pre-copy focusing on push phase~\cite{clark2005live}, (2) post-copy using pull phase~\cite{hines2009post}, and (3) hybrid migration~\cite{sahni2012hybrid} utilizing both push and pull phases. On the other hand, we can categorize multiple migration mechanisms into mainly two types, namely homogeneous and heterogeneous strategies.
	Furthermore, pre-migration and post-migration phases are handling the computing and network configuration. During the pre-migration phase, migration management software creates instance's virtual interfaces (VIFs) on the destination host, updates interface or ports binding, and networking management software, such as OpenStack Neutron server, configures the logical router. During the post-migration phase, migration management software updates port or interface states and rebinds the port with networking management software and the VIF driver unplugs the instance's virtual ports on the source host.

	\paragraph{Cold Migration:}
	Compared to the live migration, cold memory and data migrations are data transmission of only one snapshot of VM's memory and disk or the dump file of one container checkpoint from one physical host to another. In other words, pure stop-and-copy migration fits into this category.
	Although provides simplicity over the live migration solution, the cold migration bears the disadvantage that both migration time and downtime are proportional to the amount of physical memory allocated to the VM. It suffers the significant VM downtime and service disruption.
	
	\begin{figure}[t]
		\begin{minipage}{0.6\textwidth}
			\centering
			\includegraphics[width=\linewidth]{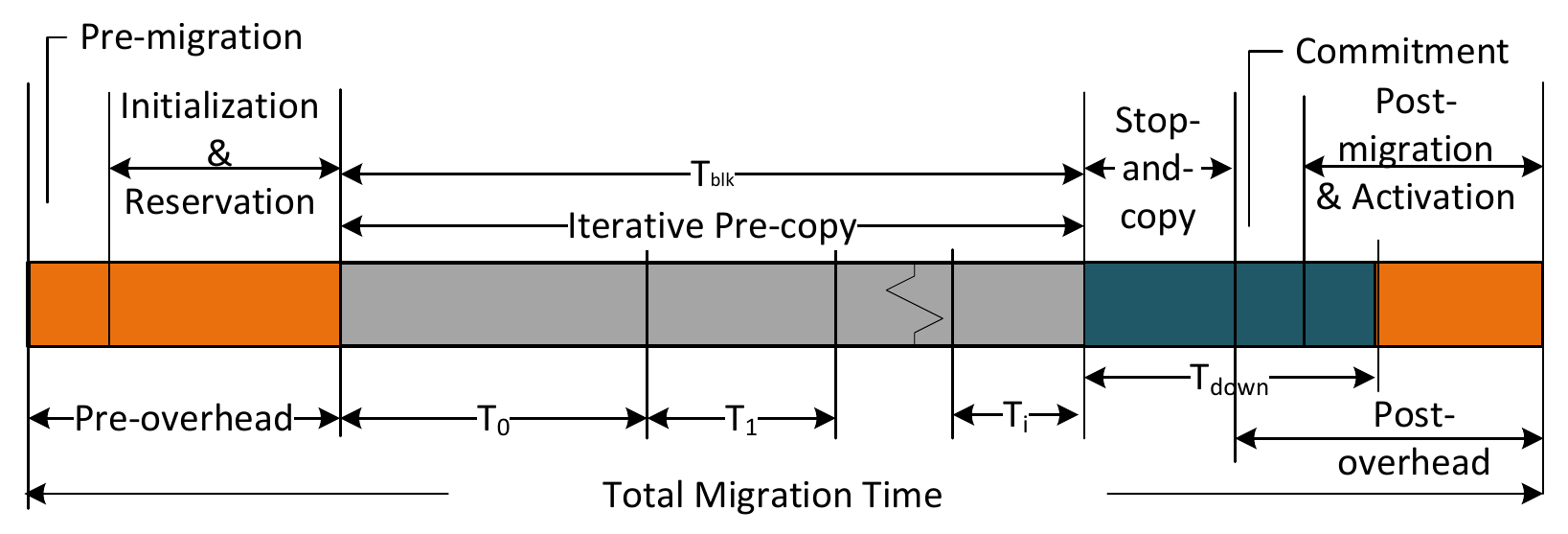}
			\caption{Pre-copy Live Migration~\cite{he2019performance}}\label{fig-chapter1: live-migration}
		\end{minipage}\hfill
		\begin{minipage}{0.38\textwidth}
			\centering
			\subfigure[LAN]{\includegraphics[width=0.45\textwidth]{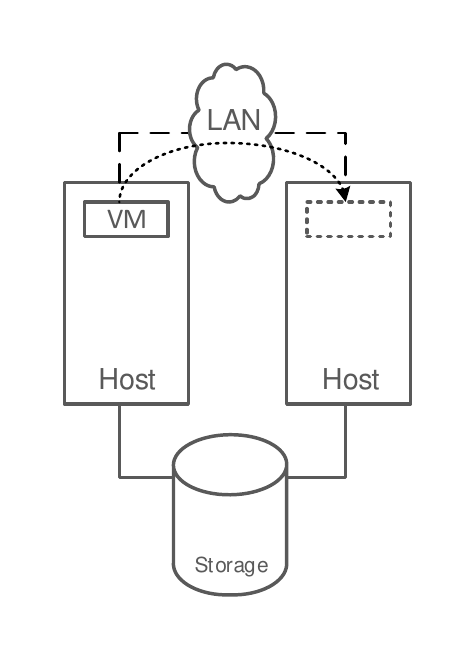}\label{fig: chapter1-migspan-lan}}
			\hfil
			\subfigure[WAN]{\includegraphics[width=0.45\textwidth]{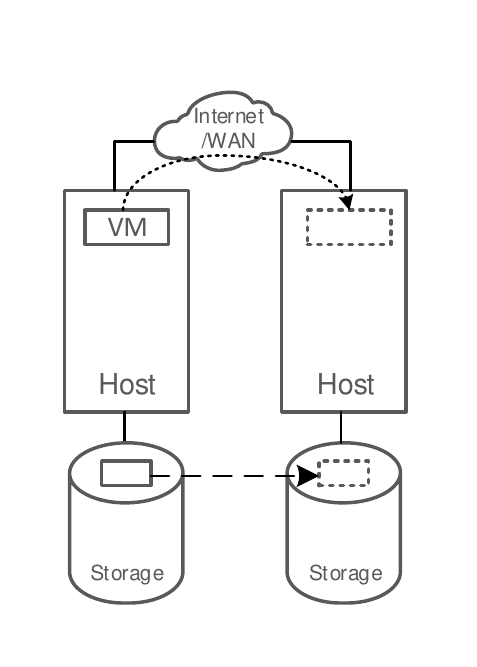}\label{fig: chapter1-migspan-wan}} 
			\caption{Migration span in LAN and WAN}
			\label{fig: mig-span}
		\end{minipage}
	\end{figure}
	
	\paragraph{Pre-copy Migration:}
	During the pre-copy migration~\cite{clark2005live}, dirty memory pages are iteratively copied from the running instance at the source host to the instance container in the destination host.
	Figure~\ref{fig-chapter1: live-migration} illustrates the pre-copy migration phases.
	Generally, the pre-copy migration of instance memory can be categorized into several phases:
	\begin{itemize}
		\item \textit{Initialization}: preselects the destination host to accelerate the future migration process.
		\item \textit{Reservation}: setups the shared file server (optional), and initializes a container of the instance for the reserved resources on the destination host.
		\item \textit{Iterative pre-copy}: For pre-copy migration, send dirty pages that are modified in the previous iteration round to the destination host. The initial memory states are copied in the first round.
		\item \textit{Ctop-and-Copy}: the VM is stopped in the source host in the last iteration round according to the downtime threshold. 
		\item \textit{Commitment}: source host gets the commitment of the successfully copied instance from the destination host 
		\item \textit{Activation}: reserved resources are assigned to the new instance and the old instance is deleted.
	\end{itemize}
	
	\paragraph{Post-copy Migration:}
	For the post-copy live migration~\cite{hines2009post}, it first suspends the instance at the source host and resumes it at the target host by migrating a minimal subset of VM execution states. Then, the source instance pro-actively pushes the remained pages to the resumed VM.
	A page fault may happen when the instance attempts to access the un-transferred pages solved by fetching these pages from the instance in the source.
	Post-copy strategy can reduce the live migration time and downtime compared with pre-copy migration. However, it is not widely adopted due to the unstable and robustness issue~\cite{fernando2019live}.
	When the running instance crashes during the post-copy migration, the service will also be crashed as there is no running instance in the original host with full memory states and data. For certain services and applications, the instance needs to constantly pull faulted pages from the source host which can degrade the QoS for an extensive period.

	\paragraph{Hybrid Live Migration:}
	Hybrid post copy~\cite{sahni2012hybrid} aims to reach the balance point by leveraging all three phases.
	It can also be considered as an optimization technique based on pre-copy and post-copy migrations.
	It starts with the pre-copy model that iteratively copies dirty pages from source to destination. The post-copy migration will be activated when the memory copy iteration does not achieve a certain percentage increase compared with the last iteration. In certain situations, it will reduce the migration time with slightly increased downtime.
	However, it bears the same disadvantages of post-copy migration that pulling faulted pages slow down the processing speed which may degrade the QoS, and VM reboot may occur when the network is unstable.
	
	\paragraph{Multiple Migration Type:}
	From the perspective of multiple live migrations, there is the standard homogeneous solution that each migration type is identical and heterogeneous solution~\cite{shribman2012pre,wang2017ada,wang2019ada} that performs pre-copy, post-copy, or hybrid copy migrations for multiple migrations at the same time. The heterogeneous strategy aims to improve network utilization and reduce the total migration time of multiple migrations while meeting the requirements of different services with various characteristics.
	
	\subsection{Migration Network Environment}
	Migration Span indicates the geographic environment where live migrations are performed. It is critical to analyze the migration span since various computing and networking settings and configurations directly affect migration management.
	In this section, we categorize live migration based on the migration span into LAN (Layer-2), such as intra-data center, and WAN (Layer-3) environment, such as inter-data center and edge-cloud migrations (Fig.~\ref{fig: mig-span}).

	\paragraph{Intra-Cloud:}
	Live migrations are the cornerstone of cloud management, such as load balancing and consolidation. The source and destination hosts of intra-cloud migration are in the same LAN environment.
	In the intra-data center environment, hosts are often shared the data via Network-Accessed Storage (NAS), excluding the need for live storage transmission (Fig.~\ref{fig: chapter1-migspan-lan}). In addition, for the share-nothing data center architecture, live migration flows are separated from the tenant data network to alleviate the network overheads of migrations on other services.

	\paragraph{Inter-Clouds \& Edge-Cloud:}  Live migration is widely adopted for inter-data center management due to various purposes, such as managing cost, emergency recovery, energy consumption, performance and latency, data transmission reduction, and regulation adoption based on the administrative domains~\cite{nagin2011inter,toosi2014interconnected}.
	For the migrations between edge and cloud data centers, strategies often need to consider the trade-off between processing time, network delay, energy, and economy.
	For example, VNF migration from edge to cloud to reduce the processing time and migrate service from cloud to edge to minimize the network delay~\cite{cziva2018dynamic}.
	
	For the migrations across WAN, such as inter-data center and edge-cloud migrations, there is no shared storage and dedicated migration network between the data center sites (Fig.~\ref{fig: chapter1-migspan-wan}). Therefore, in addition to the live instance migration focusing on memory synchronization, live storage migration is necessary. It also applies to the architecture without shared storage in LAN. The main challenges of migration in WAN are optimizing data transmission and handling the network continuity~\cite{zhang2018survey}.

	\paragraph{Edge Computing:} Edge computing includes both LAN and WAN architecture. The motivations and use cases of dynamic resource management in edge computing are similar to those in cloud computing environments. On the other hand, live migration at edges is often referred to as service migration focusing on the mobility-induced migrations in Mobile Edge Computing~\cite{wang2018survey,rejiba2019survey}. 
	In the edge WAN solutions, edge data centers are connected through WAN links as the traditional inter-data center architectures.
	With the emerging cloud-based 5G solution~\cite{huaweiwhitepaper}, edge data centers can be connected through dedicated backbone links and shared the regional cloud data center and network storage.

	\paragraph{SDN-enabled Solution:}
	By decoupling the networking software and hardware, SDN can simplify traffic management and improve the performance and throughput of both intra-data centers (SDN-enabled data centers) and inter-data centers (SD-WAN) including migration networking management.
	As a result, SDN can improve traffic throughput of live migration and reduce the networking overheads of live migration on other services in both LAN and WAN environments~\cite{wang2017virtual}.

	\begin{figure}[t]
		\centering
		\includegraphics[width=0.8\linewidth]{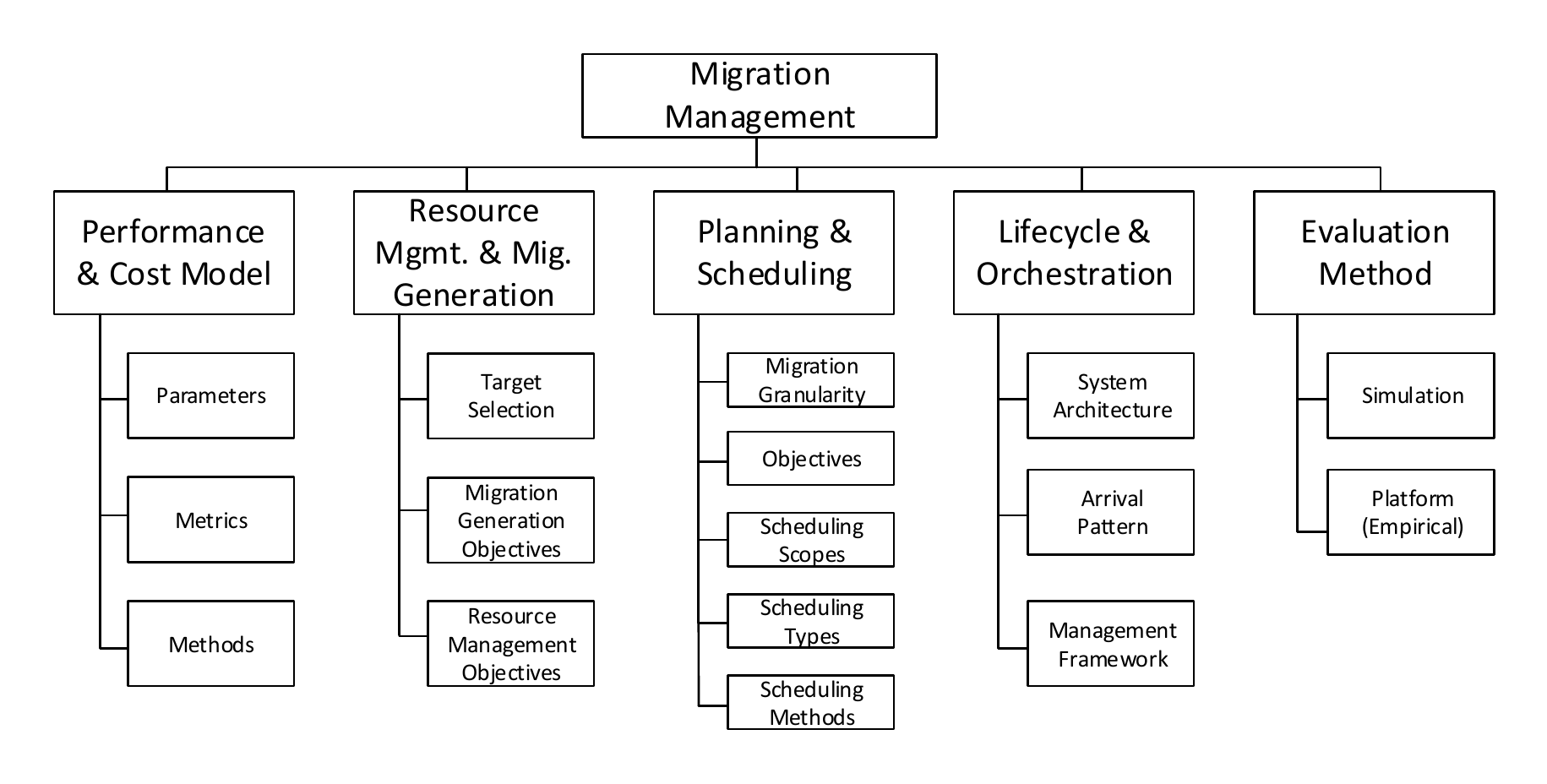}
		\caption{Taxonomy of migration management}
		\label{fig: tax-mig-manage}
	\end{figure}
	
	\section{Taxonomy of Migration Management} \label{sect: management-tax}
	Based on the existing literature, we categorize migration management into five aspects, as shown in Fig. \ref{fig: tax-mig-manage}, namely: (1) performance and cost model; (2) policy and migration generation; (3) migration planning and scheduling; (4) migration lifecycle and orchestration and (5) evaluation method. We introduce the details of migration planning and scheduling aspect and evaluation method in Section \ref{sect: tax-schedule} and Section \ref{sect: evaluation-methods}, respectively.
	
	\subsection{Migration Performance and Cost Modeling}
	The migration performance and cost modeling is the fundamental element for migration management to evaluate and predict the total cost and scheduling performance of multiple migration requests. 
	Based on the related literature and our observation, this section identifies the \textit{parameters}, \textit{metrics}, and \textit{modeling methods} involved in the performance and cost model of live migrations.

	\subsubsection{Parameters}\label{sect:mig-parameters}
	
	\begin{table}[h]
		\caption{Parameters of the live migration model}
		\label{tab: parameters}
		\resizebox{\linewidth}{!}{%
			\begin{tabular}{|l|l|l|l|l|l|}
				\hline
				Category           & \multicolumn{5}{c|}{Parameters}                                                                                  \\ \hline
				Computing          & CPU utilization  & memory utilization   & I/O interface      & WSS size                      &                   \\ \hline
				Networking         & bandwidth        & interfaces           & routing            & delay/distance                         & layers/hops       \\ \hline
				Storage            & sharing          & data size            & storage type       & read/write speed              &                   \\ \hline
				Single migration   & dirty page rate  & iterations threshold & downtime threshold & configuration delay & priority          \\ \hline
				Multiple migration & migration impact & concurrency          & migration time     & routing                       & scheduling window \\ \hline
		\end{tabular}}
	\end{table}
	
	Table \ref{tab: parameters} illustrates the parameters involved in live migration under three categories: computing, networking, and storage resources. Moreover, we also identify the migration parameters in the granularity aspect: single and multiple migrations. 
	The \textit{computing resource} parameters include CPU load and utilization, memory load and utilization, memory size,
	dirty page rate, WSS size as frequent updating memory pages for live migration optimization, and I/O interfaces (i.g. cache interface, host network card interface, inter-CPU interfaces). 
	The \textit{networking resource} parameters include the migration routing, available bandwidth (link and routing bandwidth for single and multiple paths), migration distance (the number of network hops), the number of involved network layers, network delay (link delay and end-to-end latency).
	In addition, the \textit{storage-related} parameters include writing and reading speed and storage data size.

	The single live migration parameters include dirty page rate, the threshold of pre-copy iteration rounds, downtime threshold for pre-copy migration, configuration delay in pre and post-migration processes, and the priority of the migration request.
	On the other hand, we need to consider several parameters of multiple migration scheduling, such as The migration impact on other services, the running and subsequent migrations in computing and networking aspects,
	the concurrency for multiple migration scheduling as resource contention among migrations, the single migration time in multiple migration scheduling, the migration routing considering the traffic of other services and migrations, and the scheduling window of each migration with various priorities and urgencies.
	
	\subsubsection{Metrics}
	\begin{table}[t]
		\caption{Metrics of live migration performance}
		\label{tab: metrics}
		\resizebox{0.6\linewidth}{!}{%
		\begin{tabular}{|l|lll|}
			\hline
			Category & \multicolumn{3}{c|}{Metric}                                      \\ \hline
			Time     & migration time       & downtime       & deadline/ violation time \\ \hline
			Data     & dirty memory         & storage        & stop-and-copy size       \\ \hline
			QoS      & response time        & network delay  & available resource       \\ \hline
			Energy   & physcial host        & network device & cooling                  \\ \hline
			SLA      & service availability & success ratio  & policy performance       \\ \hline
		\end{tabular}
	}
	\end{table}
	
	Many works have investigated the metrics of single migration performance. 
	However, few works are focusing on the performance metrics of multiple migrations. 
	Therefore, we also extend the investigation of the single live migration metrics to the multiple migrations in these categories.
	As shown in Table \ref{tab: metrics}, we categorize these metrics into different categories, namely time related, data, QoS, energy, and SLA.
	
	\paragraph{Migration Time:} Migration time is another key parameter used to evaluate the single migration performance and overhead. A large migration time normally results in a large overhead on both computing and networking resources for the migration VMs and other services.
	
	\paragraph{Downtime:} Downtime is one of the two main parameters used to evaluate the single migration performance. During the downtime caused by migration, the service is not available to the users.
	
	\paragraph{Iteration Number:} For migration types utilizing the pre-copy strategy, the number of iteration rounds is a direct parameter affecting the migration converging hence the migration time and downtime.
	
	\paragraph{Data Transmission Size:} It is the key parameter to judge the network overhead during the migration across the network. For pre-copy migration, it is highly positively correlated to the migration time. The total amount of data transmission is the sum of the data amount of each instance. It can be divided into two aspects: memory data and storage data.

	\paragraph{Total Migration Time:} The total migration time of multiple migrations is the time interval between the start of the first migration and the completion of the last migration. This is the key parameter to evaluate the multiple migration performance and overheads.
	
	\paragraph{Average Migration Time:}
	The average migration time is the average value of the sum of migration time of all instances within the time interval. With the continuous arrival migration requests, the average migration time is preferable to the total migration time. The total migration time of a bunch of instances is only a suitable parameter for the periodically triggered source management strategies.
	
	\paragraph{Average Downtime:} Similar to the average migration time, the average downtime is the mean value of the sum of downtime of all instance migrations within a time interval. Time unit, such as millisecond (ms) and second (s), is used for migration time and downtime.
	
	\paragraph{Energy Consumption:}
	Energy consumption consists of the electricity cost, green energy cost, cooling cost, physical host, networking devices cost. It is a critical metric of live migration overheads used for green energy algorithms and data centers. Joule (J) is used as a unit of energy and Wh or KWh is used in electrical devices.

	\paragraph{Deadline Violation:}
	Migration request or task is the key element for the multiple migration scheduling algorithms.
	Service migrations with different time requirements will have the corresponding deadline and scheduling window.
	As a result, the number of deadline violations for migrations with different priorities and urgency is the key metric to evaluate a deadline-aware or priority-aware migration scheduling algorithm.
	
	\paragraph{Resource Availability:}
	The remaining computing, networking, storage resources during and after the single migration and multiple migrations. It is critical for migration success as there should be sufficient resources for the new instance in the destination. Furthermore, resource availability during the migration affects the performance of the subsequent migrations. Resource availability after the migration is also a metric for the resource reconfiguration evaluation for various policies.
	
	\paragraph{Quality of Service:} Response time, end-to-end delay, network delay, and network bandwidth during and after the single migration or each migration during multiple migration scheduling for the migrating service and other services (co-located VMs, shared-resource VMs, connected VMs).

	\paragraph{Service Level Agreement:} Migration may cause service unavailable due to migration-related issues, such as downtime, network continuity, network robustness, and migration robustness. Therefore, cloud providers provide the SLA to subscribers and tenants as the availability rate for the services with and without migrations. Therefore, SLA violations are another critical metric.

	\subsubsection{Modeling Methods}
	This section introduces the different modeling methods for live migration performance and overheads, including \textit{theoretical modeling} and \textit{experimental modeling (profiling and prediction)}.
	
	\paragraph{Theoretical:}
	In theoretical modeling, the system behaviors are described mathematically using formulas and equations based on the correlation of each identified parameters~\cite{liu2009live,akoush2010predicting,breitgand2010cost,wei2011energy,kikuchi2011performance,kikuchi2011performance,liu2013performance,callegati2013live,xu2013iaware,huang2014prediction,aldossary2018performance}.  Some works only model the migration costs and performance based on the parameters correlation. On the other hand, works follow the behaviors of live migration, such as iterative copying dirty page rate to model, the performance, and overheads in finer granularity.
	
	\paragraph{Profiling:}
	Experimental modeling methods are based on measurements with controlled parameters.
	Empirical running analysis, such as Monte Carlo, are relied on repeated random sampling and time-series monitoring and recording for migration performance, overheads, and energy consumption profiling~\cite{akoush2010predicting, strunk2013does, hu2013quantitative}.
	For both overhead and performance modeling, empirical experiment profiling can also derive the coefficient parameters in the model equations ~\cite{huang2011power,liu2013performance,strunk2013lightweight,huang2014prediction}.
	Regression algorithms are also used to model the cost and performance based on the measurement~\cite{elsaid2014live, aldossary2018performance}.

	\paragraph{Prediction:}
	Generally, mathematic cost models can be used to estimate migration performance and overheads.
	Performance estimation and prediction  algorithms~\cite{akoush2010predicting,strunk2012costs,strunk2013lightweight,liu2013performance,li2016optimizing,aldossary2018performance} are proposed to simulate migrations processes to minimize the prediction error. Furthermore, Machine Learning (ML)-based modeling~\cite{jo2017machine,elsaid2019machine} are adopted to generalize parameters in various resources to obtain a more comprehensive cost and performance prediction model.

	\subsubsection{Cost and Overhead Modeling}
	The cost and overhead models of live migration are integrated with the policy objectives for modeling and optimizing the dynamic resource management problems.
	Similar to the migration parameters, the cost and overhead modeling can be categorized into
	computing, networking, storage, and energy caused by virtualization and live migration (see survey~\cite{xu2013managing}), migration influence on subsequent migrations and co-located services~\cite{ghorbani2012,xu2013iaware,bari2014cqncr,rybina2014analysing,fernando2020sdn}, and migration networking competitions with each other in the multiple migration context~\cite{bari2014cqncr, wang2017virtual,tsakalozos2017live}.

	\subsubsection{Performance Modeling}
	For resource management and migration scheduling algorithms, the performance models of single and multiple migrations are used to maximize the migration performance whiling achieving the objectives of resource management. The category of performance model is similar to the performance metrics classification. In other words, performance models can be categorized as migration success rate, migration time, downtime, transferred data size, iteration number, deadline violation~\cite{xu2013managing}. With the complexity of modeling multiple migrations, it is difficult to model the performance of multiple migration directly on total migration time. Therefore, multiple migration performance and cost models are focusing on the single migration performance based on the currently available resources during multiple migration scheduling~\cite{tsakalozos2017live,fernando2020sdn} or the involved parameters for the sharing resources, such as shared network routing, shared links, shared CPU, memory, network interfaces, and the number of migrations in the same host~\cite{ghorbani2012,sarker2013performance,bari2014cqncr,wang2017virtual}.

	\subsection{Migration Generation in Resource Management}
\begin{figure}[t!]
		\centering 
		\captionsetup{justification=centering}
		\includegraphics[width=0.8\linewidth]{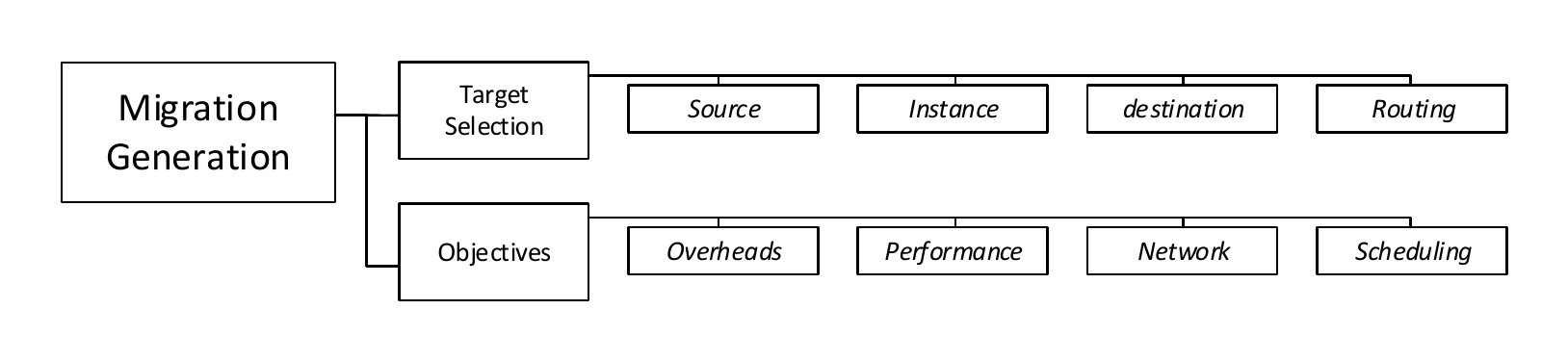}
		\caption{Categories of migration generation in dynamic resource management}
		\label{fig: tax-miggen}
	\end{figure}

In this section, we discuss migration request generation of resource management algorithms. 
We investigate how the migration performance and overhead models integrated with the policy affect the optimization problems and migration generations in two aspects: \textit{migration target selection} and \textit{migration generation objectives}. Figure~\ref{fig: tax-miggen} illustrates the details of each aspect.
	
	\subsubsection{Migration Target Selection}
	The targets of one migration include the selections of source, destination, instance, and network routing. During the migration generation for resource management policies, the targets of migrations can be selected simultaneously or individually. 
	For simultaneous solutions, such as approximation algorithms, several migration instances, source or destination hosts, and network routings can be generated at the same time. For individual solutions, such as heuristic algorithms, each migration request can be generated once at a time in each algorithm loop.
	
	\paragraph{Source Selection:}
	The source host or site of one migration request is selected based on the objectives of resource management policies, such as failure, resource utilization, energy consumption, over-subscripted host, and under-subscripted host.
	
	\paragraph{Instance Selection:}
	During the instance selection for migration request, the migration generation algorithm needs to consider the various objectives of resource management policy, the availability of resources in potential destinations, and the overheads of live migration, such as dirty page rate and the number of allowed migrations. For use cases, such as gang migration, disaster recovery, software update, and hardware maintenance, all instances within the source hosts or sites will be selected. In scenarios such as mobility-induced migration, there is no need to select the instance.
	
	\paragraph{Destination Selection:}
	Many works considered the selection of migration destination as a bin packing problem, where items with different volumes need to be packed into a finite number of bins with fixed available volumes in a way to minimize the number of bins used. There are several online algorithms to solve the problems, such as Next-Fit (NF), First-Fit (FF), Best-Fit (BF), Worst-Fit (WF), etc.
	By considering both objectives of various resource management, such as energy consumption and networking consolidation, and migration overheads and performance, heuristic and approximation algorithms are proposed.

	\paragraph{Flow Routing:}
	The available bandwidth and network delay are critical for migration performance, such as migration time and downtime. The migration flows of pre-copy migration are elephant flows and post-copy migrations are sensitive to the network delay. Meanwhile, in the network architecture where the migration traffic and service traffic share network links, the migration traffic can significantly affect the QoS of other services.  In the SDN-enabled data centers, the allocated bandwidth and routing of both migration and services traffic can be dynamically configured to improve the performance of live migration and minimize the migration network overheads.

	\subsubsection{Migration Generation Objectives}
In this section, we summarize the different objectives of migration selections in resource management policies from the perspective of live migration management, including \textit{migration overhead}, \textit{performance}, \textit{network}, and \textit{scheduling awareness}. Many works are proposed in dynamic resource management for various objectives, such as performance, networking~\cite{heller2010elastictree}, energy~\cite{heller2010elastictree,cao2014nice,cziva2018dynamic}, QoS, and disaster recovery~\cite{kokkinos2016survey}. Most of the works only consider the memory footprint (memory size) and available bandwidth. Without the proper live migration modeling, the selected instance migration requests for reallocation will result in unacceptable scheduling performance and service degradations. 
In addition, these works with migration performance and energy models~\cite{strunk2012costs,liu2013performance,zhang2017network,shi2019memory,flores2020pam} only consider the individual migration performance and computing and networking overheads during the migration generation or instance placement selection for the dynamic management policy. The total migration cost, as a result, is a linear model, which can not reflect the concurrency among migrations and resource contentions between migrations and services. As a result, the solution is only optimized for sequential migration scheduling solutions.	
	
	\paragraph{Overheads-aware:} 
	Most works of resource management policies focus on minimizing the cost or overheads of migrations and modeling the total cost as a sum of the individual live migration cost while achieving objectives of resource management, such as energy consumption~\cite{beloglazov2012managing,cao2014nice,witanto2018adaptive} and network flow consolidation~\cite{cui2016synergistic,flores2020pam}. 
	As shown in Sect \ref{sect:mig-parameters}, the overheads of the migrating instance can be categorized as computing, networking, and storage overhead, including the total number of migrations, number of co-located instances, memory size, dirty page rate, CPU workloads, data compression rate, bandwidth allocation, and migration flow routing of the migrating instance.
	
	\paragraph{Performance-aware:}
	Some works are focusing on the migration performance optimization integrating with the objectives of resource management, such as the cost and performance of migrations~\cite{sarker2013performance,xu2013iaware,cui2016synergistic,cui2016plan,flores2020pam}. However, providing multiple migration requests, existing resource management solutions through live migrations do not consider and guarantee migration scheduling performance.

	\paragraph{Network-aware:}
	Apart from network routing and bandwidth allocation,
	the network contentions among migrations and between migration and applications, as well as instance connectivity need to be considered to optimize the networking throughput and communication cost during or after the migrations~\cite{mann2012remedy,kakadia2013network,tso2013implementing,zhang2017network,cui2017traffic,eramo2017approach}. For instance, two migrations may content the same network path which can lead to migration performance degradation, which increases the total and individual migration time. On the other hand, without a dedicated migration network, the migration and application flows can also compete for the same network link leading to a QoS degradation. For instance connectivity, one migration completion can free up more bandwidth due to flow consolidation. As a result, subsequent migration can be converged faster with more available bandwidth.

	\paragraph{Scheduling-aware:} 
	Current works do not consider the migration scheduling performance~\cite{bari2014cqncr,wang2017virtual,he2021sla} with the linear model of migration cost and interfaces.
	In the migration generation phase of resource management policies, we can optimize the performance of single or multiple migration scheduling and guarantee the optimal or the near-optimal performance of the objectives of the original resource management policy. Compared to policy-aware migration management, it is more adaptive without the need for specific modeling and design. It has less impact on the enormous amounts of existing dynamic resource management policies.

	\subsection{Migration Lifecycle and Orchestration} \label{sect: framework}
	\begin{figure}[t!]
		\centering 
		\captionsetup{justification=centering}
		\includegraphics[width=0.8\linewidth]{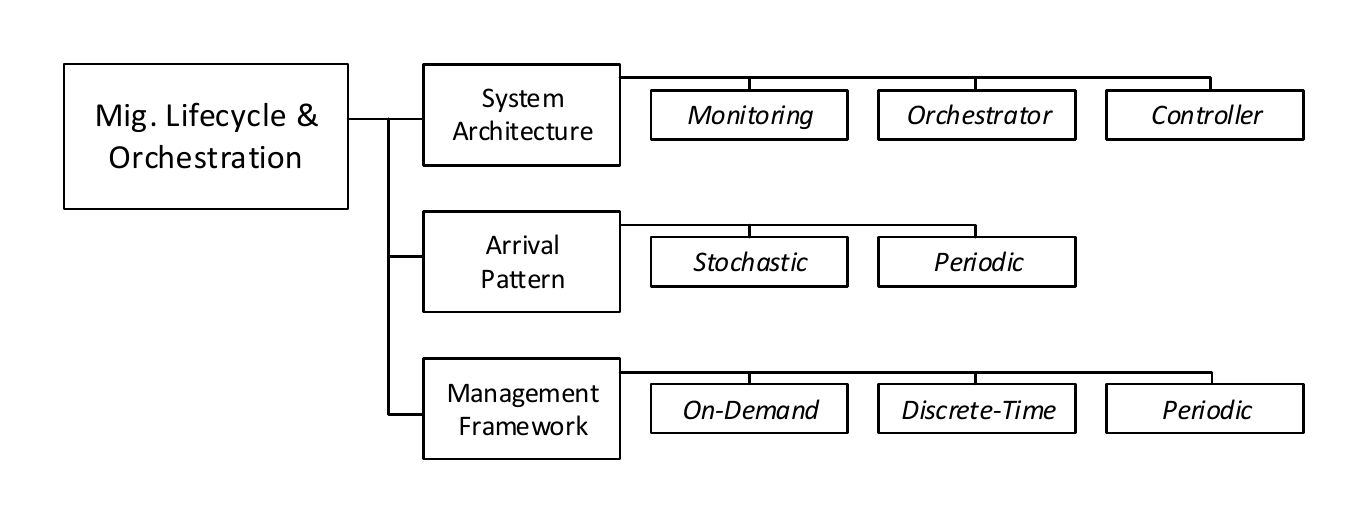}
		\caption{Categories of migration lifecycle management and orchestration}
		\label{fig: tax-migorch}
	\end{figure}
	
	Based on various scenarios of dynamic resource management through live migrations, it is critical to investigate the migration lifecycle and orchestration layer including migration arrival patterns and corresponding management framework.
	Therefore, this section summarizes the migration lifecycle management and orchestration in the several aspects: \textit{arrival pattern}, \textit{orchestration}, \textit{monitoring}, and \textit{management framework}. Figure~\ref{fig: tax-migorch} illustrates the details of each aspect.

	\subsubsection{Arrival Patterns}
	Arrival patterns of migration requests based on various paradigms and objectives in fog, edge, and cloud computing environments can be categorized as \textit{periodic} and \textit{stochastic} patterns.
	For existing dynamic resource management algorithms, the migration generation and arrival patterns are periodic due to the overhead of live migrations. The dynamic resource management, such as regular maintenance and updates, load balancing, and energy-aware consolidation algorithms, triggered the reallocation periodically.
	On the other hand, arrival patterns of event-driven migrations are often stochastic due to the nature of the service. For example, the mobility-induced migration in edge computing is based on the user behaviors and movement~\cite{wang2018survey,rejiba2019survey}.

	\subsubsection{Migration Orchestration}
	The system architecture of SDN-enabled edge and cloud data centers consists of an orchestration layer, a controller layer, and a physical resource layer.
    Several orchestration systems have been proposed to control computing and network resources in fog, edge, and cloud computing environments~\cite{martini2016design,cziva2016sdn,mayoral2017sdn,fichera2017experimenting}.
	The policy orchestrator provides the algorithms and strategies for joint computing and networking resource provisioner, network engineering server, and migration manager (lifecycle manager, migration generator, planner, and scheduler). The network topology discovery, cloud and network monitors are essential for computing and networking provisioning. Combined with cloud manager and network engineering server, live VM Management software can efficiently control the lifecycle of single migration and schedule migrations in a finer granularity by jointly provisioning computing and networking resources.

	The controller layer facilitates the allocation of computing and networking resources. The cloud controller (such as OpenStack) and the container control plane (such as Kubernetes) are responsible for the autonomous deployment, monitoring, scaling, and management of VMs and containerized applications based on provided strategies in the orchestration layer. On the other hand, the SDN controller manages the OpenFlow switches and flow forwarding entries and collects the device and flow statistics through southbound interfaces. Networking applications through SDN controller northbound interfaces perform topology discovery, network provisioning, and network engineering for both application and migration flows.

	\subsubsection{Management Framework}
	Therefore, based on the characteristic of resource management policies in various scenarios and use cases,
	the triggered pattern of migration planning can be categorized into periodic, discrete-time, and on-demand types.
	
	\paragraph{On-demand:}
	In the on-demand framework, the migration will be planned and scheduled whenever the request arrives~\cite{wang2018survey,rejiba2019survey}. The on-demand framework can be applied to the scenarios that individuals are the subjects to trigger the migration from one host to another, such as mobility-induced migration and migration requests by public cloud subscribers and tenants.
	
	\paragraph{Periodic:}
	In the periodic planning framework~\cite{zhang2018survey}, the migration plans are calculated by periodically obtaining the migration request from the dynamic resource management strategy. The time interval between each migration planning can be configured as the value of the strategy management interval.
	Within each time interval, multiple migration requests for the reallocation strategies will not be affected by the migrations from the previous round. With a large time interval, all migration instances can be completed before the reallocation time for the planning of new migration requests.
	
	\paragraph{Discrete-Time:}
	In the discrete-time framework, the arrival migration requests are put into queues to regulate the migration arrival speed and processed migration number.
	The migration planner will read multiple migration requests as the input from the entire or part of the plan wait queue and migration from the schedule waiting queue periodically calculate the migration plan based on the configured time interval, such as one second.
	For each input with multiple migration requests, the planner will calculate the migration plan based on the states of computing and networking resources. The plane of these migrations will be put into the queue for migrations waiting to be scheduled by the migration scheduler.
	Within the small time interval for each input, some of the instances from the previous migration plan are not started by the scheduler yet could affect the decision on the current planning round.
	The discrete-time framework suits the scenarios, such as mobility-induced migration in edge computing that migrations can arrive stochastically and much more frequently than the traditional dynamic resource management strategy.

	\section{Migration Planning and Scheduling} \label{sect: tax-schedule}
\begin{figure}[t!]
		\centering 
		\captionsetup{justification=centering}
		\includegraphics[width=0.8\linewidth]{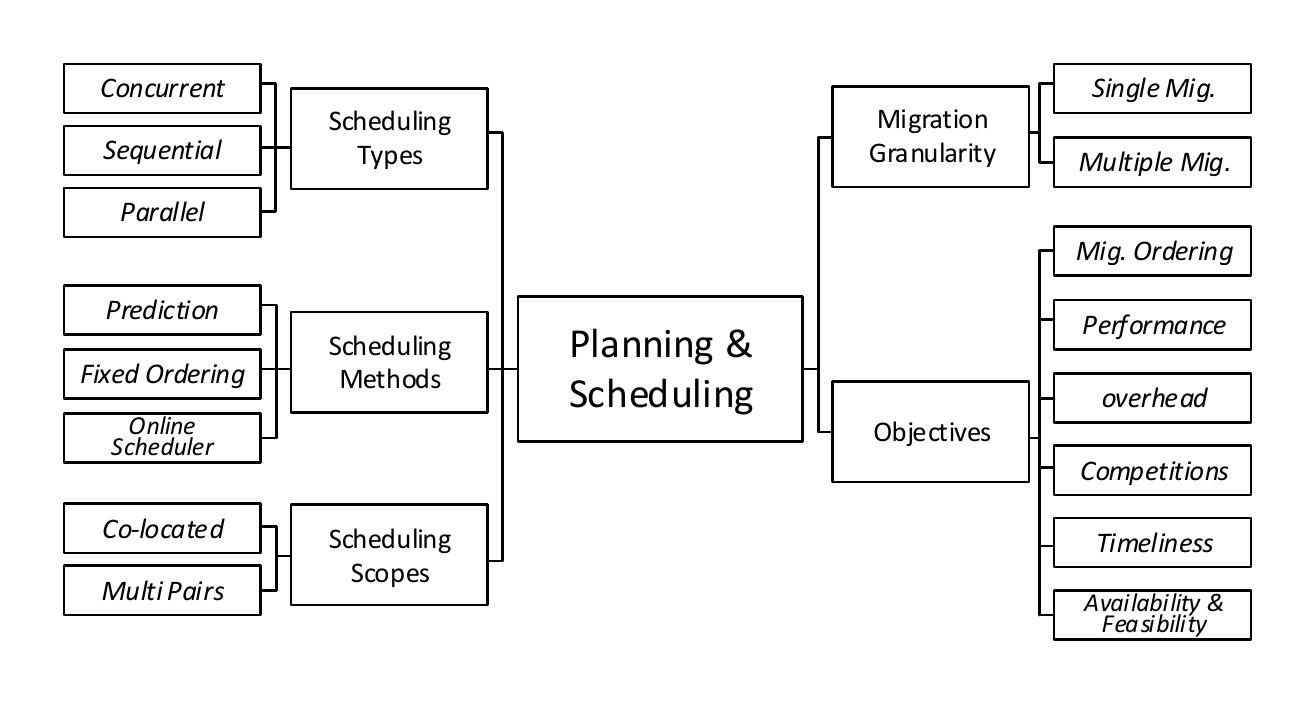}
		\caption{Categories of migration planning and scheduling algorithms}
		\label{fig: tax-migschedule}
	\end{figure}

	In this section, we introduce the taxonomy for migration planning and scheduling, including \textit{migration granularity}, \textit{scheduling objectives}, \textit{scopes}, \textit{types}, and \textit{methods}. Figure~\ref{fig: tax-migschedule} illustrates the details of each category.
	Compared to real-time task scheduling, there is enough time to perform more complex migration scheduling, which further improves multiple migration performance and alleviates migration overheads. When it comes to multiple migrations, based on the objectives of live migration, the migration planning algorithm needs to calculate and optimize the sequence of each migration. In other words, the planning algorithm needs to consider the following issues, including availability, concurrency, correlation, and objective.
	We review the related state-of-art works in Section~\ref{sect: review-schedule}.
	
	\subsection{Migration Granularity}
	The migration granularity in the context of migration planning and scheduling can be categorized into \textit{single migration} and \textit{multiple migration}.
	
	\paragraph{Single Migration:} 
	In single migration, only one instance is migrated at the same time. The research scope of single migration focuses on the performance and overhead of individual migration, including migration mechanisms, optimization techniques in both computing and networking aspects. The key metrics of single migration performance are migration time and downtime. The overheads of single migration include network interfaces, CPU, and memory stress, migration process overheads in the source and destination hosts (i.g. dirty memory tracing overheads), service-level parameters (i.g. response time), and available bandwidth for the migration service and other services in the data center.

	\paragraph{Multiple Migrations:}
	In multiple migration, multiple instances are considered to be migrated simultaneously.
	Multiple migration can be divided into various aspects based on the instance and service locations and connectivity, including co-located instances (gang migration) and cluster migrations (same source and destination pair), and related instances (connected services and applications, VNFs in SFC, entire virtual network, VMs in the same virtual data center).
	The overheads and performance of multiple migrations need to be evaluated and modeled in the migration generation, planning, and scheduling phases.
	
	The overheads of multiple migrations can be categorized into service-level and system-level overheads. The service-level overheads include multiple migration influences on the migrating service, subsequent migrations performance, and other services in the data centers, and the system-level overheads include the multiple migration influence on the entire system, such as availability of networking and computing resources.
	On the other hand, the performance of multiple migration can be divided into global and individual performance.
	Total migration time, downtime, and transferred data size are the key metrics for the global migration performance of multiple migration. Performance metrics, such as average migration time, average downtime, and deadline violation, are used for the individual performance in multiple migration.
	
	\subsection{Objectives and Scopes}
	In this section, we summarize the objectives of migration planning algorithms and scheduling strategies. It can be categorized into \textit{migration ordering}, \textit{migration competitions}, \textit{overhead and cost}, \textit{migration performance}, \textit{migration timeliness}, and \textit{migration availability and feasibility}.
	We also categorized the scheduling scope in two aspects: co-located scheduling and multiple migration scheduling with multiple source-destination pairs (multi pairs).
	
	\paragraph{Migration Ordering:} The works of migration ordering problems focus on the feasibility of multiple migrations~\cite{ghorbani2012,sarker2013performance} and migration ordering of co-located instances~\cite{rybina2014analysing,fernando2020sdn} in the one-by-one scheduling solutions. In other words, given a group of migration requests, the feasibility problem is solving the problem that whether given migrations can be scheduled and the scheduling ordering of these requests due to the resource deadlock. The performance problem in the migration ordering context is finding an optimized order to migrate the co-located instance in order to minimize the overheads of multiple migrations and maximize the performance of multiple migration in the one-by-one scheduling manner.

	\paragraph{Migration Competitions:} Resource competition problems include the competitions among migrations and between migrations and other services during the simultaneous migration scheduling~\cite{bari2014cqncr,wang2017virtual,tsakalozos2017live}. Since migrations and services are sharing computing and networking resources, it is essential to determine the start sequence of migrations in both sequential and concurrent scheduling manner to minimize the resource throttling and maximize the resource utilization with respect to both QoS and migration performance. The resource dependencies and competitions among migrations and services need to be considered in both the migration generation phase and the planning and scheduling phase in order to improve the performance of multiple migration and minimize the overheads of multiple migration.

	\paragraph{Overhead and Cost:}
	It is critical to minimize migration overheads and costs to alleviate the QoS degradations and guarantee the SLA during live migration scheduling. The computing overheads on CPU, memory and I/O interfaces affect the co-located instances negatively.
	The migrations also share the same network links with other services. It may lead to QoS degradations due to the lower bandwidth allocation. As a result, a longer migration process leads to larger computing (CPU and memory) and networking overheads (network interfaces and available bandwidth). Network management policies and routing algorithms are adopted to dynamically allocate the bandwidth and network routing to migrations. Furthermore, the migration downtime also needs to be managed to avoid unacceptable application response time and SLA violations. 
	
	\paragraph{Migration Performance:}
	The performance of multiple migration scheduling is one of the major objectives of migration planning and scheduling algorithms. The total migration time is highly relative to the final management performance.
	In other words, a smaller migration time leads to a quicker optimization convergence. Furthermore, in green data center solutions, the energy consumptions induced by live migration need to be modeled properly.
	
	\paragraph{Timeliness:}
	The timeliness of the migration schedule is also critical to resource management performance~\cite{zhang2014delay,tsakalozos2017live}.
	For the migration with various priorities and urgencies, inefficient migration planning or scheduling algorithms may result in migration deadline violations, which leads to QoS degradations and SLA violations.
	For example, some VNF needs to be migrated as soon as possible to maintain low end-to-end latency requirements. On the other hand, some migration requests of web services with high latency tolerance and robustness are configured with a much larger scheduling window.
	
	\paragraph{Availability and Feasibility:}
	Migration availability and feasibility problems are also considered in the planning and scheduling algorithms~\cite{ghorbani2012,sarker2013performance}.
	There should be reserved resources in the destination hosts and sites in order to host the new instance. In the context of multiple migration, resource deadlock and network inconsistency may also affect the migration success ratio. Intermediate migration host and efficient migration ordering algorithms are proposed to solve the migration feasibility issue.

	\paragraph{Scheduling Scopes:}
	\begin{figure}
		\centering 
		\subfigure[Co-located]{\includegraphics[width=0.25\textwidth]{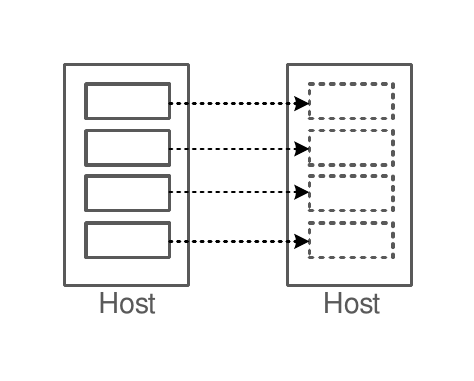}\label{fig: chapter2-gange}} 
		\subfigure[Multi pairs]{\includegraphics[width=0.3\textwidth]{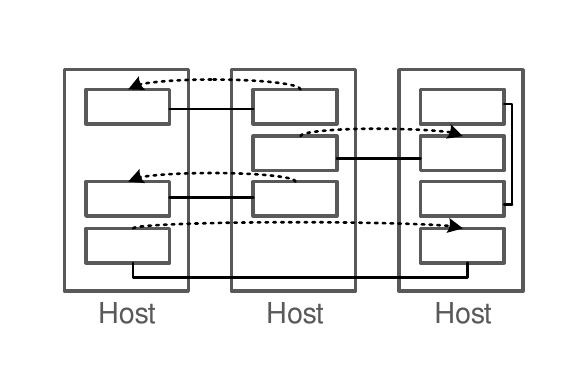}\label{fig: chapter2-multipairs}} 
		\caption{Scheduling Scopes of multiple migration scheduling: co-located migrations and multiple migration with various source and destination pairs}
		\label{fig: chapter2-scheduling-scope}
	\end{figure}
	The multiple migration scheduling and planning algorithms can be divided into co-located and migrations with multiple source-destination pairs (Fig.~\ref{fig: chapter2-scheduling-scope}). In co-located instance migrations, such as gang migration~\cite{deshpande2013gang,fernando2020sdn}, the solution only focus on one source and destination pair. On the other hand, multiple instances migration involves a various source and destination hosts or sites~\cite{bari2014cqncr,wang2017virtual,he2021sla}.
	For the migrations across dedicated network links, networking contentions between migration and services are omitted. In the data center network without dedicated migration networks, some works consider the virtual network connectivity among applications during the migration schedule.

	\subsection{Scheduling Types}
	\begin{figure}[t!]
		\centering 
		\centering
		\subfigure[Sequential]{\includegraphics[width=0.4\textwidth]{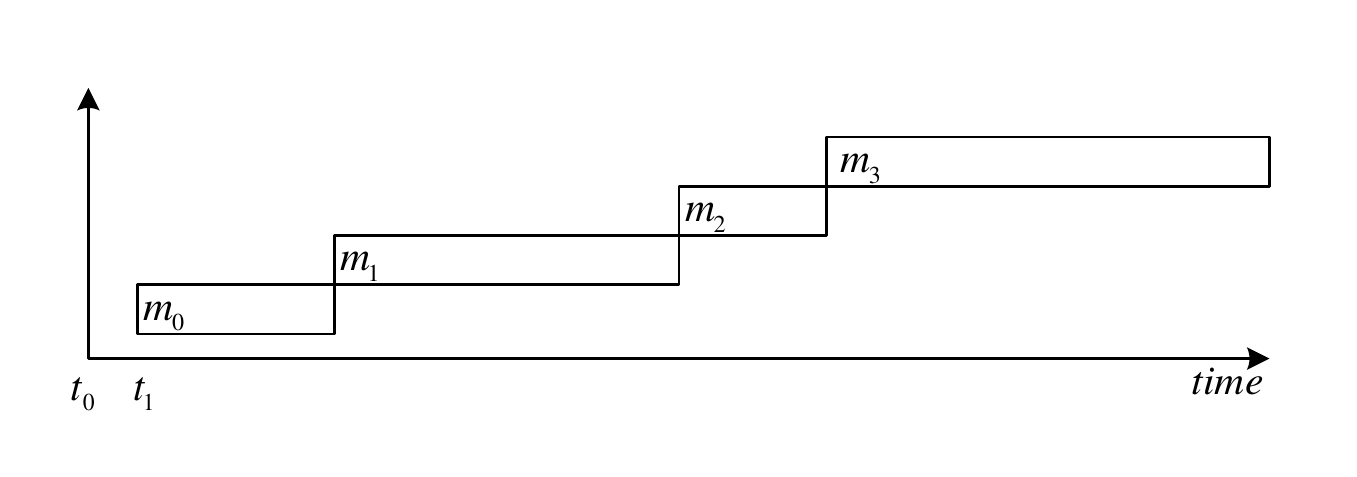}\label{fig: chapter2-sequential}} 
		\hfil
		\subfigure[Parallel]{\includegraphics[width=0.25\textwidth]{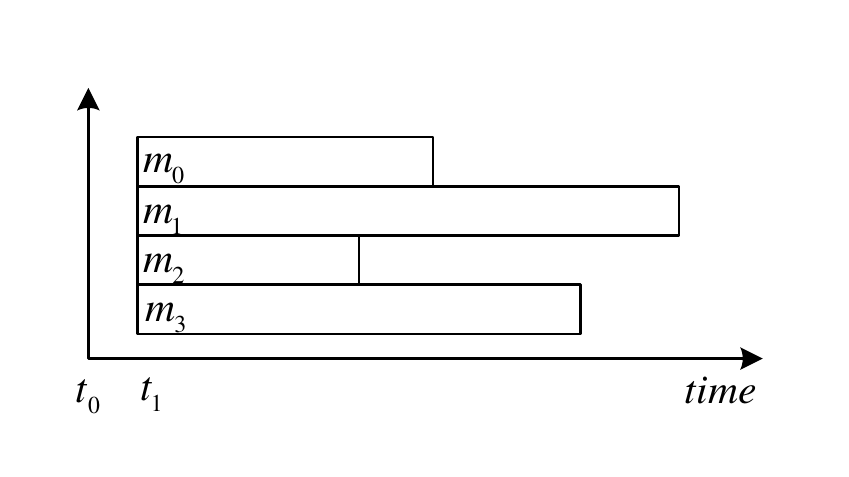}\label{fig: chapter2-parallel}} 
		\hfil
		\subfigure[Concurrent]{\includegraphics[width=0.25\textwidth]{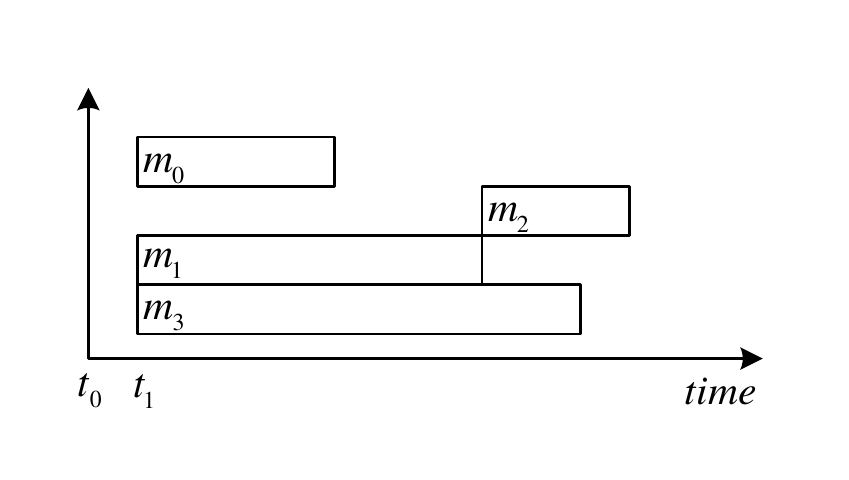}\label{fig: chapter2-concurrent}}
		\caption{Scheduling types of multiple migration scheduling}
		\label{fig: chapter2-scheduling-type}
	\end{figure}

	The migration scheduling types of multiple migration can be categorized as \textit{sequential} multiple migration, \textit{parallel} multiple migration, and \textit{concurrent} multiple migration.
	
	\paragraph{Sequential:} In the sequential multiple migration solution depicted in Fig.~\ref{fig: chapter2-sequential}, migration requests are scheduled in the one-by-one manner~\cite{ye2011live,ghorbani2012,rybina2014analysing,kang2014design,lu2014clique,lu2015vhaul,fernando2020sdn}. In most scenarios of live VM migration, the network bandwidth is not sufficient for the networking requirement of live migration when sharing network links with other migrations. Therefore, sequential migration scheduling for networking resource-dependent migrations is the optimal solution. It is used to the migration scheduling in co-located multiple migration scheduling and planning.
		
	\paragraph{Parallel:} In the parallel or synchronous multiple migration solution depicted in Fig.~\ref{fig: chapter2-parallel}, migration requests start simultaneously~\cite{ye2011live,zheng2014comma,liu2014vmbuddies,kang2014design,lu2015vhaul,sun2016new,deshpande2017traffic}.
	For the migrations with network link sharing, parallel migration scheduling is preferred only when the networking overheads induced by dirty pages and the memory footprint of migrating instances are smaller than the migration computing overheads. In other words, in most scenarios, the parallel migration may result in longer total and individual migration time and downtime. On the other hand, for the migrations without network links sharing across dedicated migration networks, the minimum total and individual migration time can be achieved.
	Some solutions are mixing the sequential and parallel migration solution that groups of migrations are started at the same time as in parallel solution and these groups of migrations are scheduled sequentially.

	\paragraph{Concurrent:} Furthermore, concurrent or asynchronous migration planning and scheduling algorithms~\cite{bari2014cqncr,wang2017virtual,he2021sla} are proposed to schedule multiple migration requests efficiently by calculating and scheduling the start time of each migration independently (Fig.~\ref{fig: chapter2-concurrent}). The migrations contend the network resources with other migrations and services. Furthermore, the service traffic relocation induced by migration completion may affect the subsequent migrations.  Therefore, it is essential to manage the dependency and concurrency among migrations during multiple migration scheduling.

	\subsection{Scheduling Methods}
	After the phase of multiple migration planning, the calculated migration plan is going to be scheduled in the migration scheduling phase. We categorize scheduling methods into three types, namely \textit{prediction}, \textit{fixed ordering}, and \textit{online scheduler}.
	To schedule migrations, the migration manager and scheduler need to know the time when a migration has finished or needs to be started. 
	
	\paragraph{Prediction:}
	Based on the prediction model and current available computing and networking resources, the start time of each migration is configured during the planning phase~\cite{wang2017virtual}. Furthermore, the bandwidth allocated to each migration is also configured based on the available bandwidth at the time of migration planning.
	
	\paragraph{Fixed ordering:}
	Multiple migration tasks are scheduled based on the order calculated by migration planning algorithms~\cite{bari2014cqncr}. In other words, one migration is started as soon as possible when one or several specific migrations are finished. The fixed ordering model of multiple migration requests is similar to the dependent tasks, which can be modeled as a Directed Acyclic Graph (DAG).

	\paragraph{Online Scheduler:}
	The states of the networking environment, such as network topology, available links and interfaces, available bandwidth, and network delay, are constantly changed. The computing resources, such as memory, vCPU, storage, destination hosts, and sites, may also become unavailable during the multiple migration scheduling.
	Integrating with dynamic computing and networking management, an online migration scheduler can dynamically start the migrations based on current states of computing and networking~\cite{he2021sla}. The resource may not be available based on the predicted scheduling time or orders. The online scheduler can dynamically adjust the migration plan to efficiently schedule multiple migrations Furthermore, by balancing the allocated bandwidth to migration and application traffic, the online scheduler can guarantee both QoS and migration performance.

\section{Current Research on Migration Management}\label{sect: review-management}

In this section, we review the current research on migration management by focusing on the migration generation during dynamic resource management and the migration planning and scheduling algorithms. Each work can involve several categories of live migration management. Therefore, we choose the major category to organize and present the reviews in a more straight forwarding way.

\subsection{Migration Generation}
First, we review and summarize representative works on migration generation based on the resource management objectives, such as load balancing, energy-saving, network communication optimization, and migration-aware solutions.

\begin{table}[t!]
	\caption{Characteristics of Migration Generation in Dynamic Resource Management}
	\label{tab:chapter2-mig-generation}
	\resizebox{\textwidth}{!}{%
		\begin{tabular}{l|ccc|cccc|ccc|cccc}
			\hline
			Reference                                        & \multicolumn{3}{c|}{Mig. Com.}                                                      & \multicolumn{4}{c|}{Mig. Net.}                             & \multicolumn{3}{c|}{Mig. Obj.}              & \multicolumn{4}{c}{Res. Obj.}                             \\
			& mem          & \begin{tabular}[c]{@{}c@{}}dirty\\ rate\end{tabular} & cpu          & bw           & route        & hop          & layer        & perf.        & cost         & num          & comp.        & net.         & energy       & QoS          \\ \hline
			Singh et al.~\cite{singh2008server}              &              &                                                      & $\checkmark$ & $\checkmark$ &              &              &              &              & $\checkmark$ &              & $\checkmark$ &              &              & $\checkmark$ \\
			Verma et al.~\cite{verma2008pmapper}             & $\checkmark$ &                                                      &              & $\checkmark$ &              &              &              &              & $\checkmark$ &              & $\checkmark$ &              &              & $\checkmark$ \\
			Wood et al.~\cite{wood2009sandpiper}             & $\checkmark$ &                                                      &              &              &              &              &              &              & $\checkmark$ &              & $\checkmark$ &              &              &              \\
			Mann et al.~\cite{mann2012remedy}                & $\checkmark$ & $\checkmark$                                         &              & $\checkmark$ &              &              &              &              & $\checkmark$ &              &              & $\checkmark$ &              &              \\
			Forsman et al.~\cite{forsman2015algorithms}      & $\checkmark$ & $\checkmark$                                         &              & $\checkmark$ &              &              &              &              & $\checkmark$ &              & $\checkmark$ &              &              &              \\
			Xiao et al.~\cite{xiao2012dynamic}               & $\checkmark$ &                                                      &              &              &              &              &              &              & $\checkmark$ &              & $\checkmark$ &              & $\checkmark$ & $\checkmark$ \\
			Beloglazov et al.~\cite{beloglazov2012energy}    &              &                                                      &              &              &              &              &              &              &              & $\checkmark$ & $\checkmark$ &              & $\checkmark$ &              \\
			Beloglazov et al.~\cite{beloglazov2012optimal}   & $\checkmark$ &                                                      &              & $\checkmark$ &              &              &              &              & $\checkmark$ &              & $\checkmark$ &              & $\checkmark$ &              \\
			Witanto et al.~\cite{witanto2018adaptive}        &              &                                                      &              &              &              &              &              &              & $\checkmark$ &              & $\checkmark$ &              &              &              \\
			Li et al.~\cite{li2017holistic}                  &              &                                                      &              &              &              &              &              &              &              &              &              &              &              &              \\
			Piao et al.~\cite{piao2010network}               &              &                                                      &              &              &              &              &              &              &              &              &              & $\checkmark$ &              & $\checkmark$ \\
			Tso et al.~\cite{tso2013implementing}            & $\checkmark$ &                                                      &              &              &              &              & $\checkmark$ &              & $\checkmark$ &              &              & $\checkmark$ &              &              \\
			Cao et al.~\cite{cao2014nice}                    & $\checkmark$ &                                                      &              &              &              & $\checkmark$ &              &              & $\checkmark$ &              &              & $\checkmark$ & $\checkmark$ &              \\
			Cui et al.~\cite{cui2016synergistic,cui2016plan} & $\checkmark$ & $\checkmark$                                         &              & $\checkmark$ &              &              &              &              & $\checkmark$ &              &              & $\checkmark$ &              &              \\
			Xu et al.~\cite{xu2013iaware}                    & $\checkmark$ & $\checkmark$                                         & $\checkmark$ & $\checkmark$ &              &              &              & $\checkmark$ & $\checkmark$ & $\checkmark$ &              &              &              & $\checkmark$ \\
			Cui et al.~\cite{cui2017traffic}                 & $\checkmark$ &                                                      &              &              & $\checkmark$ & $\checkmark$ &              &              & $\checkmark$ &              &              & $\checkmark$ &              &              \\
			Flores et al.~\cite{flores2020pam}               & $\checkmark$ &                                                      &              & $\checkmark$ & $\checkmark$ & $\checkmark$ &              & $\checkmark$ & $\checkmark$ &              &              & $\checkmark$ &              & $\checkmark$ \\ \hline
		\end{tabular}%
	}
	\begin{tablenotes}
		\small
		\item[1]  
		\textbf{Mig. Com.}: Migration Computing parameters - mem (memory size), cpu (CPU load and utilization);
		\textbf{Mig. Net.}: Migration Networking parameters - bw (bandwidth), route (migration traffic routing), hop (migration distance), layer (involved network layers); \textbf{Mig. Obj.}: Migration Objectives - perf. (migration performance), cost (migration cost and overheads), num (migration number); 
		\textbf{Res. Obj.}: Resource management Objectives - comp. (computing), net. (networking), energy (device and cooling energy cost), QoS (response time and SLA violations).
	\end{tablenotes}
\end{table}

Many works have studied the load balancing problem in dynamic resource management through live migrations~\cite{singh2008server,verma2008pmapper,wood2009sandpiper,mann2012remedy,forsman2015algorithms}.
Singh et al.~\cite{singh2008server} propose a multi-layer virtualization
system HARMONY. It uses VMs and data
migration to mitigate hotspots on servers, network devices, and storage
nodes. The load balancing algorithm is a variant of
multi-dimensional knapsack problem based on the evenness indicator, i.e. Extended Vector Product (EVP). It considers the single live migration impact on application performance
based on CPU congestion and network overheads.
Verma et al.~\cite{verma2008pmapper} estimate the migration cost based on the
deduction of application throughput. The proposed algorithm (pMapper) selects the smallest
memory size VMs from the over-utilized hosts and assigns
them to under-utilized hosts in the First Fit Decreasing
(FFD) order.
Wood et al.~\cite{wood2009sandpiper} propose a load balancing algorithm Sandpiper
that selects the smallest memory size VM from one of the most overloaded hosts to minimize the migration overheads.
Mann et al.~\cite{mann2012remedy} (Remedy) focus on the VM and destination selection for the load balance of application network flows by considering the single migration cost model based on the
dirty page rate, memory size, and available bandwidth.
Forsman et al.~\cite{forsman2015algorithms} propose a pre-copy live VM migration selection algorithm for automated load balancing strategy based on the migration cost~\cite{liu2013performance}, expected load distribution, and utilization after migration.

Dynamic VM consolidation algorithm is one of the techniques to reduce energy consumption through VM migrations. Khan et al.~\cite{khan2018dynamic} review the related works on dynamic virtual machine consolidation algorithms for energy-efficient cloud resource management.
Xiao et al.~\cite{xiao2012dynamic} investigate dynamic resource allocation through live migration. The proposed algorithm avoids the over-subscription while satisfying the resource needs of all VMs based on exponentially weighted moving average (EWMA) to predict the future loads. It also minimizes the energy consumption of physical machines by hot spot mitigation.
The cost of live migration is modeled as the memory size. The authors argue that live migration with self-ballooning causes no noticeable performance degradation.

Beloglazov et al.~\cite{beloglazov2012energy} propose Minimization of Migrations (MM) algorithm to select the minimum migration number needed to fulfill the objective of overloaded host mitigation and underloaded host consolidation. Best Fit Decreasing algorithms are used. VMs are sorted in decreasing order based on the CPU utilization and network consumption of expected allocation.
Furthermore, Beloglazov et al.~\cite{beloglazov2012optimal} (LR-MMT) focus on energy saving with local regression (LR) based on history utilization to avoid over-subscription. The proposed algorithm chooses the VM with the least memory size from the over-utilized host and the migration destination with the largest energy saving. The minimum migration time is modeled based on VM memory size and available bandwidth.

Witanto et al.~\cite{witanto2018adaptive} propose a machine-learning selector with a neural network. High-level consolidation through VM migrations may reduce energy consumption, but will result in higher SLA violations. Therefore, the proposed algorithm dynamically chooses existing consolidation algorithms and strategies to manage the trade-off between energy and SLA violation (due to VM migration downtime) based on the priority availability in the system.

For the works on networking provisioning through live migration,
Piao et al.~\cite{piao2010network} propose a virtual machine placement and migration approach to minimize the data transfer cost. The proposed migration approach is triggered when the
execution time crosses the threshold specified in the SLA.
Tso et al.~\cite{tso2013implementing} propose a distributed network-aware live VM migration scheme to dynamically reallocate VMs to minimize the overall communication footprint of active traffic flows in multi-tier data centers. The authors investigate the performance degradation issue during migration, which is caused by the congestion at the core layers of the network where bandwidth is heavily oversubscribed. The proposed distributed solution generates migration requests iteratively based on the local VM information (one-by-one migration scheduling) to localize VM traffic to the low-tier network links. The migrations are performed only when the benefits of traffic reallocation outweigh the migration cost based on VM memory and downtime. However, the work lacks a realistic comparison between migration overheads and migration benefits for communication management. With the help of SDN, the assumption of requirements that centralized approaches obtaining knowledge of global traffic dynamics are prohibitively expensive may be untenable.

Cao et al.~\cite{cao2014nice} investigate the VM consolidation problem considering the network optimization and migration cost. Migration overheads are modeled as the host power consumption (the product of power increase and migration time) and traffic cost (the produce of VM memory size and distance in hop number between source and destination).
Cziva et al.~\cite{cziva2016sdn} propose an SDN-based solution to minimize network communication cost through live VM migration in a multi-tire data center network. The authors model the communication cost as the product of the average traffic load per time unit and weight of layer link. They argue that the migration cost can be introduced into the solution by modeling the migration network traffic cost.
Similarly, Cui et al.~\cite{cui2016synergistic,cui2016plan} study the joint dynamic MiddleBox/VNF network chaining policy reconfiguration and VM migration problem in the SDN-enabled environment to find the optimal placement minimizing the total communication cost. The authors consider both VM migration time and the traffic data induced by the MiddleBox/VNF migration. However, the paper lacks the information for how the communication cost is modeled. The migration cost is considered by comparing the network rate with data per time unit with the total transferred data size of live migration.
Furthermore, only modeling the migration cost as transferred data size without considering the networking bandwidth and routing may result in poor migration scheduling performance and QoS degradations.

With the input of candidate VMs and destinations provided by existing resource management algorithms,
Xu et al.~\cite{xu2013iaware} propose a migration selector (iAware) to minimize the single migration
cost in terms of single migration execution time and host co-location
interference. It considers dirty page rate, memory
size, and available bandwidth for the single migration time.
They argue that co-location interference from a single live
migration on other VMs in the host in terms of
performance degradation is linear to the number of VMs
hosted by a physical machine in Xen. However, it only
considers one-by-one migration scheduling.

Cui et al.~\cite{cui2017traffic} propose a new paradigm for
VM migration by dynamically constructing adaptive network topologies
based on the VM demands to reduce VM migration costs and increase the communication throughput among VMs. The migration cost is modeled as the product of the number of network hops and VM's memory size. The authors argue that the general VM migration cost value can be replaced by specific cost metrics, such as migration time and downtime based on allocated bandwidth and measured dirty page rate of the VMs.

Li et al.~\cite{li2017holistic} propose a greedy-based VM scheduling algorithm to minimize the total energy consumption, including the cooling and server power consumption models. For the selection of migration requests, the proposed algorithm selects all physical machines with the temperature above the threshold proportion as source hosts. Then, it selects the VM with the minimum utilization from all selected source hosts and selects the server with the minimum power increase as the migration destination for each VM. The authors focus on energy consumption through live migrations without considering the migration cost. However, the proposed algorithm outputs multiple migration requests without the actual migration planning and scheduling algorithms.

Flores et al.~\cite{flores2020pam} propose a placement solution that integrates migration selection with data centers policies to minimize the communication cost by considering network topology, VM virtual connections, communication cost, and network hops for live migration cost. Considering the routing cost, the proposed migration scheduling algorithm migrates VMs in order to minimize the total cost of migration and VM communication. However, the cost model of migration is still linear without considering the concurrent scheduling performance of multiple migrations.

We summarize the characteristics of migration generation in Table \ref{tab:chapter2-mig-generation} based on four categories: migration computing parameters, migration network parameters, objectives of migration optimization in the solution, and objectives of resource management.

\subsection{Migration Planning and Scheduling} \label{sect: review-schedule}
\begin{table}[t!]
	\caption{Comparisons of Solutions on Multiple Migration Planning and Scheduling}
	\label{tab:chapter2-plan-review}
	\resizebox{\textwidth}{!}{%
		\begin{tabular}{l|ccc|ccc|cc|c|ccccc}
			\hline
			Reference                                                    & \multicolumn{3}{c|}{Schedule}               & \multicolumn{3}{c|}{Net}                                                                                                   & \multicolumn{2}{c|}{Scope}                                              & Heter        & \multicolumn{5}{c}{Mig Obj}                                                                                                                                                                                                                                                    \\
			& Seq.         & Parl.        & Cnrc.        & \begin{tabular}[c]{@{}c@{}}Mig.\\ net.\end{tabular} & \begin{tabular}[c]{@{}c@{}}Net.\\ mgmt.\end{tabular} & Connect.     & Co-loc.      & \begin{tabular}[c]{@{}c@{}}Multi\\ src-dst\end{tabular} &              & \begin{tabular}[c]{@{}c@{}}QoS\\ aware\end{tabular} & \begin{tabular}[c]{@{}c@{}}Mig.\\ order\end{tabular} & \begin{tabular}[c]{@{}c@{}}Mig.\\ Feas.\end{tabular} & \begin{tabular}[c]{@{}c@{}}Mig.\\ perf.\end{tabular} & \begin{tabular}[c]{@{}c@{}}Mig.\\ cost\end{tabular} \\ \hline
			Deshpande et al.~\cite{deshpande2011live}                    &              &        $\checkmark$      &              &                                                     &                                                      &              & $\checkmark$ &                                                         &              &                                                     &                                                      &                                                      &                                                      & $\checkmark$                                        \\
			Deshpande et al.~\cite{deshpande2012inter,deshpande2013gang} &              &       $\checkmark$        &              &                                                     &                                                      &              & $\checkmark$ &                                                         &              &                                                     &                                                      &                                                      &                                                      & $\checkmark$                                        \\
			Rybina et al.~\cite{rybina2014analysing}                     & $\checkmark$ &              &              &                                                     &                                                      &              & $\checkmark$ &                                                         &              &                                                     & $\checkmark$                                         &                                                      & $\checkmark$                                         & $\checkmark$                                        \\
			Fernando et al.~\cite{fernando2020sdn}                       & $\checkmark$ &              &              &                                                     & $\checkmark$                                         &              & $\checkmark$ &                                                         &              &                                                     & $\checkmark$                                         &                                                      & $\checkmark$                                         & $\checkmark$                                        \\
			Ghorbani et al.~\cite{ghorbani2012}                          & $\checkmark$ &              &              &                                                     & $\checkmark$                                         &              &              & $\checkmark$                                            &              &                                                     &                                                      & $\checkmark$                                         &                                                      &                                                     \\
			Sun et al.~\cite{sun2016new}                                 & $\checkmark$ & $\checkmark$ &              &                                                     &                                                      &              & $\checkmark$ &                                                         & $\checkmark$ &                                                     &                                                      &                                                      & $\checkmark$                                         &                                                     \\
			Deshpande et al.~\cite{deshpande2017traffic}                 &              & $\checkmark$ &              &                                                     &                                                      &              & $\checkmark$ &                                                         & $\checkmark$ & $\checkmark$                                        &                                                      &                                                      & $\checkmark$                                         & $\checkmark$                                        \\
			Zheng et al.~\cite{zheng2014comma}                           &     $\checkmark$         &   $\checkmark$           &              &                                                     & $\checkmark$                                         & $\checkmark$ &              &                                                         &              & $\checkmark$                                        &                                                      &                                                      & $\checkmark$                                         & $\checkmark$                                        \\
			Liu et al.~\cite{liu2014vmbuddies}                           &              & $\checkmark$ &              &                                                     & $\checkmark$                                         & $\checkmark$ &              &                                                         &              & $\checkmark$                                        &                                                      &                                                      & $\checkmark$                                         & $\checkmark$                                        \\
			Lu et al.~\cite{lu2015vhaul}                                 & $\checkmark$ &              &              & $\checkmark$                                        & $\checkmark$                                         & $\checkmark$ &              &                                                         &              & $\checkmark$                                        &                                                      &                                                      & $\checkmark$                                         & $\checkmark$                                        \\
			Lu et al.~\cite{lu2014clique}                                & $\checkmark$ & $\checkmark$ &              &                                                     & $\checkmark$                                         & $\checkmark$ &              &                                                         &              & $\checkmark$                                        &                                                      &                                                      & $\checkmark$                                         & $\checkmark$                                        \\
			Kang et al.~\cite{kang2014design}                            & $\checkmark$ & $\checkmark$ &              &                                                     & $\checkmark$                                         &              &              &                                                         &              &                                                     &                                                      &                                                      & $\checkmark$                                         & $\checkmark$                                        \\
			Ye et al.~\cite{ye2011live}                                  & $\checkmark$ & $\checkmark$ &              &                                                     &                                                      &              &              &                                                         &              &                                                     &                                                      &                                                      & $\checkmark$                                         & $\checkmark$                                        \\
			Sarker et al.~\cite{sarker2013performance}                   &              &              & $\checkmark$ &                                                     & $\checkmark$                                         & $\checkmark$ &              & $\checkmark$                                            &              &                                                     &                                                      & $\checkmark$                                         & $\checkmark$                                         & $\checkmark$                                        \\
			Bari et al.~\cite{bari2014cqncr}                             &              &              & $\checkmark$ &                                                     &                                                      & $\checkmark$ &              & $\checkmark$                                            &              &            $\checkmark$                                         &     $\checkmark$                                                 &                                                      & $\checkmark$                                         & $\checkmark$                                        \\
			Wang et al.~\cite{wang2017virtual}                           &              &              & $\checkmark$ & $\checkmark$                                        & $\checkmark$                                         &              &              & $\checkmark$                                            &              &                                                     &                    $\checkmark$                                  &                                                      & $\checkmark$                                         &      $\checkmark$                                                \\
			He et al.~\cite{he2021sla}                           &              &              & $\checkmark$ & $\checkmark$                                        & $\checkmark$                                         & $\checkmark$            &              & $\checkmark$                                            &             &            $\checkmark$                                          &   $\checkmark$                                                    &                                                      & $\checkmark$                                         &    $\checkmark$                                                  \\
			\hline
		\end{tabular}%
	}
	\begin{tablenotes}
		\small
		\item[1]  
		\textbf{Schedule Type}: Seq. (Sequential), Parl. (Parallel), Cnrc. (Concurrent), \textbf{Net}: Networking related - Mig. net. (Dedicated migration network), Net. mgmt. (Networking management), Connect. (instance Connectivity),  \textbf{Scope}: Co-loc. (Co-located instances), \textbf{Heter}: Heterogeneous solutions (mixing various migration types), and \textbf{Mig Obj}: migration management objectives - Mig. order (migration ordering), Mig. feas. (migration feasibility), Mig. perf (migration performance), Mig. cost (Migration cost and overheads).
	\end{tablenotes}
	
\end{table}

In this section, we review the state-of-the-art works on migration planning and scheduling algorithms. 
Studies of migration planning and scheduling focus on various aspects, such as migration feasibility, migration success or failure ratio, migration effects, scheduling deadline, application QoS, scheduling orders, migration scheduler, migration routing, and migration performance in total migration time, average migration time, downtime, and transferred data size.
As shown in Table \ref{tab:chapter2-plan-review}, we summarize the characteristics of reviewed solutions of migration planning and scheduling in various categories: scheduling type, migration networking awareness, migration scopes, heterogeneous migration types, and migration scheduling objectives.

\subsubsection{Co-located Multiple Migrations}

Deshpande et al.~\cite{deshpande2011live} consider the migration of multiple co-located VMs in the same host as the live gang migration problem. They optimize the performance of multiple live migrations of co-located VMs based on the memory deduplication and delta compression algorithm to eliminate the duplicated memory copying from the source to the destination host. Since co-located VMs share a large amount of identical memory, only identical memory pages need to be transferred during the iterative memory copying in pre-copy migration. They also employ delta compression between copied dirty pages to reduce the migration network traffic. Moreover, Deshpande et al.~\cite{deshpande2012inter,deshpande2013gang} further investigate the same problem using cluster-wide global deduplication by improving the technique from the co-located VMs in the same host to the ones in the same server rack.

Rybina et al.~\cite{rybina2014analysing} investigate the co-located resource contentions on the same physical host. The authors evaluate all possible migration orders in a sequential manner in terms of total migration time. They find that it is better to migrate the memory-intensive VM in order to alleviate the resource contentions.
Fernando et al.~\cite{fernando2020sdn} proposed a solution for the ordering of multiple VMs in the same physical host (gang migration) to reduce the resource contentions between the live migration process and the migrating VMs. The objectives of the solution are minimizing the migration performance impact on applications and the total migration time.
The migration scheduler decides the order of migrations based on different workload characteristics (CPU, Memory, network-intensive) and resource usage to minimize the total migration time and downtime.
Furthermore, an SDN-enabled network bandwidth reservation strategy is proposed that reserves bandwidth on the source and destination hosts based on the migration urgency.
When the available bandwidth in the destination can not satisfy the migration requirement, network middle-boxes~\cite{deshpande2014fast} are used as network intermediaries to temporarily buffer the dirty memory.

\subsubsection{Migration Feasibility}
For the migration scheduling feasibility, 
Ghorbani et al.~\cite{ghorbani2012} proposed a heuristic one-by-one migration sequence planning for a group of migration requests to solve the problem of transient loop and network availability during the migration. The authors consider the environments that the requirement of the virtual network must be satisfied with bandwidth over-subscription. With the bandwidth requirement of virtual links between instances, random migration sequence will lead to the failure of migrations of most of the instances. With the flow install time of current SDN controller implementation, the orders of network updates due to migration within the forwarding table may cause the transient loop issue.
The authors did not consider the concurrent VM migration scheduling with various network routings.

\subsubsection{Heterogeneous and Homogeneous Solutions}
Multiple migration can be divided into heterogeneous and homogeneous solutions. For the heterogeneous solution, different types of live migration (pre-copy and post-copy migrations) are used simultaneously.
In an environment where all migrations are sharing the same network, Sun et al.~\cite{sun2016new} considers the sequential and parallel migration performance of multiple VMs. The authors proposed an improved one-by-one migration scheduling based on the assumption that the downtime of live migration is large enough. When the first VM is stopped during the downtime of pre-copy migration, the algorithm stops and performs the post-copy on the remaining connected VMs within the same service. Furthermore, the authors proposed an m-mixed migration algorithm for parallel multiple migration started at the same time. The algorithm chooses the first m VMs to perform pre-copy migration, while the rest are performed with post-copy migration. They validate the efficiency of the proposed solutions, such as the blocking ratio and average waiting time of each migration request, based on the proposed M/M/C/C queuing models. 

The networking contentions between migrations and application traffic can increase the migration time and degrade the application QoS.
To reduce the network contentions between migrations and applications, solutions for co-located multiple VM migrations with both pre-copy and post-copy migrations are adopted. The intuition of these solutions is utilizing both inbound and outbound network interfaces. When the co-located instances are migrated using pre-copy or post-copy, the traffic between co-located instances contends with the migration traffic. Therefore, the migrating instance with post-copy in the destination host communicates to another migration instance with pre-copy in the source host, which alleviates the network contentions between the application traffic and pre-copy migration traffic. Deshpande et al.~\cite{deshpande2017traffic} proposed a traffic-sensitive live migration technique by utilizing pre-copy or post-copy migration based on the application traffic direction to reduce the migration network contention.

\subsubsection{Instance Correlation and Connectivity}
The instances in the data center are often connected through network virtual links under various application architecture, such as multi-tier web applications and Virtual Network Functions (VNFs) in Service Function Chaining (SFC).
Several studies focus on optimizing the multiple migration of multi-tier applications and network-related VMs. Research ~\cite{voorsluys2009cost,kikuchi2012impact} evaluates the impact of live migration on multi-tier web applications, such as response time.
Zheng et al.~\cite{zheng2014comma} investigate the multi-tier application migration problem and propose a communication-impact-driven coordinated approach for a sequential and parallel scheduling solution.
Liu et al.~\cite{liu2014vmbuddies} work on the correlated VM migration problem and propose an adaptive network bandwidth allocation algorithm to minimize migration cost in terms of migration time, downtime, and network traffic. In the context of multi-tier applications, the authors proposed a synchronization technique to coordinate correlated VM migrations entering the stop-and-copy phases at the same time, which reduces the network traffic between correlated applications across the inter-data center network.
Lu et al.~\cite{lu2015vhaul} also focus on the correlated VMs migration scheduling of multi-tier web applications with a dedicated network for migrations. The authors investigate the sequential and parallel migration strategies for multiple migrations which start at the same time. The proposed heuristic algorithm groups the related VMs based on the monitoring of communication traffic and sorts the sequential migration order based on migration time and resource utilization.
Expending the concept from multi-tier application connectivity, Lu et al.~\cite{lu2014clique} proposed a separation strategy by partitioning a large group of VMs into traffic-related subgroups for inter-cloud live migration. Partitioned by a mini-cut algorithm, subgroups of pre-copy migrations are scheduled sequentially and VMs in the same subgroup are scheduled in parallel to minimize the network traffic between applications across inter-data center networks.

\subsubsection{Parallel and Concurrent Scheduling}
Studies~\cite{sun2016new,kang2014design} investigate solutions for mixed sequential and parallel migrations.
Kang et al.~\cite{kang2014design} proposed a feedback-based algorithm to optimize the performance of multiple migration in both total and single migration time considering the sequential and parallel migration cost, and available network resources. It adaptively changes the migration number based on the TCP congestion control algorithm.
Ye et al.~\cite{ye2011live} investigate the multiple migration performance in sequential and parallel migrations. The authors conclude that sequential migration is the optimal solution when the available network bandwidth is insufficient.

For the multiple migration planning and scheduling algorithms of concurrent migrations, the migration planning algorithm and migration scheduler determine when and how one migration of the multiple migration requests should be performed within the time interval of the total migration time.
Sarker et al.~\cite{sarker2013performance} proposed a naive heuristic algorithm of multiple migration to minimize the migration time and downtime. The proposed scheduling algorithm starts all available migrations with minimum migration cost until there is no migration request left. The deadlock problem is solved by temporarily migrating VM to an intermediate host.
Bari et al.~\cite{bari2014cqncr} proposed a grouping-based multiple migration planning algorithm in an intra-data center environment where migrations and applications share the network links. The authors model the multiple VM migration planning based on a discrete-time model as a MIP problem. The available bandwidth may be changed after each migration due to the reconfiguration of virtual network links. The subsequent migrations are affected by the previous migration result. Considering the influence of each migration group during the scheduling, the proposed algorithm sets the group start time based on the prediction model. Migration in each group can be scheduled simultaneously if the resources occupied by the previous group are released. However, the authors neglect the influence of individual migration in their solution, which can lead to performance degradation of the total migration time.
Without considering the connectivity among applications in a WAN environment, Wang et al.~\cite{wang2017virtual} simplify the problem by maximizing the network transmission rate but directly minimizing the total migration time. With the help of SDN, the authors introduce the multipath transmission for multiple migrations. A fully polynomial-time approximation FPTAS algorithm is proposed to determine the start time of each migration.
Considering individual migration impact in terms of migration time, migration bandwidth allocation, networking routing and overheads, available resources after migration, and migration deadline and priority, He et al.~\cite{he2021sla} propose SLA-aware multiple migration planning algorithms to maximize concurrent migration groups with minimum cost and an SDN-enabled online migration scheduler to manage the migration lifecycle based on the planning groups. The experimental results show that the proposed solution can achieve the minimum total migration time while minimizing the individual migration time, migration deadline violations, SLA violations, and energy consumption.

\subsection{Summary and Comparison}
All studies covered in the survey are summarized in Table~\ref{tab:chapter2-mig-generation} and Table~\ref{tab:chapter2-plan-review} based on our taxonomy in Figure~\ref{fig: tax-mig-manage}, respectively. 
For migration generation in dynamic resource management algorithms, many studies optimize at least two of the resource objectives regarding the computing resources, networking resources, energy consumption, and application QoS.
Most researchers do not consider the migration scheduling performance in the proposal which has no tick in the table. These studies consider and model migration computing costs linearly or individually based on memory size, available bandwidth, or the total migration number. A number of works consider the actual single pre-copy live migration model based on memory size, dirty page rate, and available bandwidth. For the migration network cost modeling and management, most of the works only consider the available bandwidth or the transferred data volume, while few works consider the network routing or the migration distance on hops and layers. Integrating with existing resource management algorithms,  few works focus on migration interferences and migration scheduling performance. However, the linear models of these works are only suitable for sequential migration scheduling optimization.

For migration planning and scheduling algorithms, researchers are actively studying the migration scheduling performance and migration cost, some considering the application QoS during migrations, while others focusing on migration availability and scheduling feasibility.
For heterogeneous and homogeneous solutions,
most works are focusing on the homogeneous solution with one migration type such as pre-copy migration, while others consider both pre-copy and post-copy migrations. However, there is no concurrent planning and scheduling algorithm considering scenarios where multiple migration with mixed types.
The scheduling scope is varied based on the proposed method, early studies focus on co-located migrations with one source and destination pair, while recent proposals consider the multiple migration scheduling with various source and destination pairs. 

For networking management and efficiency,
some researchers consider the migration scheduling in dedicated migration networks or do not consider the network overheads on other services and applications. Others consider the connectivity of correlation instances with virtual network communication during migrations. Without considering the network over-subscription, the bandwidth requirements of virtual links between instances are guaranteed during migrations.
To improve migration performance and reduce migration traffic impacts, networking management algorithms are adopted to optimize the network routing for migration traffic and application traffic.

The scheduling types are varied based on the proposed solution and scheduling scopes, most of the works consider sequential migration scheduling, while others focus on the parallel migrations or apply both types. Recently, several researchers focus on concurrent migration planning and scheduling for efficient, optimal, and generic solutions.
Few works focus on the timeliness of migration with optimizations and prediction models, such as migration finishes before a given deadline without general concurrent migration scheduling. 
Furthermore, energy consumption is a critical objective in the dynamic resource management of data centers. Therefore, migration energy cost modeling also needs to be investigated.
Mathematical models, simulation platforms, and empirical methods are used for evaluation. Most of the studies only use one of the evaluation methods, while several studies use more than one method. We introduce the details of available evaluation methods and technologies in the following section.

\section{Evaluation Methods and Technologies}\label{sect: evaluation-methods}
To accelerate the research and development of dynamic resource management in cloud computing, data traces, software and tools are required for testing the migration performance and overheads in the edge and cloud data centers.
Furthermore, evaluation platforms are needed to test and evaluate the networking management algorithms based on OpenFlow protocol and SDN architecture.
Therefore, the testbed needs the capability to measure the energy consumption, downtime impacts, response time, and processing time to properly evaluate the proposed resource management policies and migration scheduling algorithms.
In this section, we introduce the related emulators, simulation tools, and empirical solutions.

\subsection{Simulation and Emulation}
Simulators are essential evaluation platforms that accelerate the innovation and implementation of proposed solutions and algorithms by providing controllable and reproducible experiment environments with ease of configuration and modification. Emulation provides an environment for the end-system to mimic the entire system, such as a network for physical hosts and application programs, and the end-system operates as if it were in a real environment.
The simulation in cloud computing can be divided into computing and networking simulation.

For the networking emulation,
Mininet~\footnote{Mininet. \url{https://github.com/mininet/mininet}} is an open-source network emulator for the rapid prototyping of the Software-Defined Networking, which emulates the entire network of hosts, switches, and links. As it utilizes network virtualization provided by the Linux kernel, Mininet can produce more accurate results in network delays and congestion at the Operating System level. It also natively supports the innovations and implements of SDN controllers and OpenFlow protocol.

Abstracting network traffic descriptions, discrete event network simulators, such as NS-3\footnote{NS-3. \url{https://www.nsnam.org/}}, OMNet++\footnote{OMNet++. \url{https://omnetpp.org/}}, and NetSim\footnote{NetSim. \url{https://www.tetcos.com/}}, provide the extensible, modular, and component-based simulation library and framework to support network simulation.
Researchers could build the SDN module extension to support the OpenFlow and SDN controller simulation. For example,
Chaves et al.~\cite{chaves2016ofswitch13} proposed OFSWITCH13 as an SDN module to support OpenFlow in NS-3. Klein et al.~\cite{klein2013openflow}implement the OpenFlow model by utilizing the INET framework of OMNet++.
NetSim does not support SDN controller and OpenFlow, but one can utilize the discrete-event mechanism to add a new OpenFlow switch module to support forwarding tables and OpenFlow events and use the real-time tunneling interaction supported by NetSim to create communication between the real SDN controller and evaluate the dynamic management algorithms.
Some works~\cite{hirofuchi2013adding} implement the migration module to generate the network traffic to simulate and evaluate the network overheads of live migration. The migration network traffic generation could be improved by profiling the live migration with various resources and applications workloads. However, it can not evaluate the computing cost and overheads induced by the live migration processes.

For the simulation platform for cloud computing,
CloudSim is a popular discrete event-driven cloud simulator implemented in Java. Various data center management algorithms and solutions can be evaluated in CloudSim, including brokering policy, VM instance placement, and dynamic resource reallocation. It also supports workload processing in VMs.  iFogSim~\cite{gupta2017ifogsim} based on CloudSim discrete event-based architecture extending the cloud data center components to fog computing components to simulate the corresponding events in the IoT context. However, CloudSim and iFogSim do not support network events in detail. The instance reallocation through live migration is only modeled as a delayed event according to the memory size.

Based on iFogSim, Myifogsim~\cite{lopes2017myifogsim} and its extension MobFogSim~\cite{puliafito2020mobfogsim} are proposed for the simulation of user mobility on the map with radio base stations (access points). It supports the evaluation of migration policies for migration request generation.
The proposed extension only simulates cold migration and post-copy migration without the support of network topology, network flow, and bandwidth emulation.
However, these proposed simulation platforms and extensions lack the capability of network simulation.
There is no widely-used pre-copy migration simulation model to minimize the downtime (downtime too high for lazy post copy), and no multiply migration simulation, and lacking capabilities for QoS simulation, such as response time of services during the migrations.

Integrating with CloudSim, CloudSimSDN~\cite{son2015cloudsimsdn} is developed and implemented to support packet-level network simulation to evaluate the network transmission and links delays in the data center network architecture including host, switches, and links.
Based on CloudSimSDN, CloudSimSDN-NFV~\cite{son2019cloudsimsdn} provides the NFV supports including automatic scaling and load balancing in SFC, and expends the network architecture from intra-data center to the inter-cloud and edge computing network.
By leveraging simulation capabilities for both computing and networking, we can extend the corresponding components based on the CloudSimSDN to simulate each phase of pre-copy live migration.

\subsection{Empirical Platforms and Prototypes}
The dynamic resource management and live migration management algorithms are located in the orchestration layer.
However, current public and research cloud platforms only support user management at the software level.
There is a lack of experimental infrastructure for cloud management through live migration since the live migration management needs to be performed at the administrator level.

OpenStackEmu~\cite{benet2017openstackemu} combines the OpenStack and SDN controller with network emulation. It enables the connection between the large-scale network and the OpenStack infrastructure hosting the actual VMs. OpenStackEmu also provides network traffic generation in the framework.
SDCon~\cite{son2018sdcon} is a orchestration framework integrating OpenStack, OpenDayLight (ODL) SDN controller, OpenVSwitch and sFlow. By enabling the NetVirt feature in ODL and setting ODL as the default SDN controller in OpenStack Neutron, an OpenStack-based private data center can be used as a testbed for evaluating various dynamic resource management policies and VM migration management algorithms.

As containers can be hosted in VMs, live container migration can be performed in the public cloud platform with virtual networking support across VMs. However, it lacks the ability to manage the networking resources as the underlying network and VMs' location in the hosts can not be guaranteed. The cross-layer operations can bring uncertainty for the migration overhead evaluation.
CloudHopper~\cite{benjaponpitak2020enabling} is a proof-of-concept live migration service for containers to hop around between Amazon Web Services, Google Cloud Platform, and Microsoft Azure.  Live container migration in CloudHopper is automated. It also supports pre-copy optimization, connection holding, traffic redirection, and multiple interdependent container migration.

\subsection{Cloud Trace Data}
Application workloads traces and cloud cluster traces support the realistic, repeatable and comparable evaluation results for various solutions and algorithms.  There are several popular traces used by the evaluation of dynamic resource management in data centers, including
PlanetLab,\footnote{Planet Lab Trace Data.
	\url{https://github.com/beloglazov/planetlab-workload-traces}}
Wikipedia 
\footnote{Wikipedia Data. \url{https://wikitech.wikimedia.org/wiki/Analytics/Archive/Data/Pagecounts-raw}},
Yahoo!,\footnote{Yahoo Benchmark. \url{https://github.com/brianfrankcooper/YCSB/wiki/Core-Workloads}}, 
Google,\footnote{Google Cluster Trace. \url{https://github.com/google/cluster-data}} 
Alibaba Cluster Trace~\footnote{Alibaba Cluster Trace. \url{https://github.com/alibaba/clusterdata}}.

PlanetLab data provides CPU utilization traces from PlanetLab VMs.
Wikipedia data provides the workloads for multi-tier web applications which is a typical application architecture in the data center.
Yahoo! cloud serving benchmark (YCSB) is a set of workloads that defines a basic benchmark for cloud data centers, including update heavy, read mostly, read-only, real least, short ranges, and real-modify-write workloads.
Google Borg cluster workload trace provides traces of workloads running on Google compute cells managed by the cluster management software (Borg).
Alibaba Cluster Trace Program provides trace data of a whole cluster running both online services and batch jobs. The trace data includes both parts of machines data and the workload of the whole cluster.

\section{GAPS Analysis and Future Directions} \label{sect: gaps}
This taxonomy covers several challenges of migration management in cloud and edge data centers. However, cloud and edge computing can be improved by addressing several key issues of live migration that require further investigation. In this section, we analyze gaps and future directions of live migration and migration management in edge and cloud computing.

\subsection{Flexible Networking Management by Network Slicing}
The containerized services allocated in the mobile edge clouds bring up the opportunity for large-scale and real-time applications to have low latency responses. Meanwhile, live container migration is introduced to support dynamic resource management and users' mobility.
The container footprint becomes much smaller compared to the VM.
As a result, the proportion of network resources limitations on migration performance is reduced. Instance-level parallelizing for multiple migration can improve the scheduling performance due to the computing cost of migration. Therefore, it is critical to investigate the networking slicing algorithms to concurrently schedule both VM and container migrations to improve multiple migration performance and
alleviate the impacts on service traffic. Holistic solutions are needed to manage the networking routing and bandwidth allocation based on the various network requirements of live VM migrations, live container migrations, and applications.

\subsection{Optimization Modeling, Compatibility and Combination}
There are continuous efforts striving to improve the performance and alleviate the overheads of live migration through system-level optimization~\cite{zhang2018survey}.
The disadvantage of the original pre-copy migration is the migration convergence.
The optimization works of improving live migration performance mainly focus on reducing the live migration time and downtime, including reducing the memory data required for transmission to the destination, speeding up the migration process in the source host, and increasing the bandwidth between the source and destination hosts.
However, there are gaps between current migration cost and performance modeling and the exiting migration optimization mechanisms in process parallelizing, compression, de-duplication, memory introspection, checkpoint and recovery, remote direct memory access (RDMA), and application awareness. The existing migration optimization is at the system level which needs to be modeled carefully and properly to reflect the nature and characteristics of the optimization.
Although there are extensive optimizations on live VM migration, each one claims a certain performance improvement compared to the default live VM migration mechanism. 
However, it is unclear the compatibility of each migration optimization technology and the performance improvement of the combination of various mechanisms.

\subsection{Live Container Migration Optimization}
There is a research interest and trend of adapting or directing imitating optimization mechanisms of live VM migration to develop and improve live container migrations~\cite{mirkin2008containers,li2015comparing,machen2017live,nadgowda2017voyager,stoyanov2018efficient}.
For example, Remote Direct Memory Access (RDMA) can be utilized as the high-speed interconnect techniques~\cite{kim2019freeflow}, which has been applied to improve the performance of live VM migration. 
RDMA enables the remote accessing of memory and disk data without the CPU and cache.
Multipath TCP (MPTCP) can be used to allow TCP connections to use multiple paths to increase the network throughput and redundancy~\cite{qiu2019experimental}.
The container layered storage can be leveraged to alleviate synchronization overheads of a file system in the architecture without shared storage~\cite{ma2018efficient}.
However, there are still gaps in this research direction and more evaluation and investigation need to be done to speed up this improvement for live container migration.

\subsection{Management Scalability}
With the expansion of edge and cloud computing networks, the complexity of the migration planning and scheduling problem becomes unavoidably large. However, the timeliness requirement of the management algorithm is also changed from cloud to edge computing. For the time-critical applications, the resource and migration management algorithms need to be scalable to suit the large-scale computing and networking environments.
Therefore, it is essential to investigate the distributed management framework and algorithms to enhance the scalability of the migration management algorithms.
Controller placement problems need to be investigated regarding the SDN controller capacity and the network latency between controller and OpenFlow switches.
The problems of edge data center placement and base station clustering need to be investigated based on the user mobility information to reduce the number of live migrations. 
Each edge manager and SDN controller needs to cover a certain area and data centers and cooperates with other controllers. A certain strategy needs to be developed to determine the size of the cluster area and the placement of each manager and controller based on the parameters such as network delay and processing capability.

\subsection{Autoscaling and Live Migration}
For instance-level scaling, scaling up and down is used for allocation and deallocation virtualized instances in cloud computing to elastically provision the resource in the same host. 
In addition, system-level scaling up~\cite{han2012lightweight} supports fine-grained scaling on the system resources, such as CPU, memory, and I/O, to reduce the considerable overhead and extra cost. 
On the other hand, scaling out is used to support application allocation in other compute nodes to increase the processing capacity of the service. The load-balancer will distribute the traffic to all running instances of the application.

Current resource management strategies of cloud providers perform live migration for both stateful and stateless services to achieve the objectives, such as energy consumption and traffic consolidation. There is no configuration and information to differentiate the instance types (stateful and stateless) across various cloud types (IaaS, PaaS, SaaS, and FaaS). The cloud or service providers can further reduce management overheads by performing autoscaling and live migration to stateful and stateless instances respectively.
Therefore, it is critical to integrate the autoscaling and live migration strategies to holistically manage the resources based on the instance types to minimize the management overhead and cost. The advantages and disadvantages of instance-level scaling, resource-level scaling, and live migration need to be investigated. In addition, the specific SLA of migration and scaling is needed for various instance types.

\subsection{Robustness and Security}
For the migration security, Shetty et al.~\cite{shetty2012survey} focus on secure live migration, control policies (DoS, Internal, Guest VM, false resource advertise, Inter VM in the same host), transmission channel (insecure and unprotected), and migration module (stack, heap, integer overflow). SDN-based network resilience management and strategies, such as traffic shaping, anomaly detection, and traffic classification, need to be investigated to tackle the network security issue for live migration.

Furthermore, migration robustness also needs to be investigated to increase the availability and accessibility of dynamic resource management.
Compared to pre-copy migration, post-copy migration starts the instance at the destination host as soon as the initial memory is copied, which makes it vulnerable to state synchronization failure. As a result, if there is a post-copy migration failure, the running processes can not be resumed and the migrating instance is shutdown.
Fernando et al.~\cite{fernando2019live} proposed a failure recovery mechanism for post-copy migration to alleviate the cost of post-copy migration failure.
Failure cost models need to be developed based on different migration types and mechanisms and applied to services with various SLA levels accordingly.

\section{Summary and Conclusions} \label{sect: conclusion}
The article presents a taxonomy of live migration management, which includes performance and cost models, migration generation in resource management policies, planning and scheduling algorithms, management lifecycle and orchestration, and evaluation methods.
Each aspect in the taxonomy is explained in detail and corresponding papers are presented.
We describe and review representative works of dynamic resource management focusing on the migration generation in migration computing parameters, networking parameters, and migration objectives.
We also categorize and review the state-of-the-art on migration planning and scheduling in migration scheduling types, migration network awareness, scheduling scope, heterogeneous and homogeneous solutions, and scheduling objectives.
Various objectives of multiple migration scheduling are explained, following the simulators, emulators, and empirical platforms.
Live migration of VM and container have been facilitated autonomous dynamic resource provision.
Although some studies have been presented in live migration optimization and dynamic resource management.
More research needs to be conducted to fill the gap between resource management and migration management in edge and cloud computing.

	\begin{acks}
	This work is partially supported by an Australian Research Council (ARC) Discovery Project (ID: DP160102414) and a China Scholarship Council - University of Melbourne PhD Scholarship.
	We thank editors and anonymous reviewers for their insightful comments and helpful suggestions on improving this manuscript.
	\end{acks}
	
	\bibliographystyle{ACM-Reference-Format}
	\bibliography{survey}


\begin{thebibliography}{135}


\ifx \showCODEN    \undefined \def \showCODEN     #1{\unskip}     \fi
\ifx \showDOI      \undefined \def \showDOI       #1{#1}\fi
\ifx \showISBNx    \undefined \def \showISBNx     #1{\unskip}     \fi
\ifx \showISBNxiii \undefined \def \showISBNxiii  #1{\unskip}     \fi
\ifx \showISSN     \undefined \def \showISSN      #1{\unskip}     \fi
\ifx \showLCCN     \undefined \def \showLCCN      #1{\unskip}     \fi
\ifx \shownote     \undefined \def \shownote      #1{#1}          \fi
\ifx \showarticletitle \undefined \def \showarticletitle #1{#1}   \fi
\ifx \showURL      \undefined \def \showURL       {\relax}        \fi
\providecommand\bibfield[2]{#2}
\providecommand\bibinfo[2]{#2}
\providecommand\natexlab[1]{#1}
\providecommand\showeprint[2][]{arXiv:#2}

\bibitem[\protect\citeauthoryear{??}{red}{2020}]%
        {redhat-criu}
 \bibinfo{year}{accessed 22 Jan 2020}\natexlab{}.
\newblock \bibinfo{title}{{Container migration with Podman on RHEL}}.
\newblock
\newblock
\urldef\tempurl%
\url{https://www.redhat.com/en/blog/container-migration-podman-rhel}
\showURL{%
\tempurl}


\bibitem[\protect\citeauthoryear{??}{goo}{2021}]%
        {google-e2}
 \bibinfo{year}{accessed 29 June 2021}\natexlab{}.
\newblock \bibinfo{title}{Dynamic resource management in E2 VMs}.
\newblock
\newblock
\urldef\tempurl%
\url{https://cloud.google.com/blog/products/
  compute/understanding-dynamic-resource-management-in-e2-vms}
\showURL{%
\tempurl}


\bibitem[\protect\citeauthoryear{Akoush, Sohan, Rice, Moore, and Hopper}{Akoush
  et~al\mbox{.}}{2010}]%
        {akoush2010predicting}
\bibfield{author}{\bibinfo{person}{Sherif Akoush}, \bibinfo{person}{Ripduman
  Sohan}, \bibinfo{person}{Andrew Rice}, \bibinfo{person}{Andrew~W Moore},
  {and} \bibinfo{person}{Andy Hopper}.} \bibinfo{year}{2010}\natexlab{}.
\newblock \showarticletitle{Predicting the performance of virtual machine
  migration}. In \bibinfo{booktitle}{\emph{2010 IEEE international symposium on
  modeling, analysis and simulation of computer and telecommunication
  systems}}. IEEE, \bibinfo{pages}{37--46}.
\newblock


\bibitem[\protect\citeauthoryear{Aldossary and Djemame}{Aldossary and
  Djemame}{2018}]%
        {aldossary2018performance}
\bibfield{author}{\bibinfo{person}{Mohammad Aldossary} {and}
  \bibinfo{person}{Karim Djemame}.} \bibinfo{year}{2018}\natexlab{}.
\newblock \showarticletitle{Performance and Energy-based Cost Prediction of
  Virtual Machines Live Migration in Clouds.}. In
  \bibinfo{booktitle}{\emph{CLOSER}}. \bibinfo{pages}{384--391}.
\newblock


\bibitem[\protect\citeauthoryear{Armbrust, Fox, Griffith, Joseph, Katz,
  Konwinski, Lee, Patterson, Rabkin, Stoica, et~al\mbox{.}}{Armbrust
  et~al\mbox{.}}{2010}]%
        {armbrust2010view}
\bibfield{author}{\bibinfo{person}{Michael Armbrust}, \bibinfo{person}{Armando
  Fox}, \bibinfo{person}{Rean Griffith}, \bibinfo{person}{Anthony~D Joseph},
  \bibinfo{person}{Randy Katz}, \bibinfo{person}{Andy Konwinski},
  \bibinfo{person}{Gunho Lee}, \bibinfo{person}{David Patterson},
  \bibinfo{person}{Ariel Rabkin}, \bibinfo{person}{Ion Stoica},
  {et~al\mbox{.}}} \bibinfo{year}{2010}\natexlab{}.
\newblock \showarticletitle{A view of cloud computing}.
\newblock \bibinfo{journal}{\emph{Commun. ACM}} \bibinfo{volume}{53},
  \bibinfo{number}{4} (\bibinfo{year}{2010}), \bibinfo{pages}{50--58}.
\newblock


\bibitem[\protect\citeauthoryear{Bari, Zhani, Zhang, Ahmed, and Boutaba}{Bari
  et~al\mbox{.}}{2014}]%
        {bari2014cqncr}
\bibfield{author}{\bibinfo{person}{Md~Faizul Bari},
  \bibinfo{person}{Mohamed~Faten Zhani}, \bibinfo{person}{Qi Zhang},
  \bibinfo{person}{Reaz Ahmed}, {and} \bibinfo{person}{Raouf Boutaba}.}
  \bibinfo{year}{2014}\natexlab{}.
\newblock \showarticletitle{CQNCR: Optimal VM migration planning in cloud data
  centers}. In \bibinfo{booktitle}{\emph{Proceedings of 2014 IFIP Networking
  Conference}}. IEEE, \bibinfo{pages}{1--9}.
\newblock


\bibitem[\protect\citeauthoryear{Beloglazov, Abawajy, and Buyya}{Beloglazov
  et~al\mbox{.}}{2012}]%
        {beloglazov2012energy}
\bibfield{author}{\bibinfo{person}{Anton Beloglazov}, \bibinfo{person}{Jemal
  Abawajy}, {and} \bibinfo{person}{Rajkumar Buyya}.}
  \bibinfo{year}{2012}\natexlab{}.
\newblock \showarticletitle{Energy-aware resource allocation heuristics for
  efficient management of data centers for cloud computing}.
\newblock \bibinfo{journal}{\emph{Future generation computer systems}}
  \bibinfo{volume}{28}, \bibinfo{number}{5} (\bibinfo{year}{2012}),
  \bibinfo{pages}{755--768}.
\newblock


\bibitem[\protect\citeauthoryear{Beloglazov and Buyya}{Beloglazov and
  Buyya}{2012a}]%
        {beloglazov2012managing}
\bibfield{author}{\bibinfo{person}{Anton Beloglazov} {and}
  \bibinfo{person}{Rajkumar Buyya}.} \bibinfo{year}{2012}\natexlab{a}.
\newblock \showarticletitle{Managing overloaded hosts for dynamic consolidation
  of virtual machines in cloud data centers under quality of service
  constraints}.
\newblock \bibinfo{journal}{\emph{IEEE transactions on parallel and distributed
  systems}} \bibinfo{volume}{24}, \bibinfo{number}{7} (\bibinfo{year}{2012}),
  \bibinfo{pages}{1366--1379}.
\newblock


\bibitem[\protect\citeauthoryear{Beloglazov and Buyya}{Beloglazov and
  Buyya}{2012b}]%
        {beloglazov2012optimal}
\bibfield{author}{\bibinfo{person}{Anton Beloglazov} {and}
  \bibinfo{person}{Rajkumar Buyya}.} \bibinfo{year}{2012}\natexlab{b}.
\newblock \showarticletitle{Optimal online deterministic algorithms and
  adaptive heuristics for energy and performance efficient dynamic
  consolidation of virtual machines in cloud data centers}.
\newblock \bibinfo{journal}{\emph{Concurrency and Computation: Practice and
  Experience}} \bibinfo{volume}{24}, \bibinfo{number}{13}
  (\bibinfo{year}{2012}), \bibinfo{pages}{1397--1420}.
\newblock


\bibitem[\protect\citeauthoryear{Benet, Nasim, Noghani, and Kassler}{Benet
  et~al\mbox{.}}{2017}]%
        {benet2017openstackemu}
\bibfield{author}{\bibinfo{person}{Cristian~Hernandez Benet},
  \bibinfo{person}{Robayet Nasim}, \bibinfo{person}{Kyoomars~Alizadeh Noghani},
  {and} \bibinfo{person}{Andreas Kassler}.} \bibinfo{year}{2017}\natexlab{}.
\newblock \showarticletitle{OpenStackEmu—A cloud testbed combining network
  emulation with OpenStack and SDN}. In \bibinfo{booktitle}{\emph{2017 14th
  IEEE Annual Consumer Communications \& Networking Conference (CCNC)}}. IEEE,
  \bibinfo{pages}{566--568}.
\newblock


\bibitem[\protect\citeauthoryear{Benjaponpitak, Karakate, and
  Sripanidkulchai}{Benjaponpitak et~al\mbox{.}}{2020}]%
        {benjaponpitak2020enabling}
\bibfield{author}{\bibinfo{person}{Thad Benjaponpitak},
  \bibinfo{person}{Meatasit Karakate}, {and} \bibinfo{person}{Kunwadee
  Sripanidkulchai}.} \bibinfo{year}{2020}\natexlab{}.
\newblock \showarticletitle{Enabling Live Migration of Containerized
  Applications Across Clouds}. In \bibinfo{booktitle}{\emph{IEEE INFOCOM
  2020-IEEE Conference on Computer Communications}}. IEEE,
  \bibinfo{pages}{2529--2538}.
\newblock


\bibitem[\protect\citeauthoryear{Bernstein}{Bernstein}{2014}]%
        {bernstein2014containers}
\bibfield{author}{\bibinfo{person}{David Bernstein}.}
  \bibinfo{year}{2014}\natexlab{}.
\newblock \showarticletitle{Containers and cloud: From lxc to docker to
  kubernetes}.
\newblock \bibinfo{journal}{\emph{IEEE Cloud Computing}} \bibinfo{volume}{1},
  \bibinfo{number}{3} (\bibinfo{year}{2014}), \bibinfo{pages}{81--84}.
\newblock


\bibitem[\protect\citeauthoryear{Breitgand, Kutiel, and Raz}{Breitgand
  et~al\mbox{.}}{2010}]%
        {breitgand2010cost}
\bibfield{author}{\bibinfo{person}{David Breitgand}, \bibinfo{person}{Gilad
  Kutiel}, {and} \bibinfo{person}{Danny Raz}.} \bibinfo{year}{2010}\natexlab{}.
\newblock \showarticletitle{Cost-aware live migration of services in the
  cloud.}
\newblock \bibinfo{journal}{\emph{SYSTOR}}  \bibinfo{volume}{10}
  (\bibinfo{year}{2010}), \bibinfo{pages}{1815695--1815709}.
\newblock


\bibitem[\protect\citeauthoryear{Callegati and Cerroni}{Callegati and
  Cerroni}{2013}]%
        {callegati2013live}
\bibfield{author}{\bibinfo{person}{Franco Callegati} {and}
  \bibinfo{person}{Walter Cerroni}.} \bibinfo{year}{2013}\natexlab{}.
\newblock \showarticletitle{Live migration of virtualized edge networks:
  Analytical modeling and performance evaluation}. In
  \bibinfo{booktitle}{\emph{2013 IEEE SDN for future networks and services
  (SDN4FNS)}}. IEEE, \bibinfo{pages}{1--6}.
\newblock


\bibitem[\protect\citeauthoryear{Cao, Gao, Chen, and Jin}{Cao
  et~al\mbox{.}}{2014}]%
        {cao2014nice}
\bibfield{author}{\bibinfo{person}{Bo Cao}, \bibinfo{person}{Xiaofeng Gao},
  \bibinfo{person}{Guihai Chen}, {and} \bibinfo{person}{Yaohui Jin}.}
  \bibinfo{year}{2014}\natexlab{}.
\newblock \showarticletitle{NICE: network-aware VM consolidation scheme for
  energy conservation in data centers}. In \bibinfo{booktitle}{\emph{2014 20th
  IEEE International Conference on Parallel and Distributed Systems (ICPADS)}}.
  IEEE, \bibinfo{pages}{166--173}.
\newblock


\bibitem[\protect\citeauthoryear{Chaves, Garcia, and Madeira}{Chaves
  et~al\mbox{.}}{2016}]%
        {chaves2016ofswitch13}
\bibfield{author}{\bibinfo{person}{Luciano~Jerez Chaves},
  \bibinfo{person}{Islene~Calciolari Garcia}, {and} \bibinfo{person}{Edmundo
  Roberto~Mauro Madeira}.} \bibinfo{year}{2016}\natexlab{}.
\newblock \showarticletitle{Ofswitch13: Enhancing ns-3 with openflow 1.3
  support}. In \bibinfo{booktitle}{\emph{Proceedings of the Workshop on ns-3}}.
  \bibinfo{pages}{33--40}.
\newblock


\bibitem[\protect\citeauthoryear{Clark, Fraser, Hand, Hansen, Jul, Limpach,
  Pratt, and Warfield}{Clark et~al\mbox{.}}{2005}]%
        {clark2005live}
\bibfield{author}{\bibinfo{person}{Christopher Clark}, \bibinfo{person}{Keir
  Fraser}, \bibinfo{person}{Steven Hand}, \bibinfo{person}{Jacob~Gorm Hansen},
  \bibinfo{person}{Eric Jul}, \bibinfo{person}{Christian Limpach},
  \bibinfo{person}{Ian Pratt}, {and} \bibinfo{person}{Andrew Warfield}.}
  \bibinfo{year}{2005}\natexlab{}.
\newblock \showarticletitle{Live migration of virtual machines}. In
  \bibinfo{booktitle}{\emph{Proceedings of the 2nd conference on Symposium on
  Networked Systems Design \& Implementation-Volume 2}}.
  \bibinfo{pages}{273--286}.
\newblock


\bibitem[\protect\citeauthoryear{CRIU}{CRIU}{2019}]%
        {criu}
\bibfield{author}{\bibinfo{person}{CRIU}.} \bibinfo{year}{2019}\natexlab{}.
\newblock \bibinfo{title}{Live migration}.
\newblock
\newblock
\urldef\tempurl%
\url{https://criu.org/Live_migration}
\showURL{%
\tempurl}


\bibitem[\protect\citeauthoryear{Cui, Cziva, Tso, and Pezaros}{Cui
  et~al\mbox{.}}{2016a}]%
        {cui2016synergistic}
\bibfield{author}{\bibinfo{person}{Lin Cui}, \bibinfo{person}{Richard Cziva},
  \bibinfo{person}{Fung~Po Tso}, {and} \bibinfo{person}{Dimitrios~P Pezaros}.}
  \bibinfo{year}{2016}\natexlab{a}.
\newblock \showarticletitle{Synergistic policy and virtual machine
  consolidation in cloud data centers}. In \bibinfo{booktitle}{\emph{IEEE
  INFOCOM 2016-The 35th Annual IEEE International Conference on Computer
  Communications}}. IEEE, \bibinfo{pages}{1--9}.
\newblock


\bibitem[\protect\citeauthoryear{Cui, Tso, Pezaros, Jia, and Zhao}{Cui
  et~al\mbox{.}}{2016b}]%
        {cui2016plan}
\bibfield{author}{\bibinfo{person}{Lin Cui}, \bibinfo{person}{Fung~Po Tso},
  \bibinfo{person}{Dimitrios~P Pezaros}, \bibinfo{person}{Weijia Jia}, {and}
  \bibinfo{person}{Wei Zhao}.} \bibinfo{year}{2016}\natexlab{b}.
\newblock \showarticletitle{Plan: Joint policy-and network-aware vm management
  for cloud data centers}.
\newblock \bibinfo{journal}{\emph{IEEE Transactions on Parallel and Distributed
  Systems}} \bibinfo{volume}{28}, \bibinfo{number}{4} (\bibinfo{year}{2016}),
  \bibinfo{pages}{1163--1175}.
\newblock


\bibitem[\protect\citeauthoryear{Cui, Yang, Xiao, Wang, and Yan}{Cui
  et~al\mbox{.}}{2017}]%
        {cui2017traffic}
\bibfield{author}{\bibinfo{person}{Yong Cui}, \bibinfo{person}{Zhenjie Yang},
  \bibinfo{person}{Shihan Xiao}, \bibinfo{person}{Xin Wang}, {and}
  \bibinfo{person}{Shenghui Yan}.} \bibinfo{year}{2017}\natexlab{}.
\newblock \showarticletitle{Traffic-aware virtual machine migration in
  topology-adaptive dcn}.
\newblock \bibinfo{journal}{\emph{IEEE/ACM Transactions on Networking}}
  \bibinfo{volume}{25}, \bibinfo{number}{6} (\bibinfo{year}{2017}),
  \bibinfo{pages}{3427--3440}.
\newblock


\bibitem[\protect\citeauthoryear{Cziva, Anagnostopoulos, and Pezaros}{Cziva
  et~al\mbox{.}}{2018}]%
        {cziva2018dynamic}
\bibfield{author}{\bibinfo{person}{Richard Cziva}, \bibinfo{person}{Christos
  Anagnostopoulos}, {and} \bibinfo{person}{Dimitrios~P Pezaros}.}
  \bibinfo{year}{2018}\natexlab{}.
\newblock \showarticletitle{Dynamic, latency-optimal vNF placement at the
  network edge}. In \bibinfo{booktitle}{\emph{Ieee infocom 2018-ieee conference
  on computer communications}}. IEEE, \bibinfo{pages}{693--701}.
\newblock


\bibitem[\protect\citeauthoryear{Cziva, Jou{\"e}t, Stapleton, Tso, and
  Pezaros}{Cziva et~al\mbox{.}}{2016}]%
        {cziva2016sdn}
\bibfield{author}{\bibinfo{person}{Richard Cziva}, \bibinfo{person}{Simon
  Jou{\"e}t}, \bibinfo{person}{David Stapleton}, \bibinfo{person}{Fung~Po Tso},
  {and} \bibinfo{person}{Dimitrios~P Pezaros}.}
  \bibinfo{year}{2016}\natexlab{}.
\newblock \showarticletitle{SDN-based virtual machine management for cloud data
  centers}.
\newblock \bibinfo{journal}{\emph{IEEE Transactions on Network and Service
  Management}} \bibinfo{volume}{13}, \bibinfo{number}{2}
  (\bibinfo{year}{2016}), \bibinfo{pages}{212--225}.
\newblock


\bibitem[\protect\citeauthoryear{Deshpande and Keahey}{Deshpande and
  Keahey}{2017}]%
        {deshpande2017traffic}
\bibfield{author}{\bibinfo{person}{Umesh Deshpande} {and} \bibinfo{person}{Kate
  Keahey}.} \bibinfo{year}{2017}\natexlab{}.
\newblock \showarticletitle{Traffic-sensitive live migration of virtual
  machines}.
\newblock \bibinfo{journal}{\emph{Future Generation Computer Systems}}
  \bibinfo{volume}{72} (\bibinfo{year}{2017}), \bibinfo{pages}{118--128}.
\newblock


\bibitem[\protect\citeauthoryear{Deshpande, Kulkarni, and Gopalan}{Deshpande
  et~al\mbox{.}}{2012}]%
        {deshpande2012inter}
\bibfield{author}{\bibinfo{person}{Umesh Deshpande}, \bibinfo{person}{Unmesh
  Kulkarni}, {and} \bibinfo{person}{Kartik Gopalan}.}
  \bibinfo{year}{2012}\natexlab{}.
\newblock \showarticletitle{Inter-rack live migration of multiple virtual
  machines}. In \bibinfo{booktitle}{\emph{Proceedings of the 6th international
  workshop on Virtualization Technologies in Distributed Computing Date}}.
  \bibinfo{pages}{19--26}.
\newblock


\bibitem[\protect\citeauthoryear{Deshpande, Schlinker, Adler, and
  Gopalan}{Deshpande et~al\mbox{.}}{2013}]%
        {deshpande2013gang}
\bibfield{author}{\bibinfo{person}{Umesh Deshpande}, \bibinfo{person}{Brandon
  Schlinker}, \bibinfo{person}{Eitan Adler}, {and} \bibinfo{person}{Kartik
  Gopalan}.} \bibinfo{year}{2013}\natexlab{}.
\newblock \showarticletitle{Gang migration of virtual machines using
  cluster-wide deduplication}. In \bibinfo{booktitle}{\emph{2013 13th IEEE/ACM
  International Symposium on Cluster, Cloud, and Grid Computing}}. IEEE,
  \bibinfo{pages}{394--401}.
\newblock


\bibitem[\protect\citeauthoryear{Deshpande, Wang, and Gopalan}{Deshpande
  et~al\mbox{.}}{2011}]%
        {deshpande2011live}
\bibfield{author}{\bibinfo{person}{Umesh Deshpande},
  \bibinfo{person}{Xiaoshuang Wang}, {and} \bibinfo{person}{Kartik Gopalan}.}
  \bibinfo{year}{2011}\natexlab{}.
\newblock \showarticletitle{Live gang migration of virtual machines}. In
  \bibinfo{booktitle}{\emph{Proceedings of the 20th international symposium on
  High performance distributed computing}}. \bibinfo{pages}{135--146}.
\newblock


\bibitem[\protect\citeauthoryear{Deshpande, You, Chan, Bila, and
  Gopalan}{Deshpande et~al\mbox{.}}{2014}]%
        {deshpande2014fast}
\bibfield{author}{\bibinfo{person}{Umesh Deshpande}, \bibinfo{person}{Yang
  You}, \bibinfo{person}{Danny Chan}, \bibinfo{person}{Nilton Bila}, {and}
  \bibinfo{person}{Kartik Gopalan}.} \bibinfo{year}{2014}\natexlab{}.
\newblock \showarticletitle{Fast server deprovisioning through scatter-gather
  live migration of virtual machines}. In \bibinfo{booktitle}{\emph{2014 IEEE
  7th International Conference on Cloud Computing}}. IEEE,
  \bibinfo{pages}{376--383}.
\newblock


\bibitem[\protect\citeauthoryear{Elsaid, Abbas, and Meinel}{Elsaid
  et~al\mbox{.}}{2019}]%
        {elsaid2019machine}
\bibfield{author}{\bibinfo{person}{Mohamed~Esam Elsaid},
  \bibinfo{person}{Hazem~M Abbas}, {and} \bibinfo{person}{Christoph Meinel}.}
  \bibinfo{year}{2019}\natexlab{}.
\newblock \showarticletitle{Machine Learning Approach for Live Migration Cost
  Prediction in VMware Environments.}. In \bibinfo{booktitle}{\emph{CLOSER}}.
  \bibinfo{pages}{456--463}.
\newblock


\bibitem[\protect\citeauthoryear{Elsaid and Meinel}{Elsaid and Meinel}{2014}]%
        {elsaid2014live}
\bibfield{author}{\bibinfo{person}{Mohamed~Esam Elsaid} {and}
  \bibinfo{person}{Christoph Meinel}.} \bibinfo{year}{2014}\natexlab{}.
\newblock \showarticletitle{Live migration impact on virtual datacenter
  performance: VMware vMotion based study}. In \bibinfo{booktitle}{\emph{2014
  International Conference on Future Internet of Things and Cloud}}. IEEE,
  \bibinfo{pages}{216--221}.
\newblock


\bibitem[\protect\citeauthoryear{Eramo, Miucci, Ammar, and Lavacca}{Eramo
  et~al\mbox{.}}{2017}]%
        {eramo2017approach}
\bibfield{author}{\bibinfo{person}{Vincenzo Eramo}, \bibinfo{person}{Emanuele
  Miucci}, \bibinfo{person}{Mostafa Ammar}, {and}
  \bibinfo{person}{Francesco~Giacinto Lavacca}.}
  \bibinfo{year}{2017}\natexlab{}.
\newblock \showarticletitle{An approach for service function chain routing and
  virtual function network instance migration in network function
  virtualization architectures}.
\newblock \bibinfo{journal}{\emph{IEEE/ACM Transactions on Networking}}
  \bibinfo{volume}{25}, \bibinfo{number}{4} (\bibinfo{year}{2017}),
  \bibinfo{pages}{2008--2025}.
\newblock


\bibitem[\protect\citeauthoryear{Fernando, Terner, Gopalan, and Yang}{Fernando
  et~al\mbox{.}}{2019}]%
        {fernando2019live}
\bibfield{author}{\bibinfo{person}{Dinuni Fernando}, \bibinfo{person}{Jonathan
  Terner}, \bibinfo{person}{Kartik Gopalan}, {and} \bibinfo{person}{Ping
  Yang}.} \bibinfo{year}{2019}\natexlab{}.
\newblock \showarticletitle{Live migration ate my vm: Recovering a virtual
  machine after failure of post-copy live migration}. In
  \bibinfo{booktitle}{\emph{IEEE INFOCOM 2019-IEEE Conference on Computer
  Communications}}. IEEE, \bibinfo{pages}{343--351}.
\newblock


\bibitem[\protect\citeauthoryear{Fernando, Yang, and Lu}{Fernando
  et~al\mbox{.}}{2020}]%
        {fernando2020sdn}
\bibfield{author}{\bibinfo{person}{Dinuni Fernando}, \bibinfo{person}{Ping
  Yang}, {and} \bibinfo{person}{Hui Lu}.} \bibinfo{year}{2020}\natexlab{}.
\newblock \showarticletitle{SDN-based Order-aware Live Migration of Virtual
  Machines}. In \bibinfo{booktitle}{\emph{IEEE INFOCOM 2020-IEEE Conference on
  Computer Communications}}. IEEE, \bibinfo{pages}{1818--1827}.
\newblock


\bibitem[\protect\citeauthoryear{Fichera, Gharbaoui, Castoldi, Martini, and
  Manzalini}{Fichera et~al\mbox{.}}{2017}]%
        {fichera2017experimenting}
\bibfield{author}{\bibinfo{person}{Silvia Fichera}, \bibinfo{person}{Molka
  Gharbaoui}, \bibinfo{person}{Piero Castoldi}, \bibinfo{person}{Barbara
  Martini}, {and} \bibinfo{person}{Antonio Manzalini}.}
  \bibinfo{year}{2017}\natexlab{}.
\newblock \showarticletitle{On experimenting 5G: Testbed set-up for SDN
  orchestration across network cloud and IoT domains}. In
  \bibinfo{booktitle}{\emph{2017 IEEE Conference on Network Softwarization
  (NetSoft)}}. IEEE, \bibinfo{pages}{1--6}.
\newblock


\bibitem[\protect\citeauthoryear{Flores, Tran, and Tang}{Flores
  et~al\mbox{.}}{2020}]%
        {flores2020pam}
\bibfield{author}{\bibinfo{person}{Hugo Flores}, \bibinfo{person}{Vincent
  Tran}, {and} \bibinfo{person}{Bin Tang}.} \bibinfo{year}{2020}\natexlab{}.
\newblock \showarticletitle{PAM \& PAL: Policy-Aware Virtual Machine Migration
  and Placement in Dynamic Cloud Data Centers}. In
  \bibinfo{booktitle}{\emph{IEEE INFOCOM 2020-IEEE Conference on Computer
  Communications}}. IEEE, \bibinfo{pages}{2549--2558}.
\newblock


\bibitem[\protect\citeauthoryear{Forsman, Glad, Lundberg, and Ilie}{Forsman
  et~al\mbox{.}}{2015}]%
        {forsman2015algorithms}
\bibfield{author}{\bibinfo{person}{Mattias Forsman}, \bibinfo{person}{Andreas
  Glad}, \bibinfo{person}{Lars Lundberg}, {and} \bibinfo{person}{Dragos Ilie}.}
  \bibinfo{year}{2015}\natexlab{}.
\newblock \showarticletitle{Algorithms for automated live migration of virtual
  machines}.
\newblock \bibinfo{journal}{\emph{Journal of Systems and Software}}
  \bibinfo{volume}{101} (\bibinfo{year}{2015}), \bibinfo{pages}{110--126}.
\newblock


\bibitem[\protect\citeauthoryear{Ge, Sun, Wang, Xu, and Wu}{Ge
  et~al\mbox{.}}{2014}]%
        {ge2014energy}
\bibfield{author}{\bibinfo{person}{Chang Ge}, \bibinfo{person}{Zhili Sun},
  \bibinfo{person}{Ning Wang}, \bibinfo{person}{Ke Xu}, {and}
  \bibinfo{person}{Jinsong Wu}.} \bibinfo{year}{2014}\natexlab{}.
\newblock \showarticletitle{Energy management in cross-domain content delivery
  networks: A theoretical perspective}.
\newblock \bibinfo{journal}{\emph{IEEE Transactions on Network and Service
  Management}} \bibinfo{volume}{11}, \bibinfo{number}{3}
  (\bibinfo{year}{2014}), \bibinfo{pages}{264--277}.
\newblock


\bibitem[\protect\citeauthoryear{Ghorbani and Caesar}{Ghorbani and
  Caesar}{2012}]%
        {ghorbani2012}
\bibfield{author}{\bibinfo{person}{Soudeh Ghorbani} {and}
  \bibinfo{person}{Matthew Caesar}.} \bibinfo{year}{2012}\natexlab{}.
\newblock \showarticletitle{Walk the line: consistent network updates with
  bandwidth guarantees}. In \bibinfo{booktitle}{\emph{Proceedings of the first
  workshop on Hot topics in software defined networks}}.
  \bibinfo{publisher}{ACM}, \bibinfo{pages}{67--72}.
\newblock
\showISBNx{1450314775}


\bibitem[\protect\citeauthoryear{Gupta, Vahid~Dastjerdi, Ghosh, and
  Buyya}{Gupta et~al\mbox{.}}{2017}]%
        {gupta2017ifogsim}
\bibfield{author}{\bibinfo{person}{Harshit Gupta}, \bibinfo{person}{Amir
  Vahid~Dastjerdi}, \bibinfo{person}{Soumya~K Ghosh}, {and}
  \bibinfo{person}{Rajkumar Buyya}.} \bibinfo{year}{2017}\natexlab{}.
\newblock \showarticletitle{iFogSim: A toolkit for modeling and simulation of
  resource management techniques in the Internet of Things, Edge and Fog
  computing environments}.
\newblock \bibinfo{journal}{\emph{Software: Practice and Experience}}
  \bibinfo{volume}{47}, \bibinfo{number}{9} (\bibinfo{year}{2017}),
  \bibinfo{pages}{1275--1296}.
\newblock


\bibitem[\protect\citeauthoryear{Han, Guo, Ghanem, and Guo}{Han
  et~al\mbox{.}}{2012}]%
        {han2012lightweight}
\bibfield{author}{\bibinfo{person}{Rui Han}, \bibinfo{person}{Li Guo},
  \bibinfo{person}{Moustafa~M Ghanem}, {and} \bibinfo{person}{Yike Guo}.}
  \bibinfo{year}{2012}\natexlab{}.
\newblock \showarticletitle{Lightweight resource scaling for cloud
  applications}. In \bibinfo{booktitle}{\emph{Proceedings of 2012 12th IEEE/ACM
  International Symposium on Cluster, Cloud and Grid Computing (CCGRID)}}.
  IEEE, \bibinfo{pages}{644--651}.
\newblock


\bibitem[\protect\citeauthoryear{Harney, Goasguen, Martin, Murphy, and
  Westall}{Harney et~al\mbox{.}}{2007}]%
        {harney2007efficacy}
\bibfield{author}{\bibinfo{person}{Eric Harney}, \bibinfo{person}{Sebastien
  Goasguen}, \bibinfo{person}{Jim Martin}, \bibinfo{person}{Mike Murphy}, {and}
  \bibinfo{person}{Mike Westall}.} \bibinfo{year}{2007}\natexlab{}.
\newblock \showarticletitle{The efficacy of live virtual machine migrations
  over the internet}. In \bibinfo{booktitle}{\emph{Proceedings of the 2nd
  International Workshop on Virtualization Technology in Distributed Computing
  (VTDC'07)}}. IEEE, \bibinfo{pages}{1--7}.
\newblock


\bibitem[\protect\citeauthoryear{He, {N. Toosi}, and Buyya}{He
  et~al\mbox{.}}{2019}]%
        {he2019performance}
\bibfield{author}{\bibinfo{person}{TianZhang He}, \bibinfo{person}{Adel {N.
  Toosi}}, {and} \bibinfo{person}{Rajkumar Buyya}.}
  \bibinfo{year}{2019}\natexlab{}.
\newblock \showarticletitle{Performance evaluation of live virtual machine
  migration in SDN-enabled cloud data centers}.
\newblock \bibinfo{journal}{\emph{J. Parallel and Distrib. Comput.}}
  \bibinfo{volume}{131} (\bibinfo{year}{2019}), \bibinfo{pages}{55--68}.
\newblock


\bibitem[\protect\citeauthoryear{He, Toosi, and Buyya}{He
  et~al\mbox{.}}{2021}]%
        {he2021sla}
\bibfield{author}{\bibinfo{person}{TianZhang He}, \bibinfo{person}{Adel~N.
  Toosi}, {and} \bibinfo{person}{Rajkumar Buyya}.}
  \bibinfo{year}{2021}\natexlab{}.
\newblock \showarticletitle{SLA-aware multiple migration planning and
  scheduling in SDN-NFV-enabled clouds}.
\newblock \bibinfo{journal}{\emph{Journal of Systems and Software}}
  \bibinfo{volume}{176} (\bibinfo{year}{2021}), \bibinfo{pages}{110943}.
\newblock


\bibitem[\protect\citeauthoryear{Heller, Seetharaman, Mahadevan, Yiakoumis,
  Sharma, Banerjee, and McKeown}{Heller et~al\mbox{.}}{2010}]%
        {heller2010elastictree}
\bibfield{author}{\bibinfo{person}{Brandon Heller}, \bibinfo{person}{Srinivasan
  Seetharaman}, \bibinfo{person}{Priya Mahadevan}, \bibinfo{person}{Yiannis
  Yiakoumis}, \bibinfo{person}{Puneet Sharma}, \bibinfo{person}{Sujata
  Banerjee}, {and} \bibinfo{person}{Nick McKeown}.}
  \bibinfo{year}{2010}\natexlab{}.
\newblock \showarticletitle{Elastictree: Saving energy in data center
  networks.}. In \bibinfo{booktitle}{\emph{Nsdi}}, Vol.~\bibinfo{volume}{10}.
  \bibinfo{pages}{249--264}.
\newblock


\bibitem[\protect\citeauthoryear{Hines, Deshpande, and Gopalan}{Hines
  et~al\mbox{.}}{2009}]%
        {hines2009post}
\bibfield{author}{\bibinfo{person}{Michael~R Hines}, \bibinfo{person}{Umesh
  Deshpande}, {and} \bibinfo{person}{Kartik Gopalan}.}
  \bibinfo{year}{2009}\natexlab{}.
\newblock \showarticletitle{Post-copy live migration of virtual machines}.
\newblock \bibinfo{journal}{\emph{ACM SIGOPS operating systems review}}
  \bibinfo{volume}{43}, \bibinfo{number}{3} (\bibinfo{year}{2009}),
  \bibinfo{pages}{14--26}.
\newblock


\bibitem[\protect\citeauthoryear{Hirofuchi, L{\`e}bre, and Pouilloux}{Hirofuchi
  et~al\mbox{.}}{2013}]%
        {hirofuchi2013adding}
\bibfield{author}{\bibinfo{person}{Takahiro Hirofuchi}, \bibinfo{person}{Adrien
  L{\`e}bre}, {and} \bibinfo{person}{Laurent Pouilloux}.}
  \bibinfo{year}{2013}\natexlab{}.
\newblock \showarticletitle{Adding a live migration model into simgrid: One
  more step toward the simulation of infrastructure-as-a-service concerns}. In
  \bibinfo{booktitle}{\emph{2013 IEEE 5th International Conference on Cloud
  Computing Technology and Science}}, Vol.~\bibinfo{volume}{1}. IEEE,
  \bibinfo{pages}{96--103}.
\newblock


\bibitem[\protect\citeauthoryear{Hong and Varghese}{Hong and Varghese}{2019}]%
        {hong2019resource}
\bibfield{author}{\bibinfo{person}{Cheol-Ho Hong} {and}
  \bibinfo{person}{Blesson Varghese}.} \bibinfo{year}{2019}\natexlab{}.
\newblock \showarticletitle{Resource management in fog/edge computing: a survey
  on architectures, infrastructure, and algorithms}.
\newblock \bibinfo{journal}{\emph{ACM Computing Surveys (CSUR)}}
  \bibinfo{volume}{52}, \bibinfo{number}{5} (\bibinfo{year}{2019}),
  \bibinfo{pages}{1--37}.
\newblock


\bibitem[\protect\citeauthoryear{Hu, Hicks, Zhang, Dow, Soni, Jiang, Bull, and
  Matthews}{Hu et~al\mbox{.}}{2013}]%
        {hu2013quantitative}
\bibfield{author}{\bibinfo{person}{Wenjin Hu}, \bibinfo{person}{Andrew Hicks},
  \bibinfo{person}{Long Zhang}, \bibinfo{person}{Eli~M Dow},
  \bibinfo{person}{Vinay Soni}, \bibinfo{person}{Hao Jiang},
  \bibinfo{person}{Ronny Bull}, {and} \bibinfo{person}{Jeanna~N Matthews}.}
  \bibinfo{year}{2013}\natexlab{}.
\newblock \showarticletitle{A quantitative study of virtual machine live
  migration}. In \bibinfo{booktitle}{\emph{Proceedings of the 2013 ACM cloud
  and autonomic computing conference}}. \bibinfo{pages}{1--10}.
\newblock


\bibitem[\protect\citeauthoryear{Hu, Patel, Sabella, Sprecher, and Young}{Hu
  et~al\mbox{.}}{2015}]%
        {hu2015mobile}
\bibfield{author}{\bibinfo{person}{Yun~Chao Hu}, \bibinfo{person}{Milan Patel},
  \bibinfo{person}{Dario Sabella}, \bibinfo{person}{Nurit Sprecher}, {and}
  \bibinfo{person}{Valerie Young}.} \bibinfo{year}{2015}\natexlab{}.
\newblock \showarticletitle{Mobile edge computing A key technology towards 5G}.
\newblock \bibinfo{journal}{\emph{ETSI white paper}} \bibinfo{volume}{11},
  \bibinfo{number}{11} (\bibinfo{year}{2015}), \bibinfo{pages}{1--16}.
\newblock


\bibitem[\protect\citeauthoryear{Huang, Gao, Wang, and Qi}{Huang
  et~al\mbox{.}}{2011}]%
        {huang2011power}
\bibfield{author}{\bibinfo{person}{Qiang Huang}, \bibinfo{person}{Fengqian
  Gao}, \bibinfo{person}{Rui Wang}, {and} \bibinfo{person}{Zhengwei Qi}.}
  \bibinfo{year}{2011}\natexlab{}.
\newblock \showarticletitle{Power consumption of virtual machine live migration
  in clouds}. In \bibinfo{booktitle}{\emph{2011 Third International Conference
  on Communications and Mobile Computing}}. IEEE, \bibinfo{pages}{122--125}.
\newblock


\bibitem[\protect\citeauthoryear{Huang, Shuang, Xu, Li, Liu, and Su}{Huang
  et~al\mbox{.}}{2014}]%
        {huang2014prediction}
\bibfield{author}{\bibinfo{person}{Qingjia Huang}, \bibinfo{person}{Kai
  Shuang}, \bibinfo{person}{Peng Xu}, \bibinfo{person}{Jian Li},
  \bibinfo{person}{Xu Liu}, {and} \bibinfo{person}{Sen Su}.}
  \bibinfo{year}{2014}\natexlab{}.
\newblock \showarticletitle{Prediction-based dynamic resource scheduling for
  virtualized cloud systems}.
\newblock \bibinfo{journal}{\emph{Journal of Networks}} \bibinfo{volume}{9},
  \bibinfo{number}{2} (\bibinfo{year}{2014}), \bibinfo{pages}{375}.
\newblock


\bibitem[\protect\citeauthoryear{Huawei}{Huawei}{2020}]%
        {huaweiwhitepaper}
\bibfield{author}{\bibinfo{person}{Huawei}.} \bibinfo{year}{2020}\natexlab{}.
\newblock \bibinfo{title}{White Paper: 5G Network Architecture - A High-Level
  Perspective}.
\newblock
\newblock
\urldef\tempurl%
\url{https://www.huawei.com/en/technology-insights/industry-insights/outlook/mobile-broadband/insights-reports/5g-network-architecture}
\showURL{%
\tempurl}


\bibitem[\protect\citeauthoryear{Jain, Kumar, Mandal, Ong, Poutievski, Singh,
  Venkata, Wanderer, Zhou, Zhu, et~al\mbox{.}}{Jain et~al\mbox{.}}{2013}]%
        {jain2013b4}
\bibfield{author}{\bibinfo{person}{Sushant Jain}, \bibinfo{person}{Alok Kumar},
  \bibinfo{person}{Subhasree Mandal}, \bibinfo{person}{Joon Ong},
  \bibinfo{person}{Leon Poutievski}, \bibinfo{person}{Arjun Singh},
  \bibinfo{person}{Subbaiah Venkata}, \bibinfo{person}{Jim Wanderer},
  \bibinfo{person}{Junlan Zhou}, \bibinfo{person}{Min Zhu}, {et~al\mbox{.}}}
  \bibinfo{year}{2013}\natexlab{}.
\newblock \showarticletitle{B4: Experience with a globally-deployed software
  defined WAN}.
\newblock \bibinfo{journal}{\emph{ACM SIGCOMM Computer Communication Review}}
  \bibinfo{volume}{43}, \bibinfo{number}{4} (\bibinfo{year}{2013}),
  \bibinfo{pages}{3--14}.
\newblock


\bibitem[\protect\citeauthoryear{Jennings and Stadler}{Jennings and
  Stadler}{2015}]%
        {jennings2015resource}
\bibfield{author}{\bibinfo{person}{Brendan Jennings} {and}
  \bibinfo{person}{Rolf Stadler}.} \bibinfo{year}{2015}\natexlab{}.
\newblock \showarticletitle{Resource management in clouds: Survey and research
  challenges}.
\newblock \bibinfo{journal}{\emph{Journal of Network and Systems Management}}
  \bibinfo{volume}{23}, \bibinfo{number}{3} (\bibinfo{year}{2015}),
  \bibinfo{pages}{567--619}.
\newblock


\bibitem[\protect\citeauthoryear{Jo, Cho, and Egger}{Jo et~al\mbox{.}}{2017}]%
        {jo2017machine}
\bibfield{author}{\bibinfo{person}{Changyeon Jo}, \bibinfo{person}{Youngsu
  Cho}, {and} \bibinfo{person}{Bernhard Egger}.}
  \bibinfo{year}{2017}\natexlab{}.
\newblock \showarticletitle{A machine learning approach to live migration
  modeling}. In \bibinfo{booktitle}{\emph{Proceedings of the 2017 Symposium on
  Cloud Computing}}. \bibinfo{pages}{351--364}.
\newblock


\bibitem[\protect\citeauthoryear{Joy}{Joy}{2015}]%
        {joy2015performance}
\bibfield{author}{\bibinfo{person}{Ann~Mary Joy}.}
  \bibinfo{year}{2015}\natexlab{}.
\newblock \showarticletitle{Performance comparison between linux containers and
  virtual machines}. In \bibinfo{booktitle}{\emph{2015 International Conference
  on Advances in Computer Engineering and Applications}}. IEEE,
  \bibinfo{pages}{342--346}.
\newblock


\bibitem[\protect\citeauthoryear{Kakadia, Kopri, and Varma}{Kakadia
  et~al\mbox{.}}{2013}]%
        {kakadia2013network}
\bibfield{author}{\bibinfo{person}{Dharmesh Kakadia}, \bibinfo{person}{Nandish
  Kopri}, {and} \bibinfo{person}{Vasudeva Varma}.}
  \bibinfo{year}{2013}\natexlab{}.
\newblock \showarticletitle{Network-aware virtual machine consolidation for
  large data centers}. In \bibinfo{booktitle}{\emph{Proceedings of the Third
  International Workshop on Network-Aware Data Management}}.
  \bibinfo{pages}{1--8}.
\newblock


\bibitem[\protect\citeauthoryear{Kang, Tsugawa, Matsunaga, Hirofuchi, and
  Fortes}{Kang et~al\mbox{.}}{2014}]%
        {kang2014design}
\bibfield{author}{\bibinfo{person}{Tae~Seung Kang},
  \bibinfo{person}{Maur{\'\i}cio Tsugawa}, \bibinfo{person}{Andr{\'e}a
  Matsunaga}, \bibinfo{person}{Takahiro Hirofuchi}, {and}
  \bibinfo{person}{Jos{\'e}~AB Fortes}.} \bibinfo{year}{2014}\natexlab{}.
\newblock \showarticletitle{Design and implementation of middleware for cloud
  disaster recovery via virtual machine migration management}. In
  \bibinfo{booktitle}{\emph{2014 IEEE/ACM 7th International Conference on
  Utility and Cloud Computing}}. IEEE, \bibinfo{pages}{166--175}.
\newblock


\bibitem[\protect\citeauthoryear{Khan, Paplinski, Khan, Murshed, and
  Buyya}{Khan et~al\mbox{.}}{2018}]%
        {khan2018dynamic}
\bibfield{author}{\bibinfo{person}{Md~Anit Khan}, \bibinfo{person}{Andrew
  Paplinski}, \bibinfo{person}{Abdul~Malik Khan}, \bibinfo{person}{Manzur
  Murshed}, {and} \bibinfo{person}{Rajkumar Buyya}.}
  \bibinfo{year}{2018}\natexlab{}.
\newblock \showarticletitle{Dynamic virtual machine consolidation algorithms
  for energy-efficient cloud resource management: a review}.
\newblock \bibinfo{journal}{\emph{Sustainable cloud and energy services}}
  (\bibinfo{year}{2018}), \bibinfo{pages}{135--165}.
\newblock


\bibitem[\protect\citeauthoryear{Kikuchi and Matsumoto}{Kikuchi and
  Matsumoto}{2011}]%
        {kikuchi2011performance}
\bibfield{author}{\bibinfo{person}{Shinji Kikuchi} {and}
  \bibinfo{person}{Yasuhide Matsumoto}.} \bibinfo{year}{2011}\natexlab{}.
\newblock \showarticletitle{Performance modeling of concurrent live migration
  operations in cloud computing systems using prism probabilistic model
  checker}. In \bibinfo{booktitle}{\emph{2011 IEEE 4th International Conference
  on Cloud Computing}}. IEEE, \bibinfo{pages}{49--56}.
\newblock


\bibitem[\protect\citeauthoryear{Kikuchi and Matsumoto}{Kikuchi and
  Matsumoto}{2012}]%
        {kikuchi2012impact}
\bibfield{author}{\bibinfo{person}{Shinji Kikuchi} {and}
  \bibinfo{person}{Yasuhide Matsumoto}.} \bibinfo{year}{2012}\natexlab{}.
\newblock \showarticletitle{Impact of live migration on multi-tier application
  performance in clouds}. In \bibinfo{booktitle}{\emph{2012 IEEE Fifth
  International Conference on Cloud Computing}}. IEEE,
  \bibinfo{pages}{261--268}.
\newblock


\bibitem[\protect\citeauthoryear{Kim, Yu, Liu, Zhu, Padhye, Raindel, Guo,
  Sekar, and Seshan}{Kim et~al\mbox{.}}{2019}]%
        {kim2019freeflow}
\bibfield{author}{\bibinfo{person}{Daehyeok Kim}, \bibinfo{person}{Tianlong
  Yu}, \bibinfo{person}{Hongqiang~Harry Liu}, \bibinfo{person}{Yibo Zhu},
  \bibinfo{person}{Jitu Padhye}, \bibinfo{person}{Shachar Raindel},
  \bibinfo{person}{Chuanxiong Guo}, \bibinfo{person}{Vyas Sekar}, {and}
  \bibinfo{person}{Srinivasan Seshan}.} \bibinfo{year}{2019}\natexlab{}.
\newblock \showarticletitle{FreeFlow: Software-based Virtual $\{$RDMA$\}$
  Networking for Containerized Clouds}. In \bibinfo{booktitle}{\emph{16th
  $\{$USENIX$\}$ Symposium on Networked Systems Design and Implementation
  ($\{$NSDI$\}$ 19)}}. \bibinfo{pages}{113--126}.
\newblock


\bibitem[\protect\citeauthoryear{Kim and Feamster}{Kim and Feamster}{2013}]%
        {kim2013improving}
\bibfield{author}{\bibinfo{person}{Hyojoon Kim} {and} \bibinfo{person}{Nick
  Feamster}.} \bibinfo{year}{2013}\natexlab{}.
\newblock \showarticletitle{Improving network management with software defined
  networking}.
\newblock \bibinfo{journal}{\emph{IEEE Communications Magazine}}
  \bibinfo{volume}{51}, \bibinfo{number}{2} (\bibinfo{year}{2013}),
  \bibinfo{pages}{114--119}.
\newblock


\bibitem[\protect\citeauthoryear{Kirkpatrick}{Kirkpatrick}{2013}]%
        {kirkpatrick2013software}
\bibfield{author}{\bibinfo{person}{Keith Kirkpatrick}.}
  \bibinfo{year}{2013}\natexlab{}.
\newblock \showarticletitle{Software-defined networking}.
\newblock \bibinfo{journal}{\emph{Commun. ACM}} \bibinfo{volume}{56},
  \bibinfo{number}{9} (\bibinfo{year}{2013}), \bibinfo{pages}{16--19}.
\newblock


\bibitem[\protect\citeauthoryear{Klein and Jarschel}{Klein and
  Jarschel}{2013}]%
        {klein2013openflow}
\bibfield{author}{\bibinfo{person}{Dominik Klein} {and}
  \bibinfo{person}{Michael Jarschel}.} \bibinfo{year}{2013}\natexlab{}.
\newblock \showarticletitle{An OpenFlow extension for the OMNeT++ INET
  framework}. In \bibinfo{booktitle}{\emph{Proceedings of the 6th International
  ICST Conference on Simulation Tools and Techniques}}.
  \bibinfo{pages}{322--329}.
\newblock


\bibitem[\protect\citeauthoryear{Kokkinos, Kalogeras, Levin, and
  Varvarigos}{Kokkinos et~al\mbox{.}}{2016}]%
        {kokkinos2016survey}
\bibfield{author}{\bibinfo{person}{Panagiotis Kokkinos},
  \bibinfo{person}{Dimitris Kalogeras}, \bibinfo{person}{Anna Levin}, {and}
  \bibinfo{person}{Emmanouel Varvarigos}.} \bibinfo{year}{2016}\natexlab{}.
\newblock \showarticletitle{Survey: Live migration and disaster recovery over
  long-distance networks}.
\newblock \bibinfo{journal}{\emph{ACM Computing Surveys (CSUR)}}
  \bibinfo{volume}{49}, \bibinfo{number}{2} (\bibinfo{year}{2016}),
  \bibinfo{pages}{1--36}.
\newblock


\bibitem[\protect\citeauthoryear{Li and Kanso}{Li and Kanso}{2015}]%
        {li2015comparing}
\bibfield{author}{\bibinfo{person}{Wubin Li} {and} \bibinfo{person}{Ali
  Kanso}.} \bibinfo{year}{2015}\natexlab{}.
\newblock \showarticletitle{Comparing containers versus virtual machines for
  achieving high availability}. In \bibinfo{booktitle}{\emph{2015 IEEE
  International Conference on Cloud Engineering}}. IEEE,
  \bibinfo{pages}{353--358}.
\newblock


\bibitem[\protect\citeauthoryear{Li, Garraghan, Jiang, Wu, and Xu}{Li
  et~al\mbox{.}}{2017}]%
        {li2017holistic}
\bibfield{author}{\bibinfo{person}{Xiang Li}, \bibinfo{person}{Peter
  Garraghan}, \bibinfo{person}{Xiaohong Jiang}, \bibinfo{person}{Zhaohui Wu},
  {and} \bibinfo{person}{Jie Xu}.} \bibinfo{year}{2017}\natexlab{}.
\newblock \showarticletitle{Holistic virtual machine scheduling in cloud
  datacenters towards minimizing total energy}.
\newblock \bibinfo{journal}{\emph{IEEE Transactions on Parallel and Distributed
  Systems}} \bibinfo{volume}{29}, \bibinfo{number}{6} (\bibinfo{year}{2017}),
  \bibinfo{pages}{1317--1331}.
\newblock


\bibitem[\protect\citeauthoryear{Li and Wu}{Li and Wu}{2016}]%
        {li2016optimizing}
\bibfield{author}{\bibinfo{person}{Ziyu Li} {and} \bibinfo{person}{Gang Wu}.}
  \bibinfo{year}{2016}\natexlab{}.
\newblock \showarticletitle{Optimizing VM live migration strategy based on
  migration time cost modeling}. In \bibinfo{booktitle}{\emph{Proceedings of
  the 2016 Symposium on Architectures for Networking and Communications
  Systems}}. \bibinfo{pages}{99--109}.
\newblock


\bibitem[\protect\citeauthoryear{Liu and He}{Liu and He}{2014}]%
        {liu2014vmbuddies}
\bibfield{author}{\bibinfo{person}{Haikun Liu} {and} \bibinfo{person}{Bingsheng
  He}.} \bibinfo{year}{2014}\natexlab{}.
\newblock \showarticletitle{Vmbuddies: Coordinating live migration of
  multi-tier applications in cloud environments}.
\newblock \bibinfo{journal}{\emph{IEEE transactions on parallel and distributed
  systems}} \bibinfo{volume}{26}, \bibinfo{number}{4} (\bibinfo{year}{2014}),
  \bibinfo{pages}{1192--1205}.
\newblock


\bibitem[\protect\citeauthoryear{Liu, Jin, Liao, Hu, and Yu}{Liu
  et~al\mbox{.}}{2009}]%
        {liu2009live}
\bibfield{author}{\bibinfo{person}{Haikun Liu}, \bibinfo{person}{Hai Jin},
  \bibinfo{person}{Xiaofei Liao}, \bibinfo{person}{Liting Hu}, {and}
  \bibinfo{person}{Chen Yu}.} \bibinfo{year}{2009}\natexlab{}.
\newblock \showarticletitle{Live migration of virtual machine based on full
  system trace and replay}. In \bibinfo{booktitle}{\emph{Proceedings of the
  18th ACM international symposium on High performance distributed computing}}.
  \bibinfo{pages}{101--110}.
\newblock


\bibitem[\protect\citeauthoryear{Liu, Jin, Xu, and Liao}{Liu
  et~al\mbox{.}}{2013}]%
        {liu2013performance}
\bibfield{author}{\bibinfo{person}{Haikun Liu}, \bibinfo{person}{Hai Jin},
  \bibinfo{person}{Cheng-Zhong Xu}, {and} \bibinfo{person}{Xiaofei Liao}.}
  \bibinfo{year}{2013}\natexlab{}.
\newblock \showarticletitle{Performance and energy modeling for live migration
  of virtual machines}.
\newblock \bibinfo{journal}{\emph{Cluster computing}} \bibinfo{volume}{16},
  \bibinfo{number}{2} (\bibinfo{year}{2013}), \bibinfo{pages}{249--264}.
\newblock


\bibitem[\protect\citeauthoryear{Lopes, Higashino, Capretz, and
  Bittencourt}{Lopes et~al\mbox{.}}{2017}]%
        {lopes2017myifogsim}
\bibfield{author}{\bibinfo{person}{M{\'a}rcio~Moraes Lopes},
  \bibinfo{person}{Wilson~A Higashino}, \bibinfo{person}{Miriam~AM Capretz},
  {and} \bibinfo{person}{Luiz~Fernando Bittencourt}.}
  \bibinfo{year}{2017}\natexlab{}.
\newblock \showarticletitle{Myifogsim: A simulator for virtual machine
  migration in fog computing}. In \bibinfo{booktitle}{\emph{Companion
  Proceedings of the10th International Conference on Utility and Cloud
  Computing}}. \bibinfo{pages}{47--52}.
\newblock


\bibitem[\protect\citeauthoryear{Lu, Xu, Cheng, Kompella, and Xu}{Lu
  et~al\mbox{.}}{2015}]%
        {lu2015vhaul}
\bibfield{author}{\bibinfo{person}{Hui Lu}, \bibinfo{person}{Cong Xu},
  \bibinfo{person}{Cheng Cheng}, \bibinfo{person}{Ramana Kompella}, {and}
  \bibinfo{person}{Dongyan Xu}.} \bibinfo{year}{2015}\natexlab{}.
\newblock \showarticletitle{vhaul: Towards optimal scheduling of live multi-vm
  migration for multi-tier applications}. In \bibinfo{booktitle}{\emph{2015
  IEEE 8th International Conference on Cloud Computing}}. IEEE,
  \bibinfo{pages}{453--460}.
\newblock


\bibitem[\protect\citeauthoryear{Lu, Stuart, Tang, and He}{Lu
  et~al\mbox{.}}{2014}]%
        {lu2014clique}
\bibfield{author}{\bibinfo{person}{Tao Lu}, \bibinfo{person}{Morgan Stuart},
  \bibinfo{person}{Kun Tang}, {and} \bibinfo{person}{Xubin He}.}
  \bibinfo{year}{2014}\natexlab{}.
\newblock \showarticletitle{Clique migration: Affinity grouping of virtual
  machines for inter-cloud live migration}. In \bibinfo{booktitle}{\emph{2014
  9th IEEE International Conference on Networking, Architecture, and Storage}}.
  IEEE, \bibinfo{pages}{216--225}.
\newblock


\bibitem[\protect\citeauthoryear{Ma, Yi, Carter, and Li}{Ma
  et~al\mbox{.}}{2018}]%
        {ma2018efficient}
\bibfield{author}{\bibinfo{person}{Lele Ma}, \bibinfo{person}{Shanhe Yi},
  \bibinfo{person}{Nancy Carter}, {and} \bibinfo{person}{Qun Li}.}
  \bibinfo{year}{2018}\natexlab{}.
\newblock \showarticletitle{Efficient live migration of edge services
  leveraging container layered storage}.
\newblock \bibinfo{journal}{\emph{IEEE Transactions on Mobile Computing}}
  \bibinfo{volume}{18}, \bibinfo{number}{9} (\bibinfo{year}{2018}),
  \bibinfo{pages}{2020--2033}.
\newblock


\bibitem[\protect\citeauthoryear{Machen, Wang, Leung, Ko, and Salonidis}{Machen
  et~al\mbox{.}}{2017}]%
        {machen2017live}
\bibfield{author}{\bibinfo{person}{Andrew Machen}, \bibinfo{person}{Shiqiang
  Wang}, \bibinfo{person}{Kin~K Leung}, \bibinfo{person}{Bong~Jun Ko}, {and}
  \bibinfo{person}{Theodoros Salonidis}.} \bibinfo{year}{2017}\natexlab{}.
\newblock \showarticletitle{Live service migration in mobile edge clouds}.
\newblock \bibinfo{journal}{\emph{IEEE Wireless Communications}}
  \bibinfo{volume}{25}, \bibinfo{number}{1} (\bibinfo{year}{2017}),
  \bibinfo{pages}{140--147}.
\newblock


\bibitem[\protect\citeauthoryear{Mann, Gupta, Dutta, Vishnoi, Bhattacharya,
  Poddar, and Iyer}{Mann et~al\mbox{.}}{2012}]%
        {mann2012remedy}
\bibfield{author}{\bibinfo{person}{Vijay Mann}, \bibinfo{person}{Akanksha
  Gupta}, \bibinfo{person}{Partha Dutta}, \bibinfo{person}{Anilkumar Vishnoi},
  \bibinfo{person}{Parantapa Bhattacharya}, \bibinfo{person}{Rishabh Poddar},
  {and} \bibinfo{person}{Aakash Iyer}.} \bibinfo{year}{2012}\natexlab{}.
\newblock \showarticletitle{Remedy: Network-aware steady state VM management
  for data centers}. In \bibinfo{booktitle}{\emph{International Conference on
  Research in Networking}}. Springer, \bibinfo{pages}{190--204}.
\newblock


\bibitem[\protect\citeauthoryear{Manvi and Shyam}{Manvi and Shyam}{2014}]%
        {manvi2014resource}
\bibfield{author}{\bibinfo{person}{Sunilkumar~S Manvi} {and}
  \bibinfo{person}{Gopal~Krishna Shyam}.} \bibinfo{year}{2014}\natexlab{}.
\newblock \showarticletitle{Resource management for Infrastructure as a Service
  (IaaS) in cloud computing: A survey}.
\newblock \bibinfo{journal}{\emph{Journal of network and computer
  applications}}  \bibinfo{volume}{41} (\bibinfo{year}{2014}),
  \bibinfo{pages}{424--440}.
\newblock


\bibitem[\protect\citeauthoryear{Marmol and Tucker}{Marmol and Tucker}{2018}]%
        {borg-criu}
\bibfield{author}{\bibinfo{person}{Victor Marmol} {and} \bibinfo{person}{Andy
  Tucker}.} \bibinfo{year}{2018}\natexlab{}.
\newblock \showarticletitle{Task Migration at Scale Using CRIU}. In
  \bibinfo{booktitle}{\emph{Linux Plumbers Conference}}.
\newblock


\bibitem[\protect\citeauthoryear{Marston, Li, Bandyopadhyay, Zhang, and
  Ghalsasi}{Marston et~al\mbox{.}}{2011}]%
        {marston2011cloud}
\bibfield{author}{\bibinfo{person}{Sean Marston}, \bibinfo{person}{Zhi Li},
  \bibinfo{person}{Subhajyoti Bandyopadhyay}, \bibinfo{person}{Juheng Zhang},
  {and} \bibinfo{person}{Anand Ghalsasi}.} \bibinfo{year}{2011}\natexlab{}.
\newblock \showarticletitle{Cloud computing—The business perspective}.
\newblock \bibinfo{journal}{\emph{Decision support systems}}
  \bibinfo{volume}{51}, \bibinfo{number}{1} (\bibinfo{year}{2011}),
  \bibinfo{pages}{176--189}.
\newblock


\bibitem[\protect\citeauthoryear{Martini, Adami, Gharbaoui, Castoldi, Donatini,
  and Giordano}{Martini et~al\mbox{.}}{2016}]%
        {martini2016design}
\bibfield{author}{\bibinfo{person}{Barbara Martini}, \bibinfo{person}{Davide
  Adami}, \bibinfo{person}{Molka Gharbaoui}, \bibinfo{person}{Piero Castoldi},
  \bibinfo{person}{Lisa Donatini}, {and} \bibinfo{person}{Stefano Giordano}.}
  \bibinfo{year}{2016}\natexlab{}.
\newblock \showarticletitle{Design and evaluation of SDN-based orchestration
  system for cloud data centers}. In \bibinfo{booktitle}{\emph{2016 IEEE
  International Conference on Communications (ICC)}}. IEEE,
  \bibinfo{pages}{1--6}.
\newblock


\bibitem[\protect\citeauthoryear{Mayoral, Vilalta, Mu{\~n}oz, Casellas, and
  Mart{\'\i}nez}{Mayoral et~al\mbox{.}}{2017}]%
        {mayoral2017sdn}
\bibfield{author}{\bibinfo{person}{Arturo Mayoral}, \bibinfo{person}{Ricard
  Vilalta}, \bibinfo{person}{Raul Mu{\~n}oz}, \bibinfo{person}{Ramon Casellas},
  {and} \bibinfo{person}{Ricardo Mart{\'\i}nez}.}
  \bibinfo{year}{2017}\natexlab{}.
\newblock \showarticletitle{SDN orchestration architectures and their
  integration with cloud computing applications}.
\newblock \bibinfo{journal}{\emph{Optical Switching and Networking}}
  \bibinfo{volume}{26} (\bibinfo{year}{2017}), \bibinfo{pages}{2--13}.
\newblock


\bibitem[\protect\citeauthoryear{Medina and Garc{\'\i}a}{Medina and
  Garc{\'\i}a}{2014}]%
        {medina2014survey}
\bibfield{author}{\bibinfo{person}{Violeta Medina} {and}
  \bibinfo{person}{Juan~Manuel Garc{\'\i}a}.} \bibinfo{year}{2014}\natexlab{}.
\newblock \showarticletitle{A survey of migration mechanisms of virtual
  machines}.
\newblock \bibinfo{journal}{\emph{ACM Computing Surveys (CSUR)}}
  \bibinfo{volume}{46}, \bibinfo{number}{3} (\bibinfo{year}{2014}),
  \bibinfo{pages}{1--33}.
\newblock


\bibitem[\protect\citeauthoryear{Merkel}{Merkel}{2014}]%
        {merkel2014docker}
\bibfield{author}{\bibinfo{person}{Dirk Merkel}.}
  \bibinfo{year}{2014}\natexlab{}.
\newblock \showarticletitle{Docker: lightweight linux containers for consistent
  development and deployment}.
\newblock \bibinfo{journal}{\emph{Linux journal}} \bibinfo{volume}{2014},
  \bibinfo{number}{239} (\bibinfo{year}{2014}), \bibinfo{pages}{2}.
\newblock


\bibitem[\protect\citeauthoryear{Mirkin, Kuznetsov, and Kolyshkin}{Mirkin
  et~al\mbox{.}}{2008}]%
        {mirkin2008containers}
\bibfield{author}{\bibinfo{person}{Andrey Mirkin}, \bibinfo{person}{Alexey
  Kuznetsov}, {and} \bibinfo{person}{Kir Kolyshkin}.}
  \bibinfo{year}{2008}\natexlab{}.
\newblock \showarticletitle{Containers checkpointing and live migration}. In
  \bibinfo{booktitle}{\emph{Proceedings of the Linux Symposium}},
  Vol.~\bibinfo{volume}{2}. \bibinfo{pages}{85--90}.
\newblock


\bibitem[\protect\citeauthoryear{Nadgowda, Suneja, Bila, and Isci}{Nadgowda
  et~al\mbox{.}}{2017}]%
        {nadgowda2017voyager}
\bibfield{author}{\bibinfo{person}{Shripad Nadgowda}, \bibinfo{person}{Sahil
  Suneja}, \bibinfo{person}{Nilton Bila}, {and} \bibinfo{person}{Canturk
  Isci}.} \bibinfo{year}{2017}\natexlab{}.
\newblock \showarticletitle{Voyager: Complete container state migration}. In
  \bibinfo{booktitle}{\emph{2017 IEEE 37th International Conference on
  Distributed Computing Systems (ICDCS)}}. IEEE, \bibinfo{pages}{2137--2142}.
\newblock


\bibitem[\protect\citeauthoryear{Nagin, Hadas, Dubitzky, Glikson, Loy,
  Rochwerger, and Schour}{Nagin et~al\mbox{.}}{2011}]%
        {nagin2011inter}
\bibfield{author}{\bibinfo{person}{Kenneth Nagin}, \bibinfo{person}{David
  Hadas}, \bibinfo{person}{Zvi Dubitzky}, \bibinfo{person}{Alex Glikson},
  \bibinfo{person}{Irit Loy}, \bibinfo{person}{Benny Rochwerger}, {and}
  \bibinfo{person}{Liran Schour}.} \bibinfo{year}{2011}\natexlab{}.
\newblock \showarticletitle{Inter-cloud mobility of virtual machines}. In
  \bibinfo{booktitle}{\emph{Proceedings of the 4th Annual International
  Conference on Systems and Storage}}. \bibinfo{pages}{1--12}.
\newblock


\bibitem[\protect\citeauthoryear{Noshy, Ibrahim, and Ali}{Noshy
  et~al\mbox{.}}{2018}]%
        {noshy2018optimization}
\bibfield{author}{\bibinfo{person}{Mostafa Noshy}, \bibinfo{person}{Abdelhameed
  Ibrahim}, {and} \bibinfo{person}{Hesham~Arafat Ali}.}
  \bibinfo{year}{2018}\natexlab{}.
\newblock \showarticletitle{Optimization of live virtual machine migration in
  cloud computing: A survey and future directions}.
\newblock \bibinfo{journal}{\emph{Journal of Network and Computer
  Applications}}  \bibinfo{volume}{110} (\bibinfo{year}{2018}),
  \bibinfo{pages}{1--10}.
\newblock


\bibitem[\protect\citeauthoryear{Piao and Yan}{Piao and Yan}{2010}]%
        {piao2010network}
\bibfield{author}{\bibinfo{person}{Jing~Tai Piao} {and} \bibinfo{person}{Jun
  Yan}.} \bibinfo{year}{2010}\natexlab{}.
\newblock \showarticletitle{A network-aware virtual machine placement and
  migration approach in cloud computing}. In \bibinfo{booktitle}{\emph{2010
  Ninth International Conference on Grid and Cloud Computing}}. IEEE,
  \bibinfo{pages}{87--92}.
\newblock


\bibitem[\protect\citeauthoryear{Puliafito, Goncalves, Lopes, Martins, Madeira,
  Mingozzi, Rana, and Bittencourt}{Puliafito et~al\mbox{.}}{2020}]%
        {puliafito2020mobfogsim}
\bibfield{author}{\bibinfo{person}{Carlo Puliafito}, \bibinfo{person}{Diogo~M
  Goncalves}, \bibinfo{person}{Marcio~M Lopes}, \bibinfo{person}{Leonardo~L
  Martins}, \bibinfo{person}{Edmundo Madeira}, \bibinfo{person}{Enzo Mingozzi},
  \bibinfo{person}{Omer Rana}, {and} \bibinfo{person}{Luiz~F Bittencourt}.}
  \bibinfo{year}{2020}\natexlab{}.
\newblock \showarticletitle{MobFogSim: Simulation of mobility and migration for
  fog computing}.
\newblock \bibinfo{journal}{\emph{Simulation Modelling Practice and Theory}}
  \bibinfo{volume}{101} (\bibinfo{year}{2020}), \bibinfo{pages}{102062}.
\newblock


\bibitem[\protect\citeauthoryear{Qiu, Lung, Ajila, and Srivastava}{Qiu
  et~al\mbox{.}}{2019}]%
        {qiu2019experimental}
\bibfield{author}{\bibinfo{person}{Yuqing Qiu}, \bibinfo{person}{Chung-Horng
  Lung}, \bibinfo{person}{Samuel Ajila}, {and} \bibinfo{person}{Pradeep
  Srivastava}.} \bibinfo{year}{2019}\natexlab{}.
\newblock \showarticletitle{Experimental evaluation of LXC container migration
  for cloudlets using multipath TCP}.
\newblock \bibinfo{journal}{\emph{Computer Networks}}  \bibinfo{volume}{164}
  (\bibinfo{year}{2019}), \bibinfo{pages}{106900}.
\newblock


\bibitem[\protect\citeauthoryear{Rejiba, Masip-Bruin, and
  Mar{\'\i}n-Tordera}{Rejiba et~al\mbox{.}}{2019}]%
        {rejiba2019survey}
\bibfield{author}{\bibinfo{person}{Zeineb Rejiba}, \bibinfo{person}{Xavier
  Masip-Bruin}, {and} \bibinfo{person}{Eva Mar{\'\i}n-Tordera}.}
  \bibinfo{year}{2019}\natexlab{}.
\newblock \showarticletitle{A survey on mobility-induced service migration in
  the fog, edge, and related computing paradigms}.
\newblock \bibinfo{journal}{\emph{ACM Computing Surveys (CSUR)}}
  \bibinfo{volume}{52}, \bibinfo{number}{5} (\bibinfo{year}{2019}),
  \bibinfo{pages}{1--33}.
\newblock


\bibitem[\protect\citeauthoryear{Ruprecht, Jones, Shiraev, Harmon, Spivak,
  Krebs, Baker-Harvey, and Sanderson}{Ruprecht et~al\mbox{.}}{2018}]%
        {ruprecht2018vm}
\bibfield{author}{\bibinfo{person}{Adam Ruprecht}, \bibinfo{person}{Danny
  Jones}, \bibinfo{person}{Dmitry Shiraev}, \bibinfo{person}{Greg Harmon},
  \bibinfo{person}{Maya Spivak}, \bibinfo{person}{Michael Krebs},
  \bibinfo{person}{Miche Baker-Harvey}, {and} \bibinfo{person}{Tyler
  Sanderson}.} \bibinfo{year}{2018}\natexlab{}.
\newblock \showarticletitle{Vm live migration at scale}.
\newblock \bibinfo{journal}{\emph{ACM SIGPLAN Notices}} \bibinfo{volume}{53},
  \bibinfo{number}{3} (\bibinfo{year}{2018}), \bibinfo{pages}{45--56}.
\newblock


\bibitem[\protect\citeauthoryear{Rybina, Patni, and Schill}{Rybina
  et~al\mbox{.}}{2014}]%
        {rybina2014analysing}
\bibfield{author}{\bibinfo{person}{Kateryna Rybina},
  \bibinfo{person}{Abhinandan Patni}, {and} \bibinfo{person}{Alexander
  Schill}.} \bibinfo{year}{2014}\natexlab{}.
\newblock \showarticletitle{Analysing the Migration Time of Live Migration of
  Multiple Virtual Machines.}
\newblock \bibinfo{journal}{\emph{CLOSER}}  \bibinfo{volume}{14}
  (\bibinfo{year}{2014}), \bibinfo{pages}{590--597}.
\newblock


\bibitem[\protect\citeauthoryear{Sahni and Varma}{Sahni and Varma}{2012}]%
        {sahni2012hybrid}
\bibfield{author}{\bibinfo{person}{Shashank Sahni} {and}
  \bibinfo{person}{Vasudeva Varma}.} \bibinfo{year}{2012}\natexlab{}.
\newblock \showarticletitle{A hybrid approach to live migration of virtual
  machines}. In \bibinfo{booktitle}{\emph{2012 IEEE International Conference on
  Cloud Computing in Emerging Markets (CCEM)}}. IEEE, \bibinfo{pages}{1--5}.
\newblock


\bibitem[\protect\citeauthoryear{Sarker and Tang}{Sarker and Tang}{2013}]%
        {sarker2013performance}
\bibfield{author}{\bibinfo{person}{Tusher~Kumer Sarker} {and}
  \bibinfo{person}{Maolin Tang}.} \bibinfo{year}{2013}\natexlab{}.
\newblock \showarticletitle{Performance-driven live migration of multiple
  virtual machines in datacenters}. In \bibinfo{booktitle}{\emph{2013 IEEE
  international conference on granular computing (GrC)}}. IEEE,
  \bibinfo{pages}{253--258}.
\newblock


\bibitem[\protect\citeauthoryear{Shetty, Anala, and Shobha}{Shetty
  et~al\mbox{.}}{2012}]%
        {shetty2012survey}
\bibfield{author}{\bibinfo{person}{Jyoti Shetty}, \bibinfo{person}{MR Anala},
  {and} \bibinfo{person}{G Shobha}.} \bibinfo{year}{2012}\natexlab{}.
\newblock \showarticletitle{A survey on techniques of secure live migration of
  virtual machine}.
\newblock \bibinfo{journal}{\emph{International Journal of Computer
  Applications}} \bibinfo{volume}{39}, \bibinfo{number}{12}
  (\bibinfo{year}{2012}), \bibinfo{pages}{34--39}.
\newblock


\bibitem[\protect\citeauthoryear{Shi and Shen}{Shi and Shen}{2019}]%
        {shi2019memory}
\bibfield{author}{\bibinfo{person}{Bin Shi} {and} \bibinfo{person}{Haiying
  Shen}.} \bibinfo{year}{2019}\natexlab{}.
\newblock \showarticletitle{Memory/disk operation aware lightweight vm live
  migration across data-centers with low performance impact}. In
  \bibinfo{booktitle}{\emph{IEEE INFOCOM 2019-IEEE Conference on Computer
  Communications}}. IEEE, \bibinfo{pages}{334--342}.
\newblock


\bibitem[\protect\citeauthoryear{Shi, Cao, Zhang, Li, and Xu}{Shi
  et~al\mbox{.}}{2016}]%
        {shi2016edge}
\bibfield{author}{\bibinfo{person}{Weisong Shi}, \bibinfo{person}{Jie Cao},
  \bibinfo{person}{Quan Zhang}, \bibinfo{person}{Youhuizi Li}, {and}
  \bibinfo{person}{Lanyu Xu}.} \bibinfo{year}{2016}\natexlab{}.
\newblock \showarticletitle{Edge computing: Vision and challenges}.
\newblock \bibinfo{journal}{\emph{IEEE internet of things journal}}
  \bibinfo{volume}{3}, \bibinfo{number}{5} (\bibinfo{year}{2016}),
  \bibinfo{pages}{637--646}.
\newblock


\bibitem[\protect\citeauthoryear{Shribman and Hudzia}{Shribman and
  Hudzia}{2012}]%
        {shribman2012pre}
\bibfield{author}{\bibinfo{person}{Aidan Shribman} {and}
  \bibinfo{person}{Benoit Hudzia}.} \bibinfo{year}{2012}\natexlab{}.
\newblock \showarticletitle{Pre-copy and post-copy vm live migration for memory
  intensive applications}. In \bibinfo{booktitle}{\emph{European Conference on
  Parallel Processing}}. Springer, \bibinfo{pages}{539--547}.
\newblock


\bibitem[\protect\citeauthoryear{Singh, Korupolu, and Mohapatra}{Singh
  et~al\mbox{.}}{2008}]%
        {singh2008server}
\bibfield{author}{\bibinfo{person}{Aameek Singh}, \bibinfo{person}{Madhukar
  Korupolu}, {and} \bibinfo{person}{Dushmanta Mohapatra}.}
  \bibinfo{year}{2008}\natexlab{}.
\newblock \showarticletitle{Server-storage virtualization: integration and load
  balancing in data centers}. In \bibinfo{booktitle}{\emph{SC'08: Proceedings
  of the 2008 ACM/IEEE conference on Supercomputing}}. IEEE,
  \bibinfo{pages}{1--12}.
\newblock


\bibitem[\protect\citeauthoryear{Son and Buyya}{Son and Buyya}{2018}]%
        {son2018sdcon}
\bibfield{author}{\bibinfo{person}{Jungmin Son} {and} \bibinfo{person}{Rajkumar
  Buyya}.} \bibinfo{year}{2018}\natexlab{}.
\newblock \showarticletitle{SDCon: Integrated control platform for
  software-defined clouds}.
\newblock \bibinfo{journal}{\emph{IEEE Transactions on Parallel and Distributed
  Systems}} \bibinfo{volume}{30}, \bibinfo{number}{1} (\bibinfo{year}{2018}),
  \bibinfo{pages}{230--244}.
\newblock


\bibitem[\protect\citeauthoryear{Son, Dastjerdi, Calheiros, Ji, Yoon, and
  Buyya}{Son et~al\mbox{.}}{2015}]%
        {son2015cloudsimsdn}
\bibfield{author}{\bibinfo{person}{Jungmin Son}, \bibinfo{person}{Amir~Vahid
  Dastjerdi}, \bibinfo{person}{Rodrigo~N Calheiros}, \bibinfo{person}{Xiaohui
  Ji}, \bibinfo{person}{Young Yoon}, {and} \bibinfo{person}{Rajkumar Buyya}.}
  \bibinfo{year}{2015}\natexlab{}.
\newblock \showarticletitle{Cloudsimsdn: Modeling and simulation of
  software-defined cloud data centers}. In \bibinfo{booktitle}{\emph{2015 15th
  IEEE/ACM International Symposium on Cluster, Cloud and Grid Computing}}.
  IEEE, \bibinfo{pages}{475--484}.
\newblock


\bibitem[\protect\citeauthoryear{Son, He, and Buyya}{Son et~al\mbox{.}}{2019}]%
        {son2019cloudsimsdn}
\bibfield{author}{\bibinfo{person}{Jungmin Son}, \bibinfo{person}{TianZhang
  He}, {and} \bibinfo{person}{Rajkumar Buyya}.}
  \bibinfo{year}{2019}\natexlab{}.
\newblock \showarticletitle{CloudSimSDN-NFV: Modeling and simulation of network
  function virtualization and service function chaining in edge computing
  environments}.
\newblock \bibinfo{journal}{\emph{Software: Practice and Experience}}
  \bibinfo{volume}{49}, \bibinfo{number}{12} (\bibinfo{year}{2019}),
  \bibinfo{pages}{1748--1764}.
\newblock


\bibitem[\protect\citeauthoryear{Stoyanov and Kollingbaum}{Stoyanov and
  Kollingbaum}{2018}]%
        {stoyanov2018efficient}
\bibfield{author}{\bibinfo{person}{Radostin Stoyanov} {and}
  \bibinfo{person}{Martin~J Kollingbaum}.} \bibinfo{year}{2018}\natexlab{}.
\newblock \showarticletitle{Efficient live migration of linux containers}. In
  \bibinfo{booktitle}{\emph{International Conference on High Performance
  Computing}}. Springer, \bibinfo{pages}{184--193}.
\newblock


\bibitem[\protect\citeauthoryear{Strunk}{Strunk}{2012}]%
        {strunk2012costs}
\bibfield{author}{\bibinfo{person}{Anja Strunk}.}
  \bibinfo{year}{2012}\natexlab{}.
\newblock \showarticletitle{Costs of virtual machine live migration: A survey}.
  In \bibinfo{booktitle}{\emph{2012 IEEE Eighth World Congress on Services}}.
  IEEE, \bibinfo{pages}{323--329}.
\newblock


\bibitem[\protect\citeauthoryear{Strunk}{Strunk}{2013}]%
        {strunk2013lightweight}
\bibfield{author}{\bibinfo{person}{Anja Strunk}.}
  \bibinfo{year}{2013}\natexlab{}.
\newblock \showarticletitle{A lightweight model for estimating energy cost of
  live migration of virtual machines}. In \bibinfo{booktitle}{\emph{2013 IEEE
  Sixth International Conference on Cloud Computing}}. IEEE,
  \bibinfo{pages}{510--517}.
\newblock


\bibitem[\protect\citeauthoryear{Strunk and Dargie}{Strunk and Dargie}{2013}]%
        {strunk2013does}
\bibfield{author}{\bibinfo{person}{Anja Strunk} {and}
  \bibinfo{person}{Waltenegus Dargie}.} \bibinfo{year}{2013}\natexlab{}.
\newblock \showarticletitle{Does live migration of virtual machines cost
  energy?}. In \bibinfo{booktitle}{\emph{2013 IEEE 27th International
  Conference on Advanced Information Networking and Applications (AINA)}}.
  IEEE, \bibinfo{pages}{514--521}.
\newblock


\bibitem[\protect\citeauthoryear{Sun, Liao, Anand, Zhao, and Yu}{Sun
  et~al\mbox{.}}{2016}]%
        {sun2016new}
\bibfield{author}{\bibinfo{person}{Gang Sun}, \bibinfo{person}{Dan Liao},
  \bibinfo{person}{Vishal Anand}, \bibinfo{person}{Dongcheng Zhao}, {and}
  \bibinfo{person}{Hongfang Yu}.} \bibinfo{year}{2016}\natexlab{}.
\newblock \showarticletitle{A new technique for efficient live migration of
  multiple virtual machines}.
\newblock \bibinfo{journal}{\emph{Future Generation Computer Systems}}
  \bibinfo{volume}{55} (\bibinfo{year}{2016}), \bibinfo{pages}{74--86}.
\newblock


\bibitem[\protect\citeauthoryear{Toosi, Calheiros, and Buyya}{Toosi
  et~al\mbox{.}}{2014}]%
        {toosi2014interconnected}
\bibfield{author}{\bibinfo{person}{Adel~Nadjaran Toosi},
  \bibinfo{person}{Rodrigo~N Calheiros}, {and} \bibinfo{person}{Rajkumar
  Buyya}.} \bibinfo{year}{2014}\natexlab{}.
\newblock \showarticletitle{Interconnected cloud computing environments:
  Challenges, taxonomy, and survey}.
\newblock \bibinfo{journal}{\emph{ACM Computing Surveys (CSUR)}}
  \bibinfo{volume}{47}, \bibinfo{number}{1} (\bibinfo{year}{2014}),
  \bibinfo{pages}{1--47}.
\newblock


\bibitem[\protect\citeauthoryear{Tsakalozos, Verroios, Roussopoulos, and
  Delis}{Tsakalozos et~al\mbox{.}}{2017}]%
        {tsakalozos2017live}
\bibfield{author}{\bibinfo{person}{Konstantinos Tsakalozos},
  \bibinfo{person}{Vasilis Verroios}, \bibinfo{person}{Mema Roussopoulos},
  {and} \bibinfo{person}{Alex Delis}.} \bibinfo{year}{2017}\natexlab{}.
\newblock \showarticletitle{Live VM migration under time-constraints in
  share-nothing IaaS-clouds}.
\newblock \bibinfo{journal}{\emph{IEEE Transactions on Parallel and Distributed
  Systems}} \bibinfo{volume}{28}, \bibinfo{number}{8} (\bibinfo{year}{2017}),
  \bibinfo{pages}{2285--2298}.
\newblock


\bibitem[\protect\citeauthoryear{Tso, Hamilton, Oikonomou, and Pezaros}{Tso
  et~al\mbox{.}}{2013}]%
        {tso2013implementing}
\bibfield{author}{\bibinfo{person}{Fung~Po Tso}, \bibinfo{person}{Gregg
  Hamilton}, \bibinfo{person}{Konstantinos Oikonomou}, {and}
  \bibinfo{person}{Dimitrios~P Pezaros}.} \bibinfo{year}{2013}\natexlab{}.
\newblock \showarticletitle{Implementing scalable, network-aware virtual
  machine migration for cloud data centers}. In \bibinfo{booktitle}{\emph{2013
  IEEE Sixth International Conference on Cloud Computing}}. IEEE,
  \bibinfo{pages}{557--564}.
\newblock


\bibitem[\protect\citeauthoryear{Tsugawa, Figueiredo, Fortes, Hirofuchi,
  Nakada, and Takano}{Tsugawa et~al\mbox{.}}{2012}]%
        {tsugawa2012use}
\bibfield{author}{\bibinfo{person}{Mauricio Tsugawa}, \bibinfo{person}{Renato
  Figueiredo}, \bibinfo{person}{Jose Fortes}, \bibinfo{person}{Takahiro
  Hirofuchi}, \bibinfo{person}{Hidemoto Nakada}, {and} \bibinfo{person}{Ryousei
  Takano}.} \bibinfo{year}{2012}\natexlab{}.
\newblock \showarticletitle{On the use of virtualization technologies to
  support uninterrupted IT services: A case study with lessons learned from the
  Great East Japan Earthquake}. In \bibinfo{booktitle}{\emph{2012 IEEE
  International Conference on Communications (ICC)}}. IEEE,
  \bibinfo{pages}{6324--6328}.
\newblock


\bibitem[\protect\citeauthoryear{Verma, Ahuja, and Neogi}{Verma
  et~al\mbox{.}}{2008}]%
        {verma2008pmapper}
\bibfield{author}{\bibinfo{person}{Akshat Verma}, \bibinfo{person}{Puneet
  Ahuja}, {and} \bibinfo{person}{Anindya Neogi}.}
  \bibinfo{year}{2008}\natexlab{}.
\newblock \showarticletitle{pMapper: power and migration cost aware application
  placement in virtualized systems}. In \bibinfo{booktitle}{\emph{Proceedings
  of the 9th ACM/IFIP/USENIX International Conference on Middleware
  (Middleware'08)}}. Springer, \bibinfo{pages}{243--264}.
\newblock


\bibitem[\protect\citeauthoryear{Verma, Pedrosa, Korupolu, Oppenheimer, Tune,
  and Wilkes}{Verma et~al\mbox{.}}{2015}]%
        {verma2015large}
\bibfield{author}{\bibinfo{person}{Abhishek Verma}, \bibinfo{person}{Luis
  Pedrosa}, \bibinfo{person}{Madhukar Korupolu}, \bibinfo{person}{David
  Oppenheimer}, \bibinfo{person}{Eric Tune}, {and} \bibinfo{person}{John
  Wilkes}.} \bibinfo{year}{2015}\natexlab{}.
\newblock \showarticletitle{Large-scale cluster management at Google with
  Borg}. In \bibinfo{booktitle}{\emph{Proceedings of the Tenth European
  Conference on Computer Systems}}. \bibinfo{pages}{1--17}.
\newblock


\bibitem[\protect\citeauthoryear{Voorsluys, Broberg, Venugopal, and
  Buyya}{Voorsluys et~al\mbox{.}}{2009}]%
        {voorsluys2009cost}
\bibfield{author}{\bibinfo{person}{William Voorsluys}, \bibinfo{person}{James
  Broberg}, \bibinfo{person}{Srikumar Venugopal}, {and}
  \bibinfo{person}{Rajkumar Buyya}.} \bibinfo{year}{2009}\natexlab{}.
\newblock \showarticletitle{Cost of virtual machine live migration in clouds: A
  performance evaluation}. In \bibinfo{booktitle}{\emph{IEEE International
  Conference on Cloud Computing}}. Springer, \bibinfo{pages}{254--265}.
\newblock


\bibitem[\protect\citeauthoryear{{Wang}, {Li}, {Zhang}, and {Jin}}{{Wang}
  et~al\mbox{.}}{2019}]%
        {wang2017virtual}
\bibfield{author}{\bibinfo{person}{H. {Wang}}, \bibinfo{person}{Y. {Li}},
  \bibinfo{person}{Y. {Zhang}}, {and} \bibinfo{person}{D. {Jin}}.}
  \bibinfo{year}{2019}\natexlab{}.
\newblock \showarticletitle{Virtual Machine Migration Planning in
  Software-Defined Networks}.
\newblock \bibinfo{journal}{\emph{IEEE Transactions on Cloud Computing}}
  \bibinfo{volume}{7}, \bibinfo{number}{4} (\bibinfo{year}{2019}),
  \bibinfo{pages}{1168--1182}.
\newblock


\bibitem[\protect\citeauthoryear{Wang, Xu, Zhang, and Liu}{Wang
  et~al\mbox{.}}{2018}]%
        {wang2018survey}
\bibfield{author}{\bibinfo{person}{Shangguang Wang}, \bibinfo{person}{Jinliang
  Xu}, \bibinfo{person}{Ning Zhang}, {and} \bibinfo{person}{Yujiong Liu}.}
  \bibinfo{year}{2018}\natexlab{}.
\newblock \showarticletitle{A survey on service migration in mobile edge
  computing}.
\newblock \bibinfo{journal}{\emph{IEEE Access}}  \bibinfo{volume}{6}
  (\bibinfo{year}{2018}), \bibinfo{pages}{23511--23528}.
\newblock


\bibitem[\protect\citeauthoryear{Wang, Sun, Xue, Qian, Li, and Li}{Wang
  et~al\mbox{.}}{2019}]%
        {wang2019ada}
\bibfield{author}{\bibinfo{person}{Zhong Wang}, \bibinfo{person}{Daniel Sun},
  \bibinfo{person}{Guangtao Xue}, \bibinfo{person}{Shiyou Qian},
  \bibinfo{person}{Guoqiang Li}, {and} \bibinfo{person}{Minglu Li}.}
  \bibinfo{year}{2019}\natexlab{}.
\newblock \showarticletitle{Ada-Things: An adaptive virtual machine monitoring
  and migration strategy for internet of things applications}.
\newblock \bibinfo{journal}{\emph{J. Parallel and Distrib. Comput.}}
  \bibinfo{volume}{132} (\bibinfo{year}{2019}), \bibinfo{pages}{164--176}.
\newblock


\bibitem[\protect\citeauthoryear{Wang, Xue, Qian, Liu, Li, Cao, and Yu}{Wang
  et~al\mbox{.}}{2017}]%
        {wang2017ada}
\bibfield{author}{\bibinfo{person}{Zhong Wang}, \bibinfo{person}{Guangtao Xue},
  \bibinfo{person}{Shiyou Qian}, \bibinfo{person}{Gongshen Liu},
  \bibinfo{person}{Minglu Li}, \bibinfo{person}{Jian Cao}, {and}
  \bibinfo{person}{Jiadi Yu}.} \bibinfo{year}{2017}\natexlab{}.
\newblock \showarticletitle{Ada-copy: An Adaptive Memory Copy Strategy for
  Virtual Machine Live Migration}. In \bibinfo{booktitle}{\emph{2017 IEEE 23rd
  International Conference on Parallel and Distributed Systems (ICPADS)}}.
  IEEE, \bibinfo{pages}{461--468}.
\newblock


\bibitem[\protect\citeauthoryear{Wei, Lin, and Kong}{Wei et~al\mbox{.}}{2011}]%
        {wei2011energy}
\bibfield{author}{\bibinfo{person}{Bing Wei}, \bibinfo{person}{Chuang Lin},
  {and} \bibinfo{person}{Xiangzhen Kong}.} \bibinfo{year}{2011}\natexlab{}.
\newblock \showarticletitle{Energy optimized modeling for live migration in
  virtual data center}. In \bibinfo{booktitle}{\emph{Proceedings of 2011
  International Conference on Computer Science and Network Technology}},
  Vol.~\bibinfo{volume}{4}. IEEE, \bibinfo{pages}{2311--2315}.
\newblock


\bibitem[\protect\citeauthoryear{Witanto, Lim, and Atiquzzaman}{Witanto
  et~al\mbox{.}}{2018}]%
        {witanto2018adaptive}
\bibfield{author}{\bibinfo{person}{Joseph~Nathanael Witanto},
  \bibinfo{person}{Hyotaek Lim}, {and} \bibinfo{person}{Mohammed Atiquzzaman}.}
  \bibinfo{year}{2018}\natexlab{}.
\newblock \showarticletitle{Adaptive selection of dynamic VM consolidation
  algorithm using neural network for cloud resource management}.
\newblock \bibinfo{journal}{\emph{Future generation computer systems}}
  \bibinfo{volume}{87} (\bibinfo{year}{2018}), \bibinfo{pages}{35--42}.
\newblock


\bibitem[\protect\citeauthoryear{Wood, Ramakrishnan, Shenoy, Van~der Merwe,
  Hwang, Liu, and Chaufournier}{Wood et~al\mbox{.}}{2014}]%
        {wood2014cloudnet}
\bibfield{author}{\bibinfo{person}{Timothy Wood}, \bibinfo{person}{KK
  Ramakrishnan}, \bibinfo{person}{Prashant Shenoy}, \bibinfo{person}{Jacobus
  Van~der Merwe}, \bibinfo{person}{Jinho Hwang}, \bibinfo{person}{Guyue Liu},
  {and} \bibinfo{person}{Lucas Chaufournier}.} \bibinfo{year}{2014}\natexlab{}.
\newblock \showarticletitle{CloudNet: Dynamic pooling of cloud resources by
  live WAN migration of virtual machines}.
\newblock \bibinfo{journal}{\emph{IEEE/ACM Transactions On Networking}}
  \bibinfo{volume}{23}, \bibinfo{number}{5} (\bibinfo{year}{2014}),
  \bibinfo{pages}{1568--1583}.
\newblock


\bibitem[\protect\citeauthoryear{Wood, Shenoy, Venkataramani, and Yousif}{Wood
  et~al\mbox{.}}{2009}]%
        {wood2009sandpiper}
\bibfield{author}{\bibinfo{person}{Timothy Wood}, \bibinfo{person}{Prashant
  Shenoy}, \bibinfo{person}{Arun Venkataramani}, {and} \bibinfo{person}{Mazin
  Yousif}.} \bibinfo{year}{2009}\natexlab{}.
\newblock \showarticletitle{Sandpiper: Black-box and gray-box resource
  management for virtual machines}.
\newblock \bibinfo{journal}{\emph{Computer Networks}} \bibinfo{volume}{53},
  \bibinfo{number}{17} (\bibinfo{year}{2009}), \bibinfo{pages}{2923--2938}.
\newblock


\bibitem[\protect\citeauthoryear{Xia, Wen, Foh, Niyato, and Xie}{Xia
  et~al\mbox{.}}{2014}]%
        {xia2014survey}
\bibfield{author}{\bibinfo{person}{Wenfeng Xia}, \bibinfo{person}{Yonggang
  Wen}, \bibinfo{person}{Chuan~Heng Foh}, \bibinfo{person}{Dusit Niyato}, {and}
  \bibinfo{person}{Haiyong Xie}.} \bibinfo{year}{2014}\natexlab{}.
\newblock \showarticletitle{A survey on software-defined networking}.
\newblock \bibinfo{journal}{\emph{IEEE Communications Surveys \& Tutorials}}
  \bibinfo{volume}{17}, \bibinfo{number}{1} (\bibinfo{year}{2014}),
  \bibinfo{pages}{27--51}.
\newblock


\bibitem[\protect\citeauthoryear{Xiao, Song, and Chen}{Xiao
  et~al\mbox{.}}{2012}]%
        {xiao2012dynamic}
\bibfield{author}{\bibinfo{person}{Zhen Xiao}, \bibinfo{person}{Weijia Song},
  {and} \bibinfo{person}{Qi Chen}.} \bibinfo{year}{2012}\natexlab{}.
\newblock \showarticletitle{Dynamic resource allocation using virtual machines
  for cloud computing environment}.
\newblock \bibinfo{journal}{\emph{IEEE transactions on parallel and distributed
  systems}} \bibinfo{volume}{24}, \bibinfo{number}{6} (\bibinfo{year}{2012}),
  \bibinfo{pages}{1107--1117}.
\newblock


\bibitem[\protect\citeauthoryear{Xu, Liu, Jin, and Vasilakos}{Xu
  et~al\mbox{.}}{2013a}]%
        {xu2013managing}
\bibfield{author}{\bibinfo{person}{Fei Xu}, \bibinfo{person}{Fangming Liu},
  \bibinfo{person}{Hai Jin}, {and} \bibinfo{person}{Athanasios~V Vasilakos}.}
  \bibinfo{year}{2013}\natexlab{a}.
\newblock \showarticletitle{Managing performance overhead of virtual machines
  in cloud computing: A survey, state of the art, and future directions}.
\newblock \bibinfo{journal}{\emph{Proc. IEEE}} \bibinfo{volume}{102},
  \bibinfo{number}{1} (\bibinfo{year}{2013}), \bibinfo{pages}{11--31}.
\newblock


\bibitem[\protect\citeauthoryear{Xu, Liu, Liu, Jin, Li, and Li}{Xu
  et~al\mbox{.}}{2013b}]%
        {xu2013iaware}
\bibfield{author}{\bibinfo{person}{Fei Xu}, \bibinfo{person}{Fangming Liu},
  \bibinfo{person}{Linghui Liu}, \bibinfo{person}{Hai Jin}, \bibinfo{person}{Bo
  Li}, {and} \bibinfo{person}{Baochun Li}.} \bibinfo{year}{2013}\natexlab{b}.
\newblock \showarticletitle{iAware: Making live migration of virtual machines
  interference-aware in the cloud}.
\newblock \bibinfo{journal}{\emph{IEEE Trans. Comput.}} \bibinfo{volume}{63},
  \bibinfo{number}{12} (\bibinfo{year}{2013}), \bibinfo{pages}{3012--3025}.
\newblock


\bibitem[\protect\citeauthoryear{Yamada}{Yamada}{2016}]%
        {yamada2016survey}
\bibfield{author}{\bibinfo{person}{Hiroshi Yamada}.}
  \bibinfo{year}{2016}\natexlab{}.
\newblock \showarticletitle{Survey on mechanisms for live virtual machine
  migration and its improvements}.
\newblock \bibinfo{journal}{\emph{Information and Media Technologies}}
  \bibinfo{volume}{11} (\bibinfo{year}{2016}), \bibinfo{pages}{101--115}.
\newblock


\bibitem[\protect\citeauthoryear{Ye, Jiang, Huang, Chen, and Wang}{Ye
  et~al\mbox{.}}{2011}]%
        {ye2011live}
\bibfield{author}{\bibinfo{person}{Kejiang Ye}, \bibinfo{person}{Xiaohong
  Jiang}, \bibinfo{person}{Dawei Huang}, \bibinfo{person}{Jianhai Chen}, {and}
  \bibinfo{person}{Bei Wang}.} \bibinfo{year}{2011}\natexlab{}.
\newblock \showarticletitle{Live migration of multiple virtual machines with
  resource reservation in cloud computing environments}. In
  \bibinfo{booktitle}{\emph{2011 IEEE 4th International Conference on Cloud
  Computing}}. IEEE, \bibinfo{pages}{267--274}.
\newblock


\bibitem[\protect\citeauthoryear{Zhang, Liu, Fu, and Yahyapour}{Zhang
  et~al\mbox{.}}{2018}]%
        {zhang2018survey}
\bibfield{author}{\bibinfo{person}{Fei Zhang}, \bibinfo{person}{Guangming Liu},
  \bibinfo{person}{Xiaoming Fu}, {and} \bibinfo{person}{Ramin Yahyapour}.}
  \bibinfo{year}{2018}\natexlab{}.
\newblock \showarticletitle{A survey on virtual machine migration: Challenges,
  techniques, and open issues}.
\newblock \bibinfo{journal}{\emph{IEEE Communications Surveys \& Tutorials}}
  \bibinfo{volume}{20}, \bibinfo{number}{2} (\bibinfo{year}{2018}),
  \bibinfo{pages}{1206--1243}.
\newblock


\bibitem[\protect\citeauthoryear{Zhang, Ren, and Lin}{Zhang
  et~al\mbox{.}}{2014}]%
        {zhang2014delay}
\bibfield{author}{\bibinfo{person}{Jiao Zhang}, \bibinfo{person}{Fengyuan Ren},
  {and} \bibinfo{person}{Chuang Lin}.} \bibinfo{year}{2014}\natexlab{}.
\newblock \showarticletitle{Delay guaranteed live migration of virtual
  machines}. In \bibinfo{booktitle}{\emph{IEEE INFOCOM 2014-IEEE Conference on
  Computer Communications}}. IEEE, \bibinfo{pages}{574--582}.
\newblock


\bibitem[\protect\citeauthoryear{Zhang, Han, He, and Chen}{Zhang
  et~al\mbox{.}}{2017}]%
        {zhang2017network}
\bibfield{author}{\bibinfo{person}{Weizhe Zhang}, \bibinfo{person}{Shuo Han},
  \bibinfo{person}{Hui He}, {and} \bibinfo{person}{Huixiang Chen}.}
  \bibinfo{year}{2017}\natexlab{}.
\newblock \showarticletitle{Network-aware virtual machine migration in an
  overcommitted cloud}.
\newblock \bibinfo{journal}{\emph{Future Generation Computer Systems}}
  \bibinfo{volume}{76} (\bibinfo{year}{2017}), \bibinfo{pages}{428--442}.
\newblock


\bibitem[\protect\citeauthoryear{Zheng, Ng, Sripanidkulchai, and Liu}{Zheng
  et~al\mbox{.}}{2014}]%
        {zheng2014comma}
\bibfield{author}{\bibinfo{person}{Jie Zheng}, \bibinfo{person}{Tze Sing~Eugene
  Ng}, \bibinfo{person}{Kunwadee Sripanidkulchai}, {and}
  \bibinfo{person}{Zhaolei Liu}.} \bibinfo{year}{2014}\natexlab{}.
\newblock \showarticletitle{Comma: Coordinating the migration of multi-tier
  applications}. In \bibinfo{booktitle}{\emph{Proceedings of the 10th ACM
  SIGPLAN/SIGOPS international conference on Virtual execution environments}}.
  \bibinfo{pages}{153--164}.
\newblock


\end{thebibliography}

\end{document}